\documentclass[12pt]{article}
\usepackage[margin=0.92in]{geometry}
\usepackage[small]{titlesec}
\usepackage{natbib,amsgen,amsthm,amsmath,amstext,amsbsy,amsopn,amssymb,latexsym}
\usepackage{mathtools}
\usepackage[usenames]{color}
\usepackage{array}
\usepackage{multirow}
\usepackage{graphicx}
\usepackage{amssymb}
\usepackage{setspace}
\usepackage{hyperref}
\usepackage{bbm}
\hypersetup{colorlinks,linkcolor={blue},citecolor={blue},urlcolor={red}}  
\usepackage{bibunits}

\usepackage{float}
\usepackage{tikz}
\usepackage{subcaption} 
\usepackage{pgfplots}
\usepackage{algorithm, algorithmicx, algpseudocode}

\let\emptyset\varnothing

\newtheorem{Proposition}{Proposition}
\newtheorem{Lemma}{Lemma}

\newtheorem{Theorem}{Theorem}

\newtheorem{Remark}{Remark}

\newcommand{\B}{{{\mathbf B}}}

\newcommand{\MM}{{\mathbb{M}}}

\newcommand{\Cov}{{\bf Cov}}

\newcommand{\xnew}{w}
\newcommand{\PP}{{\mathbf{P}}}
\newcommand{\nl}{n_l}
\newcommand{\CB}{\mathbb{B}}
\newcommand{\bp}{\beta^{*}}

\newcommand{\bpt}{\beta^{*,{\rm MP}}}
\newcommand{\RR}{\mathbb{R}}
\newcommand{\R}{\mathbb{R}}
\newcommand{\E}{\mathbf{E}}

\newcommand{\cov}{{\Gamma}_{l,k}}
\newcommand{\covest}{\widehat{\Gamma}_{l,k}}
\newcommand{\cip}{\overset{p}{\to}}
\newcommand{\cid}{\overset{d}{\to}}

\newcommand{\V}{{\bf V}}
\newcommand{\T}{\mathbb{Q}}
\newcommand{\Tar}{\mathbb{T}}

\newcommand{\Zm}{Z^{[m]}}

\newcommand{\m}{[m]}
\newcommand{\Rem}{{\rm Rem}}

\newcommand{\Data}{ \mathcal{O}}
\newcommand{\err}{ {\rm err}}

\newcommand{\widesim}[2][1.5]{
 \; \mathrel{\overset{#2}{\scalebox{#1}[1]{$\sim$}}}\;
}
\DeclareMathOperator*{\argmax}{arg\,max}
\DeclareMathOperator*{\argmin}{arg\,min}

\allowdisplaybreaks[1]

\newcommand{\ignore}[1]{}

\defaultbibliography{HDRef}
\defaultbibliographystyle{chicago}

\begin{document}

\begin{titlepage}
\title{Statistical Inference for Maximin Effects:\\ Identifying Stable Associations across Multiple Studies}
\author{ Zijian Guo\thanks{Z. Guo is an associate professor at the Department of Statistics, Rutgers University. The research of Z. Guo was supported in part by the NSF DMS 1811857, 2015373 and NIH R01GM140463, R01LM013614.}}
	\date{}
	\maketitle
	\thispagestyle{empty}
	
\begin{abstract}
Integrative analysis of data from multiple sources is critical to making generalizable discoveries. Associations that are consistently observed across multiple source populations are more likely to be generalized to target populations with possible distributional shifts. In this paper, we model the heterogeneous multi-source data with multiple high-dimensional regressions and make inferences for the maximin effect (Meinshausen, B{\"u}hlmann, AoS, 43(4), 1801--1830). The maximin effect provides a measure of stable associations across multi-source data. A significant maximin effect indicates that a variable has commonly shared effects across multiple source populations, and these shared effects may be generalized to a broader set of target populations. There are challenges associated with inferring maximin effects because its point estimator can have a non-standard limiting distribution. We devise a novel sampling method to construct valid confidence intervals for maximin effects. The proposed confidence interval attains a parametric length. This sampling procedure and the related theoretical analysis are of independent interest for solving other non-standard inference problems. Using genetic data on yeast growth in multiple environments, we demonstrate that the genetic variants with significant maximin effects have generalizable effects under new environments. 


\end{abstract}

\end{titlepage}

\begin{bibunit}

\section{Introduction}


\subsection{Problem formulation}

A vital component of contemporary medical and biological research is integrating multiple studies designed to study the same scientific question. Noteworthy examples include the integration of electronic health record (EHR) data from multiple hospitals \citep{singh2021generalizability, rasmy2018study} and genetic data collected from different subpopulations or environments \citep{keys2020cross, sirugo2019missing,kraft2009replication,cai2021individual}. Synthesis of information from multiple sources enhances the model's generalizability. For instance, the associations that are consistently observed across multiple source populations are more likely to be generalized to a wide range of target populations. However, the data heterogeneity creates challenges for prediction and inference. There is a pressing need to devise practical inference tools for extracting generalizable information from heterogeneous multi-source data. 

We consider that we have access to $L$ independent training data sets $\{X^{(l)}, Y^{(l)}\}_{1\leq l\leq L}$. For $1\leq l\leq L$, we assume that the data $\{X^{(l)}_i, Y^{(l)}_i\}_{1\leq i\leq n_l}$ are i.i.d. generated following the high-dimensional model: 
\begin{equation}
Y_i^{(l)}=[X^{(l)}_{i}]^{\intercal}b^{(l)}+\epsilon^{(l)}_i \quad \text{where}\quad \E(\epsilon^{(l)}_i\mid X^{(l)}_{i})=0,
\label{eq: multi-group model}
\end{equation} 
with the outcome $Y^{(l)}_i\in \R$ and covariates $X^{(l)}_{i} \in \R^{p}$. 
To model the data heterogeneity, we allow $\{b^{(l)}\}_{1\leq l\leq L}$ and the distributions of $X^{(l)}_{i}$ and $\epsilon^{(l)}_i$ to vary with the group label $l$. Our goal is to leverage the multi-source data and construct a generalizable model for a target population. We use $\T$ to denote the distribution of the target population. The target population may have a different covariate distribution $\T_{X}$ and conditional outcome distribution $\T_{Y|X}$ from source populations. We focus on the unlabelled target population: there are no outcome observations of the target population but only covariates $X^{\T}_{i}\in \R^{p}$ for $1\leq i\leq N_{\T}$. Such unlabelled settings frequently occur in EHR analysis \citep{humbert2022strategies} or transfer learning \citep{zhuang2020comprehensive,pan2009survey}, where the outcome labels of the target population are hard to obtain due to high costs. Due to possible distributional shifts of the unlabelled target population, identification of the true $\mathbb{Q}_{Y|X}$ is generally impossible in our framework.

This paper aims to make inferences about the covariate-shift maximin effect $\bp(\T)$ defined in the following equation \eqref{eq: maximin general shift}. We generalize the definition in \citet{meinshausen2015maximin} by allowing for covariate shifts and define $\bp(\T)$ as the solution to a distributionally robust optimization problem. Particularly, we examine a wide range of target distributions that may contain the true $\T$ and define $\bp(\T)$ as a linear model guaranteeing excellent predictive performance over this class of possible target distributions. According to \citet{meinshausen2015maximin}, when the target population differs from the source populations, maximin effects provide superior predictive performance than the regression model constructed with the merged multi-source data.

\begin{figure}[htp!]
\centering
\includegraphics[scale=0.5]{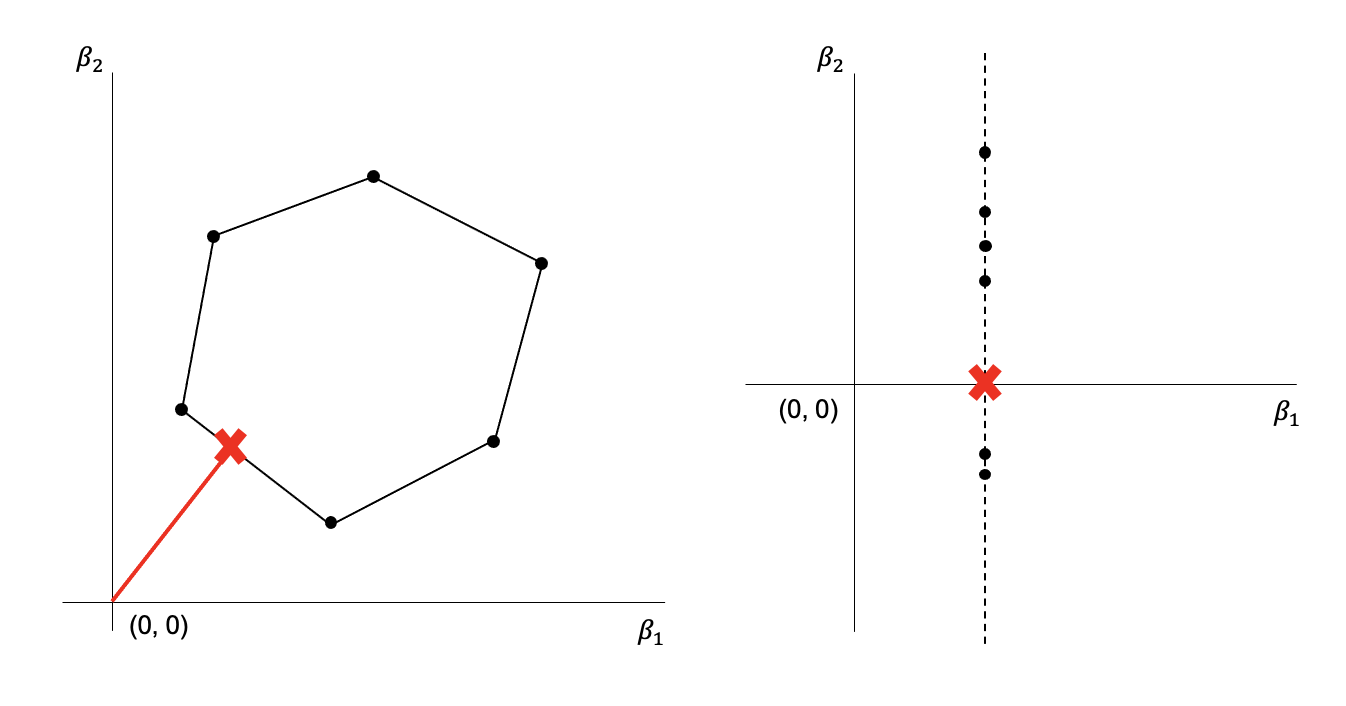}
\caption{\small Maximin effect (the red cross) for $p=2$. The left panel: $\bp(\T)$ is the convex combination of $\{b^{(l)}\}_{1\leq l\leq L}$ (black dots) having the smallest distance to the origin; the right panel: $\bp_2(\T)$ shrinks to zero when $\{b^{(l)}_2\}_{1\leq l\leq L}$ scatters around zero. The left and right figures duplicate Figure 1 in \citet{buhlmann2015magging} and Figure 2 in \citet{meinshausen2015maximin}, respectively.}
\label{fig: maximin demo}
\end{figure}

The maximin effects not only guarantee robust predictive performance over a range of target distributions but also provide a measure of stable associations shared by regression vectors $\{b^{(l)}\}_{1\leq l\leq L}$ \citep{meinshausen2015maximin}. 
Identifying variables with significant maximin effects is critical since their effects are more likely to be generalizable to new populations, even with possible distributional shifts. As shown in the following Proposition \ref{prop: identification}, $\bp(\T)$ is the convex combination of $\{b^{(l)}\}_{1\leq l\leq L}$ that has the minimum (weighted) distance to the origin; see the leftmost of Figure \ref{fig: maximin demo} for the illustration. The minimum distance ensures that $\bp(\T)$ summarizes stable associations shared across multiple source populations. As demonstrated on the rightmost of Figure \ref{fig: maximin demo}, when a variable has heterogeneous effects scattered around zero across multiple studies, its maximin effect will shrink to zero; for essential predictors with commonly shared effects across multiple data sources, the maximin effect will capture the sign of the shared effects. Moreover, the maximin effect will not be dominated by the extreme effect, only showing up in a single study. In light of the above interpretation, a significant maximin effect indicates that a predictor has commonly shared effects across various populations. 

\vspace{-3.5mm}

\vspace{-3.5mm}

Despite the importance of maximin effects, statistical inference methods for maximin effects are primarily lacking, including the construction of confidence interval (CI) and hypothesis testing. We demonstrate that the inference problem for the maximin effect is non-standard and devise a new sampling technique for solving these non-standard inference problems.

\subsection{Our results and contribution}

There are distinct challenges associated with inference for maximin effects, which occur both in low- and high-dimensional cases. Section \ref{sec: maximin challenge} illustrates several challenging settings where the maximin effect estimator may have a non-standard limiting distribution. Consequently, we cannot construct CIs for the maximin effects directly based on the asymptotic normality. We propose a novel sampling procedure to construct CIs for the maximin effects in both low and high dimensions. The main novelty is to devise a sampling method to quantify the uncertainty associated with convex weight estimation. Our proposal relies on the following intuition: after carefully sampling a large number of weight vectors, there exists at least one resampled weight vector, almost recovering the true weight vector. We provide a rigorous statement of this property in Theorem \ref{thm: sampling accuracy}. 
 Our proposed sampling CI is shown to achieve the desired coverage level and attain the parametric length.

We conduct a large-scale simulation to evaluate the finite-sample performance. When the maximin effect estimator does not have a standard limiting distribution, the CIs based on asymptotic normality, subsampling, or the m-out-of-n bootstrap undercover, but our proposed CI achieves the desired coverage; see Section \ref{sec: sim} and Section \ref{sec: non-regularity challenge} in the supplement.
In Section \ref{sec: real data}, we analyze genetic data on yeast colony growth under different growth media. The proposed inference method is compared with empirical risk minimization (ERM), which selects significant genetic variants by analyzing the merged training data. We compare our proposal and ERM by examining seven test media that were not used for training the models. The genetic variants having significant maximin effects are more generalizable to test growth media, while several genetic variants selected by ERM have no significant effects for any of these test media. 

To summarize, the contributions of the current paper are two-folded, 
\vspace{-2mm}
\begin{enumerate}
\item We propose a novel sampling approach to make inferences for maximin effects. 
The sampling method is useful for addressing other non-standard inference problems. 
\vspace{-2mm}
\item We establish the sampling property in Theorem \ref{thm: sampling accuracy} and characterize the dependence of sampling accuracy on the resampling size. The theoretical argument is new and can be of independent interest for studying other sampling methods.
\end{enumerate}

\subsection{Related works}
Distributionally robust optimization has been utilized in \citet{gao2017wasserstein,sinha2017certifying} to construct machine learning algorithms robust to the distributional shift between the training and test data. The main idea is to construct a prediction model that minimizes adversarial losses defined over a class of distributions near the source population. The current paper concerns the different settings where the prior knowledge of group information is present. When the group information is available, there has been an extensive study of the maximin effect and group distributionally robust models \citep{meinshausen2015maximin, buhlmann2015magging, sagawa2019distributionally, hu2018does}. These studies focused on estimation rather than the construction of CIs. One notable exception is that \cite{rothenhausler2016confidence} focused on the low-dimensional setting and constructed CIs for the maximin effect based on the estimator's asymptotic normality. However, we point out in Section \ref{sec: maximin challenge} that the maximin effect estimators are not necessarily asymptotically normal in challenging settings. 
The simulation results presented in Section \ref{sec: sim} demonstrate the undercoverage of CIs based on asymptotic normality.

Inference for the shared component of regression functions was considered under multiple high-dimensional linear models \citep{liu2020integrative} and partially linear models \citep{zhao2016partially}. In contrast, our proposed method does not require $\{b^{(l)}\}_{1\leq l\leq L}$ in \eqref{eq: multi-group model} to share any similarity, and our model is more flexible in modeling the heterogeneity of multi-source data. \cite{peters2016causal,rothenhausler2018anchor,arjovsky2019invariant} constructed models satisfying certain invariance principles by analyzing the heterogeneous data.

Sampling methods have a long history in statistics, such as bootstrap \citep{efron1979bootstrap,efron1994introduction}, subsampling \citep{politis1999subsampling}, generalized fiducial inference \citep{Zabell1992,xie2013confidence,Hannig2016} and repro sampling \citep{xie2022repro}. 
In contrast, instead of directly sampling from the original data, we resample the estimator of the regression covariance matrix, which makes our proposed sampling method computationally efficient; see Remark \ref{rem: efficiency}. Inference in a single high-dimensional linear model was actively investigated in the recent decade
\citep{zhang2014confidence,van2014asymptotically,javanmard2014confidence,belloni2014inference,chernozhukov2015valid,farrell2015robust,chernozhukov2018double,cai2015regci,athey2018approximate,zhu2018linear}. Inference for maximin effects has the challenge of being a non-standard inference problem, which requires novel methods and theories; see more discussions in Section \ref{sec: maximin challenge}.

\noindent {\bf Notations.} Define $n=\min_{1\leq l\leq L}\{n_{l}\}$. For $1\leq j\leq p,$ let $e_j$ denote the $j$-th Euclidean basis. We use $c$ and $C$ to denote generic positive constants that may vary from place to place.
For positive sequences $a_n$ and $b_n$, $a_n \ll b_n$ if $\limsup_{n\rightarrow\infty} {a_n}/{b_n}=0$. 
The $\ell_q$ norm of a vector $x$ is defined as $\|x\|_{q}=\left(\sum_{l=1}^{p}|x_l|^q\right)^{\frac{1}{q}}$ for $q \geq 0$ with $\|x\|_0=\left|\{1\leq l\leq p: x_l \neq 0\}\right|$ and $\|x\|_{\infty}=\max_{1\leq l \leq p}|x_l|$. For a vector $x\in \R^{p}$, a matrix $X$, and a subset $S\subset[p]$, $x_{S}$ is the sub-vector of $x$ with indices in $S$ and $X_{S}$ denotes the sub-matrix of $X$ with row indices belonging to $S.$ 
For a symmetric matrix $A\in \R^{L\times L}$ with eigendecomposition $A=U\Lambda U^{\intercal}$, we use $\lambda_{\max}(A)$ and $\lambda_{\min}(A)$ to denote its maximum and minimum eigenvalues, respectively; define $A_{+}=U\Lambda_{+} U^{\intercal}$ with $(\Lambda_{+})_{l,l}=\max\{\Lambda_{l,l},0\}$ for $1\leq l\leq L.$ For a semi-positive matrix $A$, define $A^{1/2}=U\Lambda^{1/2} U^{\intercal}$ with $(\Lambda^{1/2})_{l,l}=\sqrt{\Lambda_{l,l}}$ for $1\leq l\leq L.$ We use $A^{-1/2}$ to denote the inverse of $A^{1/2}.$
We use ${\rm vecl}(A)\in \R^{L(L+1)/2}$ to denote the vector stacking the columns of the lower triangle part of $A$. We define the one-to-one index mapping $\pi$, 
\begin{equation}
\pi(l,k)=\frac{(2L-k)(k-1)}{2}+l \quad \text{for} \quad (l,k)\in \mathcal{I}_{L}\coloneqq\{(l,k): 1\leq k\leq l\leq L\},
\label{eq: index map}
\end{equation}
which maps from the matrix index of the lower triangle part of $A$ to ${\rm vecl}(A).$ For $(l,k)\in \mathcal{I}_{L},$ we have
$[{\rm vecl}(A)]_{\pi(l,k)}=A_{l,k}.$ 

\section{Maximin Effects: Distributional Robustness and Identification} 
\label{sec: definition}

\subsection{Multi-source data setup}
\label{sec: multi-source}
We introduce the setting in the following and present the definition and identification of the covariate-shift maximin effect in Sections \ref{sec: DRO} and \ref{sec: identification}, respectively.
We consider the training data $\{X^{(l)}, Y^{(l)}\}_{1\leq l\leq L}$ collected from $L$ sources (e.g., $L$ healthcare centers). 
For $1\leq l\leq L$, let $\mathbb{P}^{(l)}_{X}$ denote the distribution of $X^{(l)}_{i}\in \R^{p}$ and $\mathbb{P}^{(l)}_{Y|X}$ denote the conditional distribution of the outcome $Y^{(l)}_{i}$ given $X^{(l)}_{i}.$ We write 
\begin{equation}
X^{(l)}_{i}{\widesim{\rm i.i.d.}} \mathbb{P}^{(l)}_{X}, \quad Y^{(l)}_{i}\mid X^{(l)}_{i}{\widesim{\rm i.i.d.}} \mathbb{P}^{(l)}_{Y|X}\quad \text{for} \quad 1\leq i\leq n_l.
\label{eq: source data no shift}
\end{equation}
Heterogeneity may exist when the multi-source data are collected for different subpopulations or under different environments. To model this, we allow $\{\mathbb{P}^{(l)}_{X},\mathbb{P}^{(l)}_{Y|X}\}_{1\leq l\leq L}$ to be different from each other. For the target population (e.g., a new healthcare center), we use $\T_{X}$ and $\T_{Y|X}$ to respectively denote the covariate and conditional outcome distribution and write
\begin{equation}
X^{\T}_{i}{\widesim{\rm i.i.d.}} \T_{X}, \quad Y^{\T}_{i}\mid X^{\T}_{i}{\widesim{\rm i.i.d.}} \T_{Y|X}\quad \text{for} \quad 1\leq i\leq N_{\T}.
\label{eq: target data no shift}
\end{equation}
We use $\{X^{(l)}_{i}\}_{1\leq l\leq L}$ and $X^{\T}_{i}$ to denote the measurement of the same set of covariates across different subpopulations or under different environments; $\{Y^{(l)}_{i}\}_{1\leq l\leq L}$ and $Y^{\T}_{i}$ to denote the measurement of the same outcome variable across different subpopulations or under different environments.  
This paper allows for the co-existence of covariate shifts and posterior drifts between the source and target populations, where the covariate shift stands for the covariate distribution $\T_{X}$ differing from any of $\{\mathbb{P}^{(l)}_{X}\}_{1\leq l\leq L}$ and the posterior drift stands for the conditional outcome distribution $\T_{Y|X}$ differing from any of $\{\mathbb{P}^{(l)}_{Y|X}\}_{1\leq l\leq L}$.  

We focus on the regime where the target population does not have outcome labels, that is, the covariates $\{X^{\T}_{i}\}_{1\leq i\leq N_{\T}}$ are observed, but the outcome labels $\{Y^{\T}_{i}\}_{1\leq i\leq N_{\T}}$ are missing. Such an unlabelled setting is common in EHR data analysis \citep{humbert2022strategies} and transfer learning applications \citep{zhuang2020comprehensive,pan2009survey}. For example, due to the high costs, a new hospital might not have the outcome labels.

\subsection{Maximin effects: generalizability via distributionally robust optimization}
\label{sec: DRO}

When the target population does not have outcome observations and $\T_{Y|X}$ is allowed to differ from any of $\{\mathbb{P}^{(l)}_{Y|X}\}_{1\leq l\leq L}$, $\T_{Y|X}$ is in general not identifiable. Instead of making inferences for the true $\T_{Y|X}$, we introduce in the following a new inference target, the covariate-shift maximin effect, as a solution to a distributionally robust optimization problem. We define the following class of joint distributions which might contain the true $\T$,
\begin{equation}
\mathcal{C}(\T_{X})\coloneqq\left\{\Tar=(\T_{X}, \Tar_{Y|X}): \Tar_{Y|X}=\sum_{l=1}^{L} q_l \cdot \mathbb{P}^{(l)}_{Y|X} \;\;\; \text{with}\;\;\; q\in \Delta^{L} \right\},
\label{eq: distribution class shift}
\end{equation}
where $\Delta^{L}=\{q\in \R^{L}:\sum_{l=1}^{L}q_{l}=1, \min_{l}q_l\geq 0\}$ denotes the $L$-dimension simplex. In \eqref{eq: distribution class shift}, the covariate distribution is fixed at $\T_{X}$ since it is identifiable with the data $\{X^{\T}_{i}\}_{1\leq i\leq N_{\T}}$; however, since the $\T_{Y|X}$ is not identifiable, we consider the conditional outcome distribution as any convex combination of $\{\mathbb{P}^{(l)}_{Y|X}\}_{1\leq l\leq L}$. When the true $\T_{Y|X}$ lies in the convex combination of $\{\mathbb{P}^{(l)}_{Y|X}\}_{1\leq l\leq L},$ the distribution class $\mathcal{C}(\T_{X})$ contains the true target population $\T=\left(\T_{X},\T_{Y|X}\right).$

We now define the maximin effect as a model optimizing the worst-case reward associated with the distribution class $\mathcal{C}(\T_{X})$. For a generic model $\beta\in \R^{p}$, the reward function $\E_{(X_i,Y_i)\sim \Tar} \left[Y_i^2-(Y_i-X_i^{\intercal}\beta)^2\right]$ measures the variance explained by $X_i^{\intercal}\beta$ when the test data $\{X_i, Y_i\}$ are generated following the distribution $\Tar$. We define the worst-case reward of the model $\beta$ as 
$R_{\T}(\beta)=\min_{\Tar\in \mathcal{C}(\T_{X})} \E_{(X_i,Y_i)\sim \Tar} \left[Y_i^2-(Y_i-X_i^{\intercal}\beta)^2\right],$ which examines every population $\Tar$ belonging to $\mathcal{C}(\T_{X}).$
The covariate-shift maximin effect $\bp(\T)$ is defined to optimize the worst-case reward, 
 \begin{equation}
\bp(\T):=\argmax_{\beta\in \R^{p}}R_{\T}(\beta) \quad \text{with}\quad R_{\T}(\beta)=\min_{\Tar\in \mathcal{C}(\T_{X})} \E_{(X_i,Y_i)\sim \Tar} \left[Y_i^2-(Y_i-X_i^{\intercal}\beta)^2\right],
\label{eq: maximin general shift}
\end{equation}
where $\mathcal{C}(\T_{X})$ is defined in \eqref{eq: distribution class shift} and $\E_{(X_i,Y_i)\sim \Tar}$ denotes the expectation with respect to the distribution $\Tar$. The definition of $\bp(\T)$ can be interpreted from a two-side game perspective \citep{meinshausen2015maximin}: we select a model $\beta$, and the counter agent searches over $\mathcal{C}(\T_{X})$ and generates the most challenging target population for this $\beta$. $\bp(\T)$ guarantees the optimal prediction accuracy for such an adversarially generated target population. 

The robust prediction model 
$\bp(\T)$ guarantees excellent predictive performance for a broad class of target populations belonging to $\mathcal{C}(\T_{X}).$ This explains the generalizability of $\bp(\T)$ since it is not designed to optimize the predictive performance for a single target population but over many possible target populations. $\bp(\T)$ is typically different from the best linear approximation derived from the true $\T_{Y|X},$ which is not identifiable under our framework.

The definition \eqref{eq: maximin general shift} falls into the general category of distributionally robust optimization \citep[e.g.]{sagawa2019distributionally, hu2018does, rothenhausler2018anchor,gao2017wasserstein,sinha2017certifying,jakobsen2022distributional}, who proposed to achieve the distributional robustness by investigating the predictive performance for a class of target distributions. We have provided a distributional robustness interpretation of the maximin effect in \citet{meinshausen2015maximin} and generalized its definition by allowing for distributional shifts among $\T_{X}$ and $\{\mathbb{P}^{(l)}_{X}\}_{1\leq l\leq L}$. When there is no covariate shift, let $\mathbb{P}_{X}$ denote the shared covariate distribution of the source and target populations. $\bp(\T)$ in \eqref{eq: maximin general shift} is equivalent to the maximin effect defined in \citet{meinshausen2015maximin}. The maximin effect in \eqref{eq: maximin general shift} can be expressed as an equivalent minimax estimator $\bp(\T):=\argmin_{\beta\in \R^{p}}\max_{\Tar\in \mathcal{C}(\T_{X})} \left\{\E_{\Tar} \ell(Y_i, X_i^{\intercal}\beta)\right\}$ with $\ell(Y_i, X_i^{\intercal} \beta)=(Y_i-X_i^{\intercal}\beta)^2- Y_i^2$,
which is in the form of the group distributionally robust optimization \citep{sagawa2019distributionally, hu2018does}.

The maximin or minimax optimizations have essential applications to minimax group fairness \citep{martinez2020minimax, diana2021minimax} and the maximin projection  \citep{shi2018maximin}. In the supplement, we provide more detailed discussions in Sections \ref{sec: minimax fairness} and \ref{sec: maximin proj}.

\begin{Remark} \rm A collection of transfer learning algorithms are designed to leverage the assumption $\T_{Y|X}\approx \mathbb{P}^{(l)}_{Y|X}$ for some $1\leq l\leq L$ and estimate $\T_{Y|X}$; see \cite{liu2020integrative,zhao2016partially,li2020transfer,tian2021transfer} for examples. In contrast, our framework does not impose such similarity conditions. Our goal is to construct a generalizable prediction model over a range of target populations instead of recovering the true $\T_{Y|X}.$ 
\end{Remark}


\subsection{Identification and interpretation of $\bp(\T)$}
\label{sec: identification}
In the following, we present the identification of $\bp(\T)$ defined in \eqref{eq: maximin general shift} and emphasize that $\bp(\T)$ summarizes the stable associations shared by multi-source data. We focus on the multiple linear models as in \eqref{eq: multi-group model} for the remaining of this paper. The linear models in \eqref{eq: multi-group model} can be extended to handle non-linear conditional expectation if $X^{(l)}_{i}$ contains the basis transformation of the covariates. Under \eqref{eq: multi-group model}, we simplify the definition of $\bp(\T)$ in \eqref{eq: maximin general shift} as
\begin{equation}
\begin{aligned}
\bp(\T)=\argmax_{\beta\in \R^{p}} R_{\T}(\beta) \quad \text{with} \quad R_{\T}(\beta)=\min_{b\in \mathbb{B}} \left[2 b^{\intercal}\Sigma^{\T}\beta- \beta^{\intercal}\Sigma^{\T} \beta\right],
\end{aligned}
\label{eq: general maximin}
\end{equation}
where $\mathbb{B}=\{b\in \R^{p}: b=\sum_{l=1}^{L}q_l\cdot b^{(l)} \; \text{with}\; q\in \Delta^{L}\}$ and $\Sigma^{\T}=\E X_{1}^{\T} (X_{1}^{\T})^{\intercal}.$ 

The following proposition shows how to identify the maximin effect $\bp(\T).$ 
\begin{Proposition} If the model \eqref{eq: multi-group model} holds and $\lambda_{\min}(\Sigma^{\T})>0$, $\bp(\T)$ defined in \eqref{eq: maximin general shift} is identified as
\begin{equation}
\bp(\T)=\sum_{l=1}^{L}[\gamma^{*}(\T)]_{l} b^{(l)} \quad \text{with}\quad \gamma^{*}(\T)\coloneqq \argmin_{\gamma \in \Delta^{L}} \gamma^{\intercal}\Gamma^{\T} \gamma
\label{eq: optimal weight}
\end{equation}
where $\Gamma_{lk}^{\T}=(b^{(l)})^{\intercal}\Sigma^{\T} b^{(k)}$ for $1\leq l,k \leq L$ and $\Delta^{L}=\{\gamma\in \RR^{L}: \gamma_j\geq0,\; \sum_{j=1}^{L}\gamma_j=1\}$ is the simplex over $\R^{L}$. Furthermore, $\max_{\beta \in \R^{p}} R_{\T}(\beta)= [\bp(\T)]^{\intercal}\Sigma^{\T}\bp(\T).$
\label{prop: identification}
\end{Proposition}
Proposition \ref{prop: identification} provides an explicit way of computing $\bp(\T)$, which is a generalization of Theorem 1 in \cite{meinshausen2015maximin}. For any $\gamma$, $\gamma^{\intercal}\Gamma^{\T} \gamma=\E \left[(X^{\T}_i)^{\intercal}\sum_{l=1}^{L} {\gamma}_l b^{(l)}\right]^2$ represents the second-order moment of the predicted values $(X^{\T}_i)^{\intercal}\sum_{l=1}^{L} {\gamma}_l b^{(l)}$ evaluated on the target population. The optimal weight is defined to minimize this second-order moment. Geometrically speaking, $\bp(\T)$ represents the point on the convex hull of $\{b^{(l)}\}_{1\leq l\leq L}$ that is closest to the origin \citep{meinshausen2015maximin}. This interpretation ensures that the maximin effect summarizes the stable associations shared by the heterogeneous regression vectors $\{b^{(l)}\}_{1\leq l\leq L}$. As illustrated in Figure \ref{fig: maximin demo}, when $\{b_j^{(l)}\}_{1\leq l\leq L}$ have different signs across different sources, $\bp_j$ will shrink to zero due to the cancelation in \eqref{eq: optimal weight}. However, if $\{b_j^{(l)}\}_{1\leq l\leq L}$ share the same sign, the convex combination in \eqref{eq: optimal weight} ensures that the maximin effect shares the same sign. When there is no confusion, we write $\Gamma^{\T}, \bp(\T),\gamma^{*}(\T)$ as $\Gamma, \bp,\gamma^{*}$, respectively.
 
It is important to conduct statistical inference for $\xnew^{\intercal}\bp$ with $\xnew$ denoting a pre-specified loading. With $w=e_j$, the test of $\xnew^{\intercal}\bp=0$ is reduced to the maximin significance test $H_{0,j}:\bp_j= 0$ for $1\leq j\leq p$, which is crucial for scientific discovery and robust prediction model construction. The maximin significance of the $j$-th covariate indicates that its effect is homogeneously positive or negative across multiple environments; moreover, it indicates that the $j$-th covariate is likely to have a similar effect for a new environment. The non-zero maximin effect also suggests that it can be helpful to include the $j$-th covariate in the prediction model for the target population. 
Additionally, with $\xnew$ denoting a future covariate observation, statistical inference for $\xnew^{\intercal}\bp$ is well motivated by constructing an optimal treatment regime with heterogeneous data \citep{shi2018maximin}. We provide more discussions in Section \ref{sec: maximin proj} in the supplement.

\begin{Remark} \rm
In the covariate shift setting, a collection of works \citep[e.g.]{tsuboi2009direct, shimodaira2000improving,sugiyama2007covariate} were focused on the misspecified conditional outcome models. In contrast, we focus on the correctly specified conditional outcome model \eqref{eq: multi-group model}. Consequently, the regression vectors $\{b^{(l)}\}_{1\leq l\leq L}$ do not change with the target population $\T_{X}$. However, the maximin effect $\bp(\T)$ changes with the target population since the weight $\gamma^{*}(\T)$ is determined by the target covariate distribution $\T_{X}$. \end{Remark}
\section{Statistical Inference Challenges: Non-regularity and Instability}
\label{sec: maximin challenge}
In the following, we demonstrate the inference challenges for the maximin effect and will devise a novel sampling approach in Section \ref{sec: method} to address these challenges. 
The inference challenges arise from that estimators of $\gamma^{*}(\T)$ and $\bp(\T)$ may have a non-standard limiting distribution. To demonstrate the challenges, we consider the special case $L=2$ and obtain the solution of \eqref{eq: optimal weight} as $\gamma^{*}(\T)=(\gamma^{*}_1,1-\gamma^{*}_1)^{\intercal}$ with
 \begin{equation}
\gamma^{*}_1=\min\left\{\max\left\{\frac{\Gamma^{\T}_{22}-\Gamma^{\T}_{12}}{\Gamma^{\T}_{11}+\Gamma^{\T}_{22}-2\Gamma^{\T}_{12}},0\right\},1\right\}.
\label{eq: true value}
\end{equation}
We construct an approximately unbiased estimator $\widehat{\Gamma}^{\T}$ for $\Gamma^{\T}$  in the following equation \eqref{eq: cov est general} and then estimate $\gamma^{*}_1$ by 
$
\widehat{\gamma}_1=\min\left\{\max\left\{\bar{\gamma}_1,0\right\},1\right\}$ with $\bar{\gamma}_1=\frac{\widehat{\Gamma}^{\T}_{22}-\widehat{\Gamma}^{\T}_{12}}{\widehat{\Gamma}^{\T}_{11}+\widehat{\Gamma}^{\T}_{22}-2\widehat{\Gamma}^{\T}_{12}}$. 

We illustrate two challenging settings where $\widehat{\gamma}_1$ may not have a standard limiting distribution even if $\widehat{\Gamma}^{\T}$ is asymptotically normal.   The first is the non-regularity setting due to the boundary effect. The estimation error $\widehat{\gamma}_1-\gamma^{*}_1$ is decomposed as a mixture distribution,
$
\sqrt{n}(\bar{\gamma}_1-{\gamma}^{*}_1)\cdot {\bf 1}\{0<\bar{\gamma}_1<1\}+ (-\sqrt{n}{\gamma}^{*}_1)\cdot {\bf 1}\{\bar{\gamma}_1\leq 0\}+ \sqrt{n}(1-{\gamma}^{*}_1)\cdot {\bf 1}\{\bar{\gamma}_1\geq 1\},
$
where the last two terms appear due to the boundary constraint $0\leq \gamma^*_1\leq 1.$ 
 It is well known that boundary constraints lead to estimators with non-standard limiting distributions; see \cite{self1987asymptotic,andrews1999estimation,drton2009likelihood} and the references therein. Similarly, the boundary effect leads to a non-standard or non-regular distribution for the corresponding maximin effect estimator. For non-regular settings due to the boundary effect, inference methods based on asymptotic normality or bootstrap fail to work \citep[e.g.]{andrews2000inconsistency}. 
 
 The second challenge is instability, which occurs when some of $\{b^{(l)}\}_{1\leq l\leq L}$ are similar to each other. For $L=2,$ if $b^{(1)}\approx b^{(2)}$, then $\Gamma^{\T}_{11}+\Gamma^{\T}_{22}-2\Gamma^{\T}_{12}$ in \eqref{eq: true value} is close to zero. It is hard to accurately estimate $\gamma^*_1$ since a small error in estimating $\Gamma^{\T}_{11}+\Gamma^{\T}_{22}-2\Gamma^{\T}_{12}$ may lead to a large error of estimating $\gamma^*_1.$ In Section \ref{sec: sim} and Section \ref{sec: non-regularity challenge} in the supplement, we illustrate that CIs assuming the asymptotic normality and by $m$ out of $n$ bootstrap or subsampling fail to provide valid inference for the maximin effect in the presence of non-regularity or instability.

\section{Sampling Inference Methods for Maximin Effects}
\label{sec: method}
We devise a novel sampling approach to make inference for $\xnew^{\intercal}\bp$ with $\xnew$ denoting the pre-specified loading vector. As an important example, $\xnew^{\intercal}\bp$ becomes $\bp_j$ with $\xnew=e_j.$ In Section \ref{sec: init estimation}, we construct the estimators $\{\widehat{\xnew^{\intercal}b^{(l)}}\}_{1\leq l\leq L}$ and $\widehat{\Gamma}^{\T}$ and employ Proposition \ref{prop: identification} to construct the point estimator of $\xnew^{\intercal}\bp$ as \begin{equation}
\widehat{\xnew^{\intercal}\bp}=\sum_{l=1}^{L} \widehat{\gamma}_{l}\cdot \widehat{\xnew^{\intercal}b^{(l)}} \quad \text{with}\quad \widehat{\gamma}:=\arg\min_{\gamma\in \Delta^{L}} 
\gamma^{\intercal}\widehat{\Gamma}^{\T} \gamma.
\label{eq: point estimator}
\end{equation}
In Section \ref{sec: sampling method}, we propose a novel sampling method to quantify the uncertainty of $\widehat{\xnew^{\intercal}\bp}$ defined in \eqref{eq: point estimator} and construct the CI for $\xnew^{\intercal}\bp$.

\subsection{Point estimation of $\xnew^{\intercal}\bp$}
\label{sec: init estimation}
The point estimator in \eqref{eq: point estimator} relies on good initial esitmators $\{\widehat{\xnew^{\intercal}b^{(l)}}\}_{1\leq l\leq L}$ and $\widehat{\Gamma}^{\T}.$
In the following, we consider both low- and high-dimensional settings and construct $\{\widehat{\xnew^{\intercal}b^{(l)}}\}_{1\leq l\leq L}$ and $\widehat{\Gamma}^{\T}$ satisfying 
\begin{equation}
(\widehat{\xnew^{\intercal}b^{(l)}}-\xnew^{\intercal}b^{(l)})/{\sqrt{\widehat{\rm V}^{(l)}_{\xnew}}}\cid \mathcal{N}(0,1), \quad {\rm vecl}(\widehat{\Gamma}^{\T}-\Gamma^{\T})\stackrel{d}{\approx}\mathcal{N}({\bf 0},\widehat{\bf V}),
\label{eq: theory dist}
\end{equation}
where ${\widehat{\rm V}^{(l)}_{\xnew}}$ and $\widehat{\bf V}$ denote the estimated covariance to be respectively specified in the following equations \eqref{eq: asymp var} and \eqref{eq: cov est}, ${\rm vecl}(\widehat{\Gamma}^{\T}-\Gamma^{\T})$ is the vector stacking the columns of the lower triangular part of the matrix $\widehat{\Gamma}^{\T}-\Gamma^{\T}$, $\cid$ stands for convergence in distribution, and $\stackrel{d}{\approx}$ stands for approximately equal in distribution. 

\vspace{-3mm}
\subsubsection{Low-dimensional setting}
We start with the low-dimensional setting and provide intuitions for the construction of $\widehat{\xnew^{\intercal}b^{(l)}}$ and $\widehat{\Gamma}^{\T}.$
For $1\leq l\leq L$, let $\widehat{b}^{(l)}_{\rm OLS}$ denote the OLS estimator computed based on $(X^{(l)}, Y^{(l)}).$ We estimate $\xnew^{\intercal}{b}^{(l)}$ and $\Gamma^{\T}$ by plugging in the OLS and sample covariance matrix. Define $$\widehat{\xnew^{\intercal}b^{(l)}}=\xnew^{\intercal}\widehat{b}^{(l)}_{\rm OLS} \quad \text{for} \quad 1\leq l\leq L$$
\begin{equation}
\widehat{\Gamma}^{\T}_{l,k}=\left[\widehat{b}^{(l)}_{\rm OLS}\right]^{\intercal} \left(\frac{1}{N_{\T}}\sum_{i=1}^{N_{\T}} X^{\T}_i [X^{\T}_i]^{\intercal}\right)\widehat{b}^{(k)}_{\rm OLS} \quad \text{for}\quad 1\leq k\leq l\leq L.
\label{eq: reg cov low-dim}
\end{equation}
 The standard regression theory guarantees that these plug-in estimators satisfy \eqref{eq: theory dist} under regularity conditions, where $
{\widehat{\rm V}^{(l)}_{\xnew}}=
{\widehat{\sigma}_{l}^2}\cdot \xnew^{\intercal}[(X^{(l)})^{\intercal}X^{(l)}]^{-1}\xnew$ with $\widehat{\sigma}_l^2=\|Y^{(l)}-X^{(l)}\widehat{b}^{(l)}_{\rm OLS}\|_2^2/(n_l-p)$ and $\widehat{\bf V}$ is presented in the following Remark \ref{rem: special case}.

\subsubsection{High-dimensional setting}
In high dimensions, $\{b^{(l)}\}_{1\leq l\leq L}$ are estimated by penalized estimators. If we simply plug in the penalized estimators, it leads to biased estimators of $\xnew^{\intercal}b^{(l)}$ and $\Gamma^{\T}$. To address this, we correct the bias of the plug-in estimators and construct debiased estimators of $\xnew^{\intercal}b^{(l)}$ and $\Gamma^{\T}$.

\medskip

\noindent{\bf The debiased estimator of $\xnew^{\intercal}{b}^{(l)}$.} For $1\leq l\leq L$, let $\widehat{b}^{(l)}$ denote the Lasso estimator \citep{tibshirani1996regression} computed based on $(X^{(l)},Y^{(l)}).$ We conduct the bias-correction step to correct the bias of the plug-in estimator $\xnew^{\intercal}\widehat{b}^{(l)}$. Particularly, we follow \cite{cai2019individualized} and construct the debiased estimator of $\xnew^{\intercal}b^{(l)}$ as
\begin{equation}
\widehat{\xnew^{\intercal}b^{(l)}}=\xnew^{\intercal}\widehat{b}^{(l)}+[\widehat{v}^{(l)}]^{\intercal}\frac{1}{\nl}(X^{(l)})^{\intercal}(Y^{(l)}-X^{(l)}\widehat{b}^{(l)}),
\label{eq: linear separate}
\end{equation}
where  $\widehat{v}^{(l)}\in \R^{p}$ is constructed as 
\begin{equation*}
\begin{aligned}
\widehat{v}^{(l)}=\argmin_{v\in \R^{p}} v^{\intercal}\frac{1}{\nl}(X^{(l)})^{\intercal}X^{(l)} v \quad \text{s.t.} \quad \max_{z\in \mathcal{F}(\xnew)}\left|\langle z, \frac{1}{\nl}(X^{(l)})^{\intercal}X^{(l)} v-\xnew\rangle\right|\leq \eta_{l}& \\
\quad \|X^{(l)} v\|_{\infty}\leq \|\xnew\|_2\tau_l&
\end{aligned}
\end{equation*}
with $\mathcal{F}(\xnew)=\{e_1,\cdots, e_{p}, \xnew/\|\xnew\|_2\}$, $\eta_{l}= c_1\|\xnew\|_2\sqrt{{\log p}/{\nl}}$, and $\tau_l=c_2 \sqrt{\log n_l}$ for some positive constants $c_1,c_2>0.$ 
\cite{cai2019individualized} has established that $\widehat{\xnew^{\intercal}b^{(l)}}$ satisfies \eqref{eq: theory dist} with 
\begin{equation}
{\widehat{\rm V}^{(l)}_{\xnew}}=
({\widehat{\sigma}_{l}^2}/{\nl^2})[\widehat{v}^{(l)}]^{\intercal}(X^{(l)})^{\intercal}X^{(l)}\widehat{v}^{(l)} \quad \text{and}\quad \widehat{\sigma}_l^2=\|Y^{(l)}-X^{(l)}\widehat{b}^{(l)}\|_2^2/n_l.
\label{eq: asymp var}
\end{equation}

\noindent{\bf The debiased estimator of ${\Gamma}^{\T}$.} We present the key idea for constructing the debiased estimator $\widehat{\Gamma}^{\T}$  by generalizing the inference methods in \citet{verzelen2018adaptive, cai2020semisupervised, guo2019group}. We will provide the full details in Section \ref{sec: high-dim Gamma} in the supplement. 
For $1\leq l\leq L$, we randomly split $(X^{(l)},Y^{(l)})$ into approximately equal-size subsamples $(X^{(l)}_{A_l},Y^{(l)}_{A_l})$ and $(X^{(l)}_{B_l},Y^{(l)}_{B_l})$, where the index sets $A_{l}$ and $B_{l}$ satisfy $A_l\cap B_l=\emptyset$, $|A_l|=\lfloor \nl/2\rfloor$ and $|B_l|=n_l-|A_l|$. 
 We randomly split $X^{\T}$ into $X_{A}^{\T}$ and $X_{B}^{\T}$, where the index sets $A$ and $B$ satisfy $A\cap B=\emptyset$, $|A|=\lfloor N_{\T}/2\rfloor$ and $|B|=N_{\T}-|A|$. For $1\leq l\leq L,$ we construct the Lasso estimator $\widehat{b}_{init}^{(l)}$  using the subsample $(Y^{(l)}_{A_l},X^{(l)}_{A_l})$. Define $\widehat{\Sigma}^{\T}=\frac{1}{|B|}\sum_{i\in B} X^{\T}_{i} (X^{\T}_{i})^{\intercal}$. 
We fix a pair of indexes $1\leq k\leq l\leq L,$ and construct the plug-in estimator $[\widehat{b}_{init}^{(l)}]^{\intercal}\widehat{\Sigma}^{\T}\widehat{b}_{init}^{(k)}$. We propose the following estimator of $\Gamma^{\T}_{l,k}$ by correcting the plug-in estimator,
{\small
\begin{equation}
{\covest}^{\T}=(\widehat{b}_{init}^{(l)})^{\intercal} \widehat{\Sigma}^{\T} \widehat{b}_{init}^{(k)}+[\widehat{u}^{(l,k)}]^{\intercal}\frac{1}{|B_l|}[X^{(l)}_{B_l}]^{\intercal}(Y^{(l)}_{B_l}-X^{(l)}_{B_l}\widehat{b}_{init}^{(l)})+[\widehat{u}^{(k,l)}]^{\intercal}\frac{1}{|B_k|}[X^{(k)}_{B_k}]^{\intercal}(Y^{(k)}_{B_k}-X^{(k)}_{B_k}\widehat{b}_{init}^{(k)})
\label{eq: cov est general}
\end{equation}}
where $\widehat{u}^{(l,k)}\in \R^{p}$ and $\widehat{u}^{(k,l)}\in \R^{p}$ are the projection directions constructed in equations \eqref{eq: constraint 1}, \eqref{eq: constraint 2} and \eqref{eq: constraint 3} in the supplement. 
We now specify the estimated covariance matrix $\widehat{\bf V}$ of the vector ${\rm vecl}(\widehat{\Gamma}^{\T})$ which stacks the columns of the lower triangular part of the matrix $\widehat{\Gamma}^{\T}.$ 
For $1\leq k_1\leq l_1\leq L$ and $1\leq k_2\leq l_2\leq L$, we estimate the covariance between $\widehat{\Gamma}^{\T}_{l_1,k_1}-{\Gamma}^{\T}_{l_1,k_1}$ and $\widehat{\Gamma}^{\T}_{l_2,k_2}-{\Gamma}^{\T}_{l_2,k_2}$ by 
\begin{equation}
\widehat{\bf V}_{\pi(l_1,k_1),\pi(l_2,k_2)}=\widehat{\bf V}^{(a)}_{\pi(l_1,k_1),\pi(l_2,k_2)}+\widehat{\bf V}^{(b)}_{\pi(l_1,k_1),\pi(l_2,k_2)},
\label{eq: cov est}
\end{equation}
where $\pi$ is defined in \eqref{eq: index map} and $\widehat{\bf V}^{(a)}_{\pi(l_1,k_1),\pi(l_2,k_2)}$ and $\widehat{\bf V}^{(b)}_{\pi(l_1,k_1),\pi(l_2,k_2)}$ measure the uncertainty of estimating the high-dimensional regression vectors $b^{(l_1)}, b^{(l_2)}, b^{(k_1)}, b^{(k_2)}$ and that of estimating $\Sigma^{\T}$, respectively. We provide their exact formula in \eqref{eq: var 1st} and \eqref{eq: var 2nd} in the supplement.  
We justify the theoretical property of $\widehat{\Gamma}^{\T}$ in Proposition \ref{prop: verification of A3} in the supplement. 

\begin{Remark}[Special cases and sampling splitting] \rm {The sample splitting is not needed for constructing $\Sigma^{\T}$ for the low-dimensional setting and the high-dimensional setting with no covariate shift. For the low-dimensional setting, we construct $\widehat{\bf V}$ by applying the no sample splitting version of \eqref{eq: cov est} with $\widehat{u}^{(l,k)}=\left(\frac{1}{n_l}\sum_{i=1}^{n_l} X^{(l)}_i [X^{(l)}_i]^{\intercal}\right)^{-1}\left(\frac{1}{N_{\T}}\sum_{i=1}^{N_{\T}} X^{\T}_i [X^{\T}_i]^{\intercal}\right)\widehat{b}^{(k)}_{\rm OLS}$. For the high-dimensional no covariate shift setting, we apply \eqref{eq: cov est general} by taking $\widehat{u}^{(l,k)}$ and $\widehat{u}^{(k,l)}$ as $\widehat{b}^{(k)}$ and $\widehat{b}^{(l)}$, respectively; see the details in Section \ref{sec: high-dim Gamma noshift} in the supplement. For the high-dimensional setting with covariate shift, the main reason of sampling splitting is to create certain independence structure between the random errors $\{\epsilon^{(l)}_{B_l},\epsilon^{(k)}_{B_k}, \widehat{\Sigma}^{\T}-\Sigma^{\T}\}$ and the projection directions $\widehat{u}^{(k,l)}, \widehat{u}^{(l,k)}$, which are constructed based on the data $X^{(l)}, Y^{(l)}_{A_l},X^{(k)}, Y^{(k)}_{A_k}$ and $X^{\T}_{A}.$  Importantly, the sample splitting is not needed for constructing $\widehat{\xnew^{\intercal}b^{(l)}}$ in \eqref{eq: linear separate} for all settings.} 
\label{rem: special case}
\end{Remark}


\subsection{Inference for $\xnew^{\intercal}\bp$: sampling and aggregation}
\label{sec: sampling method}

In this subsection, we construct CI for $\xnew^{\intercal}\bp$ by quantifying the uncertainty of the estimator $\widehat{\xnew^{\intercal}\bp}$ defined in \eqref{eq: point estimator}. As highlighted in Section \ref{sec: maximin challenge}, the main challenge is that $\widehat{\gamma}$ might not have a standard limiting distribution. To address this, we devise a novel sampling method to quantify the uncertainty of $\widehat{\gamma}$. 

We start with the intuition for the sampling method and then provide the full details right after. We sample $\{\widehat{\Gamma}^{\m}\}_{1\leq m\leq M}$ such that ${\rm vecl}(\widehat{\Gamma}^{\m}-\widehat{\Gamma}^{\T})$ approximately follows $\mathcal{N}({\bf 0},\widehat{\bf V}),$ which is the approximate distribution of ${\rm vecl}(\widehat{\Gamma}^{\T}-\Gamma^{\T})$ in \eqref{eq: theory dist}. We then construct the sampled weight vectors $\{\widehat{\gamma}^{\m}\}_{1\leq m\leq M}$ by solving the optimization problem, 
\begin{equation}
\widehat{\gamma}^{\m}=\argmin_{\gamma\in \Delta^{L}} \gamma^{\intercal} \widehat{\Gamma}^{\m}_{+} \gamma.
\label{eq: sample weight} 
\end{equation}
We show in the following Theorem \ref{thm: sampling accuracy} that with a high probability, there exists at least $1\leq m^*\leq M$ such that $\widehat{\gamma}^{[m^*]}$ is nearly the same as the true $\gamma^{*}$. Since the uncertainty of estimating $\gamma^*$ by $\widehat{\gamma}^{[m^*]}$ is almost negligible, we only need to quantify the uncertainty due to $\{\widehat{\xnew^{\intercal}b^{(l)}}\}_{1\leq l\leq L}$. 


In the following, we construct CIs for $\xnew^{\intercal}\bp$ by leveraging the estimators $\{\widehat{\xnew^{\intercal}b^{(l)}}\}_{1\leq l\leq L}$ and $\widehat{\Gamma}^{\T}$ proposed in Section \ref{sec: init estimation}. Our proposal consists of two steps.

\vspace{-3mm}
\paragraph{Step 1: Sampling the weight vectors.} 
Conditioning on the observed data, we generate i.i.d. samples $\{{\rm vecl}(\widehat{\Gamma}^{\m})\}_{1\leq m\leq M}$ as
\begin{equation}
{\rm vecl}(\widehat{\Gamma}^{\m})\sim \mathcal{N}\left({\rm vecl}(\widehat{\Gamma}^{\T}), \widehat{\V}+{d_0}/{{n}}\cdot {\bf I}\right) \quad \text{with}\quad d_0=\max\left\{\tau_0\cdot n \cdot \|\widehat{\V}\|_{\infty},1 \right\},
\label{eq: resampling}
\end{equation}
where $M$ is the sampling size (default set as $500$), $\tau_0>0$ is a positive constant (default set as $0.2$), and ${\bf I}$ is the identity matrix with a conformal dimension. The resampling in \eqref{eq: resampling} only  
specifies the lower triangular part of $\widehat{\Gamma}^{\m}$. We use symmetry to impute the upper triangular part of $\widehat{\Gamma}^{\m}$, that is, $\widehat{\Gamma}^{\m}_{l,k}=\widehat{\Gamma}^{\m}_{k,l}$ for $1\leq l<k\leq L.$  { In the resampling step \eqref{eq: resampling}, we slightly enlarge the covariance matrix $\widehat{\V}$ to $\widehat{\V}+{d_0}/{{n}}\cdot {\bf I}$, ensuring that $\widehat{\V}+{d_0}/{{n}}\cdot {\bf I}$ is positive definite even for a nearly singular $\widehat{\V}$. Since $n \cdot \|\widehat{\V}\|_{\infty}$ is of a constant order, $d_0$ is chosen at a constant level. The resampling method is effective for any positive constant $\tau_0>0$ and any sufficiently large resampling size $M$. The choice of $\tau_0$ will affect the length of our proposed CI, where a larger value of $\tau_0$ can lead to noisier resampled $\widehat{\Gamma}^{\m}$ and a longer CI. In numerical studies, we use the default values $M=500$ and $\tau_0=0.2$ and observe reliable results.} 

We further screen out a small proportion of the resampled matrices $\{\widehat{\Gamma}^{\m}\}_{1\leq m\leq M}$ if they appear on the tails of the multivariate normal distribution $\mathcal{N}\left({\rm vecl}(\widehat{\Gamma}^{\T}), \widehat{\V}+{d_0}/{{n}}\cdot {\bf I}\right).$ Particularly, we introduce the following index set $\mathbb{M}$,
 \begin{equation}
\mathbb{M}=\left\{1\leq m\leq M: \max_{1\leq k\leq l\leq L}\left|\frac{\widehat{\Gamma}^{\m}_{l,k}-\widehat{\Gamma}^{\T}_{l,k}}{\sqrt{\widehat{\V}_{\pi(l,k),\pi(l,k)}+d_0/n}}\right|\leq 1.1\cdot z_{\alpha_0/[L(L+1)]}\right\},
\label{eq: generating condition}
\end{equation}
where $z_{\alpha_0/[L(L+1)]}$ is the upper $\alpha_0/[L(L+1)]$ quantile of the standard normal distribution (default value $\alpha_0=0.01$). The index set $\mathbb{M}$ in \eqref{eq: generating condition} excludes the $m$-th resampled data if the maximum deviation between $\widehat{\Gamma}^{\m}$ and $\widehat{\Gamma}^{\T}$ exceeds the threshold level, which is  chosen to adjust for multiplicity with the Bonferroni correction. The index set $\MM$ approximately removes $\alpha_0\cdot M$ resampled data, but keeps the remaining $(1-\alpha_0)\cdot M$ resampled data. 

\vspace{-2.5mm}

\paragraph{Step 2: Aggregation.} 
For $m\in \mathbb{M}$, we use the resampled $\widehat{\Gamma}^{\m}$ to construct the sampled weight vectors $\widehat{\gamma}^{\m}$ as in \eqref{eq: sample weight}. We treat each of $\{\widehat{\gamma}^{\m}\}_{m\in \mathbb{M}}$ as being fixed and construct an interval for $\xnew^{\intercal}\bp$ by leveraging the limiting distribution of $\{\widehat{\xnew^{\intercal}b^{(l)}}\}_{1\leq l\leq L}$ in \eqref{eq: theory dist}. 
For $m \in \mathbb{M}$, we compute $
\widehat{\xnew^{\intercal}\beta}^{\m}=\sum_{l=1}^{L} \widehat{\gamma}^{\m}_{l}\cdot \widehat{\xnew^{\intercal}b^{(l)}},
$ and $
\widehat{\rm se}^{\m}(\xnew)=\sqrt{\sum_{l=1}^{L}[\widehat{\gamma}^{\m}_{l}]^{2}{\widehat{\rm V}^{(l)}_{\xnew}}
}$ with $\widehat{\rm V}^{(l)}_{\xnew}$ defined in \eqref{eq: asymp var}. Then we construct the $m$-th sampled interval as, 
\begin{equation}
{\rm Int}_{\alpha}^{\m}(\xnew)=\left(\widehat{\xnew^{\intercal}\beta}^{\m}-z_{\alpha/2} \widehat{{\rm se}}^{\m}(\xnew), \widehat{\xnew^{\intercal}\beta}^{\m}+z_{\alpha/2} \widehat{{\rm se}}^{\m}(\xnew)\right),
\label{eq: sampled interval} 
\end{equation}
with $z_{\alpha/2}$ denoting the upper $\alpha/2$ quantile of the standard normal distribution. 

\begin{figure}[H]
\centering
\includegraphics[scale=0.5]{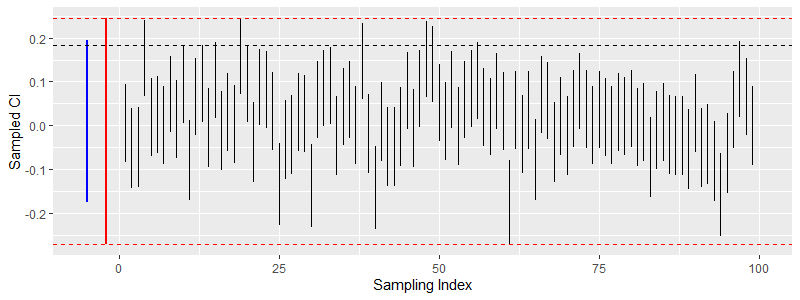}
\caption{\small Illustration of ${\rm CI}_{\alpha}\left(\xnew^{\intercal}\bp\right)$ with $M=100$ (in red) for setting 2 in Section \ref{sec: sim}. The intervals in black denote ${\rm Int}_{\alpha}^{\m}(\xnew)$ for $m\in \MM.$ The interval in blue is the oracle normality CI in \eqref{eq: normality CI}. The horizontal black dashed line represents the value of $\xnew^{\intercal}\bp.$}
\label{fig: sampling CI}
\end{figure}

 We construct the CI for $\xnew^{\intercal}\bp$ by aggregating the sampled intervals defined in \eqref{eq: sampled interval}, 
\begin{equation}
{\rm CI}_{\alpha}\left(\xnew^{\intercal}\bp\right)=\cup_{m\in \mathbb{M}} {\rm Int}_{\alpha}^{\m}(\xnew),
\label{eq: aggregation 1}
\end{equation} 
with $\mathbb{M}$ defined in \eqref{eq: generating condition} and ${\rm Int}_{\alpha}^{\m}(\xnew)$ defined in \eqref{eq: sampled interval}. In Figure \ref{fig: sampling CI}, we illustrate our proposed CI using the red interval ${\rm CI}_{\alpha}\left(\xnew^{\intercal}\bp\right)$. Note that many of $\{{\rm Int}_{\alpha}^{\m}(\xnew)\}_{m\in \MM}$ do not cover $\xnew^{\intercal}\bp$ since the uncertainty of $\widehat{\gamma}^{\m}$ is not quantified in constructing ${\rm Int}_{\alpha}^{\m}(\xnew)$.

\vspace{-3mm}
For $0<\alpha<1,$ we propose the level $\alpha$ test as $\phi_{\alpha}={\bf 1}\left\{0\not\in {\rm CI}_{\alpha}\left(\xnew^{\intercal}\bp\right)\right\}$ for the null hypothesis $H_0: \xnew^{\intercal}\bp=0.$ As an important application, we test the maximin significance of the $j$-th variable by setting $\xnew=e_j$.  For the null hypothesis $H_{0,j}: \bp_j=0$ with $1\leq j\leq p$, we follow the p-value definition in \citet{xie2013confidence} and invert ${\rm CI}_{\tau}\left(\bp_j\right)$ to compute the p-value as,
\begin{equation}
\text{p-value}\coloneqq  \min\left\{\tau\in (0,1): 0 \in {\rm CI}_{\tau}\left(\bp_j\right)\right\},
\label{eq: p-val}
\end{equation}
where ${\rm CI}_{\tau}\left(\bp_j\right)$ is defined  in \eqref{eq: aggregation 1} with $\xnew=e_j.$

We provide a few important remarks on our proposed sampling method.  
\begin{Remark}[Reasoning of sampling and screening] \rm The following decomposition reveals the effectiveness of our proposed sampling method: for any $m\in \mathbb{M},$ 
\begin{equation}
\begin{aligned}
\widehat{\xnew^{\intercal}\beta}^{\m}-\xnew^{\intercal}\bp
=\sum_{l=1}^{L} (\widehat{\gamma}^{\m}_{l}-{\gamma}^{*}_{l})\cdot\widehat{\xnew^{\intercal}b^{(l)}}+\sum_{l=1}^{L} {\gamma}^{*}_{l} \cdot (\widehat{\xnew^{\intercal}b^{(l)}}-{\xnew^{\intercal}b^{(l)}}).
\end{aligned}
\label{eq: maximin inference decomp}
\end{equation}
The following Theorem \ref{thm: sampling accuracy} shows that there exists $m^*\in \mathbb{M}$ such that ${\gamma}^{[m^*]}\approx {\gamma}^{*}$. For \eqref{eq: maximin inference decomp} with $m=m^*$, the uncertainty of $\sum_{l=1}^{L} (\widehat{\gamma}^{[m^*]}_{l}-{\gamma}^{*}_{l})\cdot\widehat{\xnew^{\intercal}b^{(l)}}$ is negligible and we just need to quantify the uncertainty of $\sum_{l=1}^{L} {\gamma}^{*}_{l} \cdot (\widehat{\xnew^{\intercal}b^{(l)}}-{\xnew^{\intercal}b^{(l)}})$. Our proposed sampling CI takes a union of sampled intervals, where each interval quantifies the uncertainty of $\{\widehat{\xnew^{\intercal}b^{(l)}}\}_{1\leq l\leq L}$ and the union step accounts for the uncertainty of $\widehat{\gamma}.$ Furthermore, we explain why the index set $\mathbb{M}$ is useful in controlling the CI length. For any sampled interval ${\rm Int}_{\alpha}^{\m}(\xnew)$, $|\widehat{\xnew^{\intercal}\beta}^{\m}-\widehat{\xnew^{\intercal}\bp}|$ measures the distance from its center to $\widehat{\xnew^{\intercal}\bp}$ and $2z_{\alpha/2} \widehat{{\rm se}}^{\m}(\xnew)$ measures its length. After taking a union, we control the interval length by 
 \begin{equation}
 \begin{aligned}
\mathbf{Leng}\left({\rm CI}_{\alpha}\left(\xnew^{\intercal}\bp\right)\right)&\leq 2\max_{m\in \MM}\left(|\widehat{\xnew^{\intercal}\beta}^{\m}-\widehat{\xnew^{\intercal}\bp}|+z_{\alpha/2} \widehat{{\rm se}}^{\m}(\xnew)\right)\\
&=2\max_{m\in \MM}\left(\left|\sum_{l=1}^{L}\left(\widehat{\gamma}^{\m}_{l}-\widehat{\gamma}_{l}\right)\cdot \widehat{\xnew^{\intercal}b^{(l)}}\right|+z_{\alpha/2} \widehat{{\rm se}}^{\m}(\xnew)\right).
\end{aligned}
\label{eq: key length intui}
\end{equation}
The index set $\mathbb{M}$ screens out the resampled $\widehat{\Gamma}^{\m}$ having a large deviation from $\widehat{\Gamma}^{\T}$, ensuring a parametric upper bound for $\max_{m\in \MM}\|\widehat{\Gamma}^{\m}-\widehat{\Gamma}^{\T}\|_F$. This further establishes a parametric upper bound for $\max_{m\in \MM}\|\widehat{\gamma}^{\m}-\widehat{\gamma}\|_2,$ which is key to establishing the parametric length of ${\rm CI}_{\alpha}\left(\xnew^{\intercal}\bp\right)$ in \eqref{eq: key length intui}; see Theorem \ref{thm: inference for linear} and its proof for the detailed argument.
\end{Remark}

\begin{Remark}[Resampling] \rm Our proposal mainly requires one of the resampled $\{\widehat{\Gamma}^{\m}\}_{m\in \mathbb{M}}$ to recover $\Gamma^{\T}$. To achieve this, we do not have to account for the dependence structure among the entries of ${\rm vecl}(\widehat{\Gamma}^{\T})$. 
We may simplify \eqref{eq: resampling} as $
{\rm vecl}(\widehat{\Gamma}^{\m})\sim \mathcal{N}\left({\rm vecl}(\widehat{\Gamma}^{\T}), {\rm diag}(\widehat{\V})+{d_0}/{{n}}\cdot {\bf I}\right) ,
$ with ${\rm diag}(\widehat{\V})$ denoting the diagonal matrix containing the diagonal elements of $\widehat{\V}$. 
Our proposed method is computationally efficient. For each $\widehat{\Gamma}^{\m}$, we solve an $L$-dimensional optimization problem in \eqref{eq: sample weight}, instead of a $p$-dimensional optimization problem. When the group number $L$ is much smaller than $p$, this significantly reduces the computation cost compared to non-parametric bootstrap, which directly samples the data and requires the implementation of high-dimensional optimization for each sampled data.  
 \label{rem: efficiency}
\end{Remark}



We summarize our proposal in Algorithm \ref{algo: SAR} and will discuss the tuning parameter selection at the beginning of Section \ref{sec: sim}.

\begin{algorithm}[H]
\caption{Maximin Effect Inference with Sampling Methods} 
\begin{flushleft}
\vspace{-2mm}
\textbf{Input:} $\{X^{(l)},Y^{(l)}\}_{1\leq l\leq L}$, $X^{\T}$; loading $\xnew\in \R^{p}$; $\alpha\in (0,1/2)$; sampling size $M$; $\tau_0>0$. \\
\textbf{Output:} Confidence interval ${\rm CI}_{\alpha}\left(\xnew^{\intercal}\bp\right)$
\vspace{-5mm}
\end{flushleft}
\begin{algorithmic}[1]    
 \State Compute $\{\widehat{\xnew^{\intercal}b^{(l)}}\}_{1\leq l\leq L}$ as in \eqref{eq: linear separate} and the variance $\{{\widehat{\rm V}^{(l)}_{\xnew}}\}_{1\leq l\leq L}$ as in \eqref{eq: asymp var};
 \State Compute $\{\widehat{\Gamma}^{\T}_{l,k}\}_{1\leq k\leq l\leq L}$ as in \eqref{eq: cov est general} and the covariance matrix $\widehat{\bf V}$ as in \eqref{eq: cov est};
 \State Construct the estimator $\widehat{\xnew^{\intercal}\bp}$ in \eqref{eq: point estimator};
\For{{$m \gets 1,2,\ldots,M$}}
 \State Sample $\widehat{\Gamma}^{\m}$ as in \eqref{eq: resampling} with $\widehat{\Gamma}^{\T}, \widehat{\V}$ and $\tau_0>0$;
 \State Construct $\widehat{\gamma}^{\m}$ in \eqref{eq: sample weight} and ${\rm Int}_{\alpha}^{\m}(\xnew)$ in \eqref{eq: sampled interval};
 \EndFor \Comment{Sampling}
 \State Construct the index set $\MM$ in \eqref{eq: generating condition};
 \State Construct ${\rm CI}_{\alpha}\left(\xnew^{\intercal}\bp\right)$ in \eqref{eq: aggregation 1}. \Comment{Aggregation}
\end{algorithmic}
\label{algo: SAR}
\end{algorithm}


\section{Theoretical Justification}
\label{sec: theory}
Before presenting the main theorems, we introduce the assumptions for the model \eqref{eq: multi-group model}. Define $s=\max_{1\leq l\leq L}\|b^{(l)}\|_0$ and $n=\min_{1\leq l\leq L} n_{l}.$ 
\begin{enumerate}
\item[(A1)] For $1\leq l\leq L$, $\{X^{(l)}_{i}, Y^{(l)}_{i}\}_{1\leq i\leq \nl}$ are i.i.d. random variables, where $X^{(l)}_{i}\in \R^{p} $ is sub-gaussian with $\Sigma^{(l)}=\E X^{(l)}_{i}[X^{(l)}_{i}]^{\intercal}$ satisfying $c_0\leq \lambda_{\min}\left(\Sigma^{(l)}\right)\leq \lambda_{\max}\left(\Sigma^{(l)}\right)\leq C_0$ for positive constants $C_0>c_0>0$; the error $\epsilon^{(l)}_i$ is sub-gaussian with $\E(\epsilon^{(l)}_i\mid X^{(l)}_{i})=0,$ $\E([\epsilon^{(l)}_i]^2\mid X^{(l)}_{i})=\sigma_{l}^2,$ and $\E([\epsilon^{(l)}_i]^{2+c}\mid X^{(l)}_{i})\leq C$ for some positive constants $c>0$ and $C>0$. $\{X^{\T}_{i}\}_{1\leq i\leq N_{\T}}$ are i.i.d. sub-gaussian with $\Sigma^{\T}=\E X^{\T}_{i}[X^{\T}_{i}]^{\intercal}$ satisfying $c_1\leq \lambda_{\min}\left(\Sigma^{\T}\right)\leq \lambda_{\max}\left(\Sigma^{\T}\right)\leq C_1$ for positive constants $C_1>c_1>0.$ 
\item[(A2)] There exists positive constants $C>0$ and $0<c<1$ such that $\max_{1\leq l\leq L}\|b^{(l)}\|_2\leq C$ and $n\geq c \cdot \max_{1\leq l\leq L} n_{l}.$ $L$ is finite and the model complexity parameters $(s,n,p, N_{\T})$ satisfy $n\gg {(s \log p)^2}$ and $ N_{\T}\gg n^{3/4}[\log \max\{N_{\T},p\}]^{2}.$
\end{enumerate}

We always consider asymptotic expressions in the limit where both $n,p \to \infty.$ 
Assumption (A1) is commonly assumed for the theoretical analysis of high-dimensional linear models; c.f. \cite{buhlmann2011statistics}. 
The positive definite $\Sigma^{(l)}$ and the sub-gaussianity of $X^{(l)}_{i}$ guarantee the restricted eigenvalue condition with a high probability \citep{bickel2009simultaneous,zhou2009restricted}. The sub-gaussian errors are generally required for the theoretical analysis of the Lasso estimator in high dimensions \cite[e.g.]{bickel2009simultaneous,buhlmann2011statistics}. The moment conditions on $\epsilon^{(l)}_i$ are needed to establish the asymptotic normality of the debiased estimators of single regression coefficients \citep{javanmard2014confidence}. Similarly, they are imposed here to establish the asymptotic normality of $\widehat{\Gamma}^{\T}$ and $\widehat{\xnew^{\intercal}b^{(l)}}.$ The model complexity condition $n\gg (s \log p)^2$ in (A2) is assumed in the CI construction for high-dimensional linear models \citep{zhang2014confidence,van2014asymptotically,javanmard2014confidence}. The condition on $N_{\T}$ is mild as there is typically a large amount of unlabelled data for the target population. As pointed out in Remark \ref{rem: special case}, our proposal can be extended to handle the special no covariate shift setting, where the required assumption on $N_{\T}$ throughout our current theoretical analysis can be removed. The boundedness assumptions on $L$ and $\|b^{(l)}\|_2$ are mainly imposed to simplify the presentation, so is the assumption $n\geq c\cdot \max_{1\leq l\leq L} n_{l}$. 


We justify the sampling step proposed in Section \ref{sec: sampling method} and show that there exists at least one sampled weight vector converging to $\gamma^{*}$ at a rate faster than $1/\sqrt{n}.$ We introduce $\err_n(M)$ to characterize the sampling accuracy:
\begin{equation}
\err_n(M)=\left[\frac{4 \log n}{C^{*}(\alpha_0)\cdot M}\right]^{\frac{2}{L(L+1)}}, 
\label{eq: sampling ratio}
\end{equation}
where $\alpha_0\in(0,0.01]$ is the pre-specified constant used in the construction of $\mathbb{M}$ in \eqref{eq: generating condition}, and $C^{*}(\alpha_0)$ is a constant defined in \eqref{eq: sampling constants} in the supplement. Note that resampling size $M$ is set by the users and the sampling accuracy $\err_n(M)\rightarrow 0$ with $M\rightarrow \infty. $
The following theorem establishes the rate of convergence for the best approximation accuracy among all sampled vectors $\{\widehat{\gamma}^{\m}\}_{m\in \mathbb{M}}.$

\begin{Theorem} Consider the model \eqref{eq: multi-group model}. Suppose $\lambda_{\min}(\Gamma^{\T})>0$ and Conditions {\rm (A1)} and {\rm (A2)} hold. Then
 \begin{equation*}
\liminf_{n,p \rightarrow \infty} \liminf_{M\rightarrow \infty}\PP\left(\min_{m\in \mathbb{M}}\|\widehat{\gamma}^{\m}-\gamma^{*}\|_2 \leq \frac{\sqrt{2} \err_n(M)}{\lambda_{\min}(\Gamma^{\T})} \cdot \frac{1}{\sqrt{n}}\right)\geq 1-\alpha_0,
 \end{equation*}
 where $\alpha_0\in(0,0.01]$ is the pre-specified constant used in the construction of $\mathbb{M}$ in \eqref{eq: generating condition}.
\label{thm: sampling accuracy}
\end{Theorem}
We discuss the implication of Theorem \ref{thm: sampling accuracy}.
When we resample a large amount of data such that $\err_n(M)$ is much smaller than $\lambda_{\min}(\Gamma^{\T})$, then there is a high chance of having a sampling index $m^*\in \mathbb{M}$ such that $\|\widehat{\gamma}^{[m^{*}]}-\gamma^{*}\|_2 \ll \frac{1}{\sqrt{n}}$. 
Theorem \ref{thm: sampling accuracy} covers the important setting with a nearly singular $\Gamma^{\T}$, that is, $\lambda_{\min}(\Gamma^{\T})>0$ for any given $p$ but $ \liminf_{p\rightarrow \infty}\lambda_{\min}(\Gamma^{\T})= 0.$ This setting will appear if some of $\{b^{(l)}\}_{1\leq l\leq L}$ are similar to each other but not exactly the same. However, in this challenging setting, we may have to choose a relatively large sampling number $M>0$ such that $\err_n(M)\ll \lambda_{\min}(\Gamma^{\T}).$ Theorem \ref{thm: sampling accuracy} is not applied to the exactly singular setting $\lambda_{\min}(\Gamma^{\T})=0$ while the ridge-type maximin effect introduced in Section \ref{sec: ridge-type} is helpful for the exactly singular setting. 
In the proof of Theorem \ref{thm: sampling accuracy}, we only require $\err_n(M) \ll \min\{1,\lambda_{\min}(\Gamma^{\T})\},$ which will be automatically satisfied with taking $M\rightarrow \infty.$  In practice, we set $M=500$ as the default value and observe reliable inference results.

 The following theorem establishes the properties of ${\rm CI}_{\alpha}(\xnew^{\intercal}\bp)$ defined in \eqref{eq: aggregation 1}.

\begin{Theorem} Suppose that the conditions of Theorem \ref{thm: sampling accuracy} hold. Then the confidence interval ${\rm CI}_{\alpha}(\xnew^{\intercal}\bp)$ defined in 
\eqref{eq: aggregation 1} satisfies 
\begin{equation}
\liminf_{n,p\rightarrow \infty}\liminf_{M\rightarrow \infty}\PP\left(\xnew^{\intercal} \bp\in {\rm CI}_{\alpha}\left(\xnew^{\intercal}\bp\right)\right)\geq 1-\alpha-\alpha_0,
\label{eq: coverage}
\end{equation}
where $\alpha\in (0,1/2)$ is the pre-specified significance level and $\alpha_0\in(0,0.01]$ is the pre-specified constant used in the construction of $\mathbb{M}$ in \eqref{eq: generating condition}. By further assuming $N_{\T}\gtrsim\max\{n,p\}$ and $\lambda_{\min}(\Gamma^{\T})\gtrsim \sqrt{\log p/\min\{n,N_{\T}\}}$, then there exists some positive constant $C>0$ such that
\begin{equation}
\lim_{n,p \rightarrow \infty} \PP\left(\mathbf{Leng}\left({\rm CI}_{\alpha}\left(\xnew^{\intercal}\bp\right)\right)\leq C \max\left\{ 1,\frac{z_{\alpha_0/[L(L+1)]}}{\lambda_{\min}(\Gamma^{\T})}\right\}\cdot \frac{\|\xnew\|_2}{\sqrt{n}}\right)=1,
\label{eq: length bound}
\end{equation}
where $\mathbf{Leng}\left({\rm CI}\left(\xnew^{\intercal}\bp\right)\right)$ denotes the interval length and $z_{\alpha_0/[L(L+1)]}$ is the upper $\alpha_0/[L(L+1)]$ quantile of the standard normal distribution. 
\label{thm: inference for linear}
\end{Theorem}

A few remarks are in order for Theorem \ref{thm: inference for linear}. Firstly, the validity of the constructed CI does not require the asymptotic normality of $\widehat{\xnew^{\intercal}\bp},$ which might not hold due to the non-regularity and instability. 
Secondly, in \eqref{eq: coverage}, we only establish one-sided coverage guarantee since we take a union over $\MM$ in our proposed sampling method. We examine the tightness of the coverage inequality \eqref{eq: coverage} in the simulation studies; see Table \ref{tab: summary} in Section \ref{sec: sim}. Thirdly, if $\lambda_{\min}(\Gamma^{\T})\geq c$ for a positive constant $c>0$, then the CI length is of the rate $\|\xnew\|_2/\sqrt{n}$. In consideration of a single high-dimensional linear model, \citet{cai2019individualized} showed that, without the knowledge of the sparsity level of $b^{(l)}$, the optimal length of CIs for $\xnew^{\intercal} b^{(l)}$ is $\|\xnew\|_2/\sqrt{n}$ if 
$\sqrt{\|\xnew\|_0}\leq C \sqrt{n}/\log p$ and $\|b^{(l)}\|_0 \leq C \sqrt{n}/\log p$ for some positive $C>0$; see Corollary 4 in \citet{cai2019individualized} for the exact details. In Section \ref{sec: sim}, we evaluate the precision properties of our proposed CI in finite samples; see Table \ref{tab: summary} in Section \ref{sec: sim} for a summary. 

\section{Stability: Ridge-type Maximin Effect} 
\label{sec: ridge-type}


We introduce a ridge-type maximin effect which ensures a more stable data integration than the maximin effect especially in the instability setting. Section \ref{sec: maximin challenge} highlights that the maximin integration suffers from the instability challenge if $\gamma^*$ is not uniquely defined. The instability setting shows up when the regression vectors $\{b^{(l)}\}_{1\leq l\leq L}$ are similar to each other. Even though our proposed sampling CI in \eqref{eq: aggregation 1} is still valid for the instability settings, CIs for the ridge-type maximin effect may be much shorter due to the more stable integration.

We provide a more stable integration by adding a ridge penalty in constructing the integration weight. For $\delta\geq 0,$  we propose the following ridge-type maximin effect 
\begin{equation}
\beta^{*}_{\delta}(\T)=\sum_{l=1}^{L}[\gamma^{*}_{\delta}(\T)]_l \cdot b^{(l)} \quad \text{with} \quad 
\gamma^{*}_{\delta}(\T)=\argmin_{\gamma \in \Delta^{L}}\left[\gamma^{\intercal}\Gamma^{\T} \gamma+{\delta}\|\gamma\|_2^2\right],
\label{eq: optimal weight ridge}
\end{equation}
which adds the ridge penalty in constructing the weight vector. When there is no confusion, we write $\bp_{\delta}(\T)$ and $\gamma^{*}_{\delta}(\T)$ as $\bp_{\delta}$ and $\gamma^{*}_{\delta}$, respectively. With $\delta=0,$ $\beta^{*}_{\delta}$ becomes $\beta^{*}$ in \eqref{eq: optimal weight}. For a positive $\delta>0$, $\beta^{*}_{\delta}$ is generally a different model from $\beta^{*}$. We establish the properties of $\beta^{*}_{\delta}$ in the following Proposition \ref{prop: DRO ridge}. Define 
${\mathbf B}=\begin{pmatrix}b^{(1)},\ldots,b^{(L)}\end{pmatrix}\in \R^{p\times L}$. We assume $p\geq L$ and the matrix ${\mathbf B}$ is of the full column rank with the SVD, ${\mathbf B}=U_{p\times L} \Lambda_{L\times L} V^{\intercal}_{L\times L}$. For $1\leq i\leq N_{\T},$ we generate the noise vector $W_i(\delta)\in \R^{p}$ as $W_i(\delta)=\sqrt{\delta} \cdot U W_i^{0}$ with $W_i^{0}\sim \mathcal{N}({\bf 0}, \Lambda^{-2})$ and $W_i^{0}$ being independent of $X^{\T}_i.$
\begin{Proposition}
The ridge-type maximin effect $\beta^{*}_{\delta}$ in \eqref{eq: optimal weight ridge} is uniquely defined for $\delta>0$ and 
\begin{equation}
R_{\T}[\bp_{\delta}]\geq R_{\T}[\bp]-2\delta(\|\gamma^{*}_{\delta}\|_\infty-\|\gamma^{*}_{\delta}\|_2^2)\geq R_{\T}(\bp)-\frac{\delta}{2}\cdot \left(1-\frac{1}{L}\right),
\label{eq: reward reduction}
\end{equation}where $R_{\T}[\cdot]$ and $\bp$ are defined in \eqref{eq: maximin general shift} and $\gamma^{*}_{\delta}$ is defined in \eqref{eq: optimal weight ridge}. In addition, $\beta^{*}_{\delta}(\T)$ is the solution to the following distributionally robust optimization problem,
\begin{equation}
\bp_{\delta}(\T):=\argmax_{\beta\in \R^{p}}\min_{\Tar\in \mathcal{C}(\T^{\delta}_{X})} \left\{\E_{\Tar} Y_i^2-\E_{\Tar}(Y_i-X_i^{\intercal}\beta)^2\right\},
\label{eq: maximin general shift ridge}
\end{equation}
where the covariate distribution $\T^{\delta}_{X}$ denotes the distribution of $X^{\T}_i+W_i(\delta)$ and the distribution class $\mathcal{C}(\T^{\delta}_{X})$ is defined in \eqref{eq: distribution class shift} with $\T_{X}$ replaced by $\T^{\delta}_{X}$.
 \label{prop: DRO ridge}
\end{Proposition}
Proposition \ref{prop: DRO ridge} controls the reward reduction $R_{\T}[\bp_{\delta}]-R_{\T}[\bp]$ if a ridge-type maximin effect is used in comparison to the maximin effect. The ridge penalty $\delta$ controls the reward reduction, which is negligible for a small positive $\delta$. We show in \eqref{eq: maximin general shift ridge} that $\beta^{*}_{\delta}$ is also the solution to a distributionally robust optimization problem, where the extra ridge penalty is equivalent to perturbing the target population's covariates $X^{\T}_i$ with the random noise $W_i(\delta).$

The proposed methods detailed in Section \ref{sec: method} can be extended to dealing with $\xnew^{\intercal}\bp_{\delta}$ for any $\delta\geq 0.$ Specifically, we generalize the point estimator \eqref{eq: point estimator} as  
\begin{equation}
\widehat{\xnew^{\intercal}\bp_{\delta}}=\sum_{l=1}^{L} [\widehat{\gamma}_{\delta}]_{l}\cdot \widehat{\xnew^{\intercal}b^{(l)}} \quad \text{with}\quad \widehat{\gamma}_{\delta}:=\arg\min_{\gamma\in \Delta^{L}} 
\left[\gamma^{\intercal}\widehat{\Gamma}^{\T} \gamma+{\delta}\|\gamma\|_2^2\right]. 
\label{eq: point estimator ridge}
\end{equation}
Regarding the CI construction for $\xnew^{\intercal}\beta^{*}_{\delta},$ we replace $\widehat{\gamma}^{\m}$ in \eqref{eq: sample weight} by
\begin{equation}
\widehat{\gamma}^{\m}_{\delta}=\argmin_{\gamma\in \Delta^{L}} \gamma^{\intercal}(\widehat{\Gamma}^{\m}+\delta \cdot {\rm I})_{+} \gamma \quad \text{for}\quad \delta\geq 0.
\label{eq: sample weight ridge} 
\end{equation}
The theoretical results in Section \ref{sec: theory} can be directly generalized with replacing $\Gamma^{\T}$ by $\Gamma^{\T}+\delta\cdot{\rm I}$.

For the instability settings with $\lambda_{\min}(\Gamma^{\T})>0$, our proposed sampling method in Section \ref{sec: method} provides valid inference for $\bp(\T)$ and $\bp_{\delta}(\T)$. The ridge penalty will reduce the uncertainty of estimating the weight vector, resulting in shorter CIs; see Figures \ref{fig: setting 1} and \ref{fig: method comparison}. 

Finally, we discuss the empirical assessment of the integration stability. 
For $L=2,$ we obtain $\gamma^{*}_{\delta}=([\gamma^{*}_{\delta}]_1,1-[\gamma^{*}_{\delta}]_1)^{\intercal}$ with
$
[\gamma^{*}_{\delta}]_1=\min\{\max\{\tfrac{\Gamma_{22}+\delta-\Gamma_{12}}{\Gamma_{11}+\Gamma_{22}+2\delta-2\Gamma_{12}},0\},1\}.  
$ This expression shows that  the maximin integration is unstable for $\Gamma_{11}+\Gamma_{22}-2\Gamma_{12}$ being near zero. 
For $L\geq 2,$ we propose a general instability measure (depending on $\delta$) as 
$
\mathbb{I}(\delta)={\sum_{m=1}^{M}\|\widehat{\gamma}^{\m}_{\delta}-\widehat{\gamma}_{\delta}\|^2_2}/{\sum_{m=1}^{M} \|\widehat{\Gamma}^{\m}-\widehat{\Gamma}^{\T}\|^2_2},$
with $\{\widehat{\Gamma}^{\m}\}_{1\leq m\leq M}$ and $\{\widehat{\gamma}^{\m}_{\delta}\}_{1\leq m\leq M}$ defined in \eqref{eq: resampling} and \eqref{eq: sample weight ridge}, respectively. A large value of $\mathbb{I}(\delta)$ indicates that the weight vector estimation is not stable; see the numerical illustrations in Table \ref{tab: complete instability} in the supplement. 
\section{Simulation Results}
\label{sec: sim}
Throughout the simulation, we make inference for $\xnew^{\intercal}\bp_{\delta}$ with $\delta\geq 0$ and set the significance level $\alpha=0.05$. We implement Algorithm \ref{algo: SAR} by replacing the weight construction in \eqref{eq: point estimator} and \eqref{eq: sample weight}
 with the corresponding ridge-type versions in \eqref{eq: point estimator ridge} and 
\eqref{eq: sample weight ridge}.  We specify how to choose the tuning parameters for constructing $\widehat{\xnew^{\intercal}b^{(l)}}$ and $\widehat{\Gamma}^{\T}$ in high dimensions. The Lasso estimators $\{\widehat{b}^{(l)}\}_{1\leq l\leq L}$ are implemented by the R-package \texttt{glmnet} \citep{friedman2010regularization} with tuning parameters chosen by cross validation; the estimator 
$\widehat{\xnew^{\intercal}b^{(l)}}$ in \eqref{eq: linear separate} is implemented using the R-package \texttt{SIHR} \citep{rakshit2021sihr} with the built-in selection of the tuning parameters $\eta_l$ and $\tau_l;$ the tuning parameter selection for $\widehat{\Gamma}^{\T}_{l,k}$ in \eqref{eq: cov est general} is presented in \eqref{eq: dual problem} in the supplement.  We believe that the sample splitting used for constructing $\widehat{\Gamma}^{\T}_{l,k}$ in \eqref{eq: cov est general} is only needed for the theoretical justification. In the supplement, we provide the numerical comparison between our proposed methods with and without sample splitting in Table \ref{tab: splitting comparison}. We observe that the procedure without sample splitting performs well and improves efficiency compared to sample splitting. We construct $\widehat{\Gamma}^{\T}$ without the sample splitting in the simulation and real data analysis. The code with the tuning parameter selection is submitted together with the current paper.

We compare our proposed CI with a normality CI of the form \begin{equation}
(\widehat{\xnew^{\intercal}\bp_{\delta}}-1.96\cdot \widehat{\rm SE},\widehat{\xnew^{\intercal}\bp_{\delta}}+1.96\cdot \widehat{\rm SE}),
\label{eq: normality CI}
\end{equation} where $\widehat{\xnew^{\intercal}\bp_{\delta}}$ is defined in \eqref{eq: point estimator ridge} and $\widehat{\rm SE}$ denotes the empirical standard deviation of $\widehat{\xnew^{\intercal}\bp_{\delta}}$ calculated based on $500$ simulations. Since $\widehat{\rm SE}$ is calculated in an oracle way, this normality CI is not a practical procedure but a favorable implementation of the CI constructed by assuming the asymptotic normality of the point estimator $\widehat{\xnew^{\intercal}\bp_{\delta}}$. Throughout the simulation, we report the average measures over 500 simulations.


We show that the normality CI in \eqref{eq: normality CI} undercovers in the presence of non-regularity and instability. We generate 11 simulation settings with $L=4$ and $p=500$. Particularly, the setting (I-0) corresponds to $b^{(1)}=\cdots=b^{(L)}$, the settings (I-1) to (I-6) correspond to instability settings with $b^{(1)}\approx \cdots \approx b^{(L)}$, the settings (I-7) to (I-9) correspond to the non-regularity settings, and (I-10) corresponds to an easier setting without non-regularity and instability. The detailed settings are reported in Section \ref{sec: I settings} in the supplement. \begin{table}[H]
\centering
\resizebox{0.75\linewidth}{!}{
\begin{tabular}[t]{|c|c|c|c|c|c|c|}
\hline
\multicolumn{1}{|c|}{} & \multicolumn{1}{c|}{} &\multicolumn{2}{c|}{Coverage} & \multicolumn{2}{c|}{Length} & \multicolumn{1}{c|}{} \\
\cline{3-4} \cline{5-7}
Setting & $\mathbb{I}(\delta)$ & normality & Proposed & normality & Proposed & Length Ratio\\
\hline
(I-0) & 1.526 & 0.925 & 0.996 & 0.225 & 0.401 & 1.783\\
\hline
(I-1) & 3.368 & 0.700 & 0.960 & 0.352 & 0.597 & 1.693\\
\hline
(I-2) & 3.707 & 0.818 & 0.978 & 0.320 & 0.543 & 1.699\\
\hline
(I-3) & 3.182 & 0.748 & 0.970 & 0.352 & 0.588 & 1.673\\
\hline
(I-4) & 1.732 & 0.770 & 0.956 & 0.520 & 0.796 & 1.532\\
\hline
(I-5) & 1.857 & 0.796 & 0.978 & 0.445 & 0.710 & 1.594\\
\hline
(I-6) & 1.987 & 0.710 & 0.980 & 0.480 & 0.832 & 1.732\\
\hline
(I-7) & 0.029 & 0.848 & 0.985 & 0.250 & 0.507 & 2.028\\
\hline
(I-8) & 0.031 & 0.758 & 0.981 & 0.262 & 0.530 & 2.020 \\
\hline
(I-9) & 0.010 & 0.830 & 0.988 & 0.690 & 1.264 & 1.832 \\
\hline
(I-10) & 0.030 & 0.940 & 0.988 & 0.232 & 0.315 & 1.354\\
\hline
\end{tabular}
}
\caption{\small High-dimensional setting with $p=500$: coverage and length of the CI in Algorithm \ref{algo: SAR} and the normality CI in \eqref{eq: normality CI} (with $\delta=0$). The column indexed with ``Coverage" and ``Length" represent the empirical coverage and average length for CIs, respectively; the columns indexed with ``normality" and ``Proposed" represent the normality CI and our proposed CI, respectively. The column indexed with ``Length Ratio" represents the ratio of the average length of our proposed CI to that of the normality CI. The column indexed with ``$\mathbb{I}(\delta)$" reports the instability measure.}
\label{tab: i-settings result}
\end{table}

We focus on the maximin effect without the ridge penalty. In Table \ref{tab: i-settings result}, except for (I-0) and (I-10), the empirical coverages of the normality CI in \eqref{eq: normality CI} are between 70\% and 85\%.  Our proposed CI achieves the desired coverage at the expense of a wider interval. The ratio of the average length of our proposed CI to the normality CI is between 1.35 and 2.02. The instability measures $\mathbb{I}(\delta)$ are large for the instability settings (I-0) to (I-6) but small for the remaining stable settings. (I-0) is special in the sense that the instability of identifying $\gamma^*$ does not create a bias for estimating the maximin effect, that is, any convex combination of unbiased estimators of $\{b^{(l)}\}_{1\leq l\leq L}$ will be unbiased. This explains why the normality CI works under (I-0).

\begin{figure}[H]
 \centering
 \includegraphics[scale=0.53]{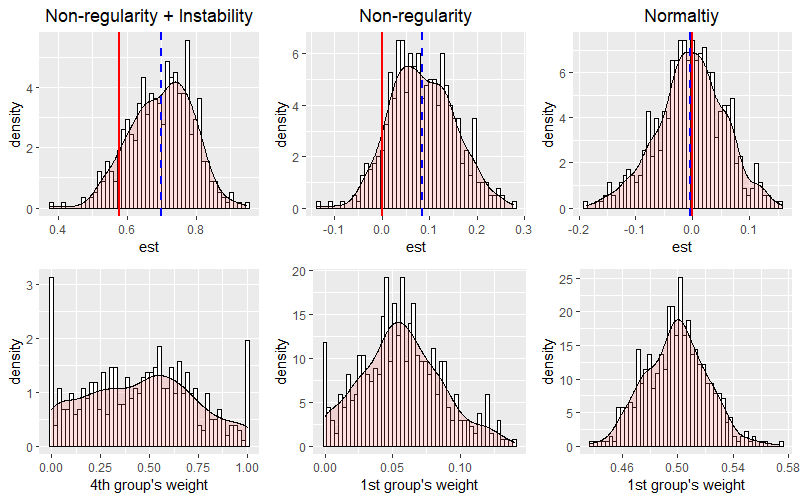}
 \caption{\small The histogram of the maximin estimator $\widehat{\xnew^{\intercal}\bp}$ (top) and one coordinate of the weight estimator (bottom) over 500 simulations. The figures from the leftmost to the rightmost correspond to settings (I-1), (I-8), and (I-10). The solid red line denotes the true value $\xnew^{\intercal}\bp$ while the blue dashed line denotes the sample average over 500 simulations.}
\label{fig: hist oracle}
\end{figure}

To further investigate the under-coverage of the normality CI, we plot in Figure \ref{fig: hist oracle} the histogram of $\widehat{\xnew^{\intercal}\bp}$ 
and $\widehat{\gamma}$ in \eqref{eq: point estimator} over 500 simulations. The leftmost panel of Figure \ref{fig: hist oracle} corresponds to the setting (I-1) with non-regularity and instability. Due to the instability, the histogram of the weight estimates has some concentrations near both 0 and 1, which results in the bias component of $\widehat{\xnew^{\intercal}\bp}$ being comparable to its standard error. Consequently, the empirical coverage of the corresponding normality CI is only 70\%. The middle panel of Figure \ref{fig: hist oracle} corresponds to the setting (I-8) with non-regularity, where the weight for the first group is left-censored at zero. This censoring at zero leads to the bias of $\widehat{\xnew^{\intercal}\bp}$ being comparable to its standard error and under-coverage of the normality CI. The rightmost panel corresponds to the favorable setting (I-10) without non-regularity and instability. The weight distributions and the maximin effect estimator are nearly normal, and the corresponding normality CI in \eqref{eq: normality CI} achieves the 95\% coverage level.

\vspace{-3mm}

\vspace{-5mm}

We investigate our proposed method over additional settings.
We generate $\{X^{(l)},Y^{(l)}\}_{1\leq l\leq L}$ following \eqref{eq: multi-group model}, where, for the $l$-th group, $\{X^{(l)}_{i}\}_{1\leq i\leq n_l}\stackrel{\rm i.i.d.}{\sim} \mathcal{N}({\bf 0},\Sigma^{(l)})$ and $\{\epsilon^{(l)}_{i} \}_{1\leq i\leq n_l}\stackrel{\rm i.i.d.}{\sim} \mathcal{N}({0},\sigma^2_l).$ For $1\leq l\leq L,$ we take $n_l=n$, $\sigma_{l}=1$ and $\Sigma^{(l)}=\Sigma,$ with $\Sigma_{j,k}=0.6^{|j-k|}$ for $1\leq j,k \leq p$. In the covariate shift setting, we generate $\{X^{\T}_{i}\}_{1\leq i\leq N_{\T}}\stackrel{\rm i.i.d.}{\sim} \mathcal{N}({\bf 0},\Sigma^{\T})$. $p$ is set as $500$ and $N_{\T}$ is set as $2000$ by default. 
We describe settings 1-3 in the following, and provides the details for settings 4-6 in Section \ref{sec: additional simulation} in the supplement.

\noindent {\it \underline{Setting 1 ($L=2$ with no covariate shift)}}. $b^{(1)}_j=j/40$ for $1\leq j\leq 10$, $b^{(1)}_j=1$ for $j=22,23$, $b^{(1)}_{j}=0.1$ for $j=499,500$, and $b^{(1)}_j=0$ otherwise;
$b^{(2)}_{j}=b^{(1)}_{j}$ for $1\leq j \leq 499$ and $b^{(2)}_{500}=0.3$; $[{\xnew}]_j=1$ for $j=500$ and $[{\xnew}]_{j}=0$ otherwise.

\noindent {\it \underline{Setting 2 ($L=2$ with covariate shift)}}. $b^{(1)}$ and $b^{(2)}$ are the same as setting 1, except for $b^{(1)}_{498}=0.5$, $b^{(1)}_j = -0.5$ for $j=499, 500,$ and $b^{(2)}_{500}=1$. 
$[{\xnew}]_{j}=1$ for $498\leq j\leq 500$, and $[{\xnew}]_{j}=0$ otherwise.
$\Sigma^{\T}_{i,i}=1.5$ for $1\leq i\leq 500$, $\Sigma^{\T}_{i,j}=0.9$ for $1\leq i\neq j\leq 5,$ $\Sigma^{\T}_{i,j}=0.9$ for $499\leq i\neq j\leq 500$ and $\Sigma^{\T}_{i,j}=\Sigma_{i,j}$ otherwise.

\noindent {\it \underline{Setting 3 ($L=2$ with/without covariate shift)}}. 
$b^{(1)}$ and $b^{(2)}$ are the same as setting 1, except for $b^{(1)}_{498}=0.5$, $b^{(1)}_j = -0.5$ for $j=499, 500,$ and $b^{(2)}_{500}=1$; $[\xnew]_{j}=1$ for $j=499,500,$ and $[{\xnew}]_{j}=0$ otherwise.
Setting 3(a) is the covariate shift setting with $\Sigma^{\T}_{i,i}=1.5$ for $1\leq i\leq 500$, $\Sigma^{\T}_{i,j}=0.6$ for $1\leq i\neq j\leq 5,$ $\Sigma^{\T}_{i,j}=-0.9$ for $499\leq i\neq j\leq 500$ and $\Sigma^{\T}_{i,j}=\Sigma_{i,j}$ otherwise; Setting 3(b) is the no covariate shift setting.

We compute
$
\text{Coverage Error}=|\text{Empirical Coverage} - 95\%|,
$
with the empirical coverage computed based on 500 simulations. We report the ratio of the average length of our proposed CI to that of the normality CI in \eqref{eq: normality CI}. For each setting, we average the coverage error and the length ratio over different combinations of $\delta\in \{0,0.1,0.5,1,2\}$ and $n\in \{200,300,500\}$. \footnote{For setting 5, instead of averaging over $\delta\in \{0,0.1,0.5,1,2\}$, we take an average with respect to the {\rm perb} parameter; see more details in Section \ref{sec: additional simulation} in the supplement. } 
In Table \ref{tab: summary}, we summarize the average coverage error and length ratio over different settings. Since our proposed CIs generally achieve 95\% for $n\geq 200$, the coverage errors mainly result from over-coverage instead of under-coverage. For settings 3(a), 4(a), and 5, the empirical coverage of our proposed CI is nearly 95\%, and the corresponding length ratios for settings 4(a) and 5 are near 1. For settings 3(b) and 6, our proposed CIs are over-coverage, but the average length ratios are at most $1.864.$ 
\begin{table}[H]
\centering
\resizebox{\linewidth}{!}{
\begin{tabular}{|c|c|c|c|c|c|c|c|c|c|}
 \hline
 Setting & 1 & 2 & 3(a) & 3(b) & 4(a) & 4(b) & 4(c) & 5 & 6 \\
 \hline
 Coverage Error & 2.60\% & 3.45\% & 0.95\% & 4.51\% & 1.64\% & 2.71\% & 3.57\% & 0.75\% & 4.25\% \\
 \hline
 Length Ratio & 1.322 & 1.607 & 1.516 &1.864 & 1.047 & 1.356 & 1.587 & 1.268 & 1.554 \\
 \hline
\end{tabular}}
\caption{Average coverage error and length ratio across different settings.}
\label{tab: summary}
\end{table}

In the following, we demonstrate the dependence of our sampling method on $n$ and $\delta$ and compare the maximin effects with and without covariate shifts. We provide more details in Section \ref{sec: additional simulation} in the supplement, including settings with a larger group number $L$, a larger dimension $p$, and the regression models with perturbed effects or opposite effects.

\begin{figure}[htp!]
\centering
\includegraphics[width=0.8\linewidth]{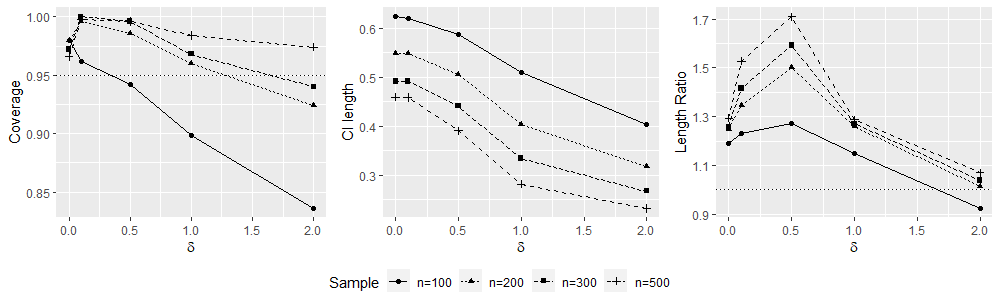}
\caption{\small Dependence on $\delta$ and $n$ (setting 1). ``Coverage" and ``CI Length" stand for the empirical coverage and the average length of our proposed CI, respectively; ``Length Ratio" represents the ratio of the average length of our proposed CI to the normality CI in \eqref{eq: normality CI}.}
\label{fig: setting 1}
\end{figure}

{\bf \noindent Dependence on $n$ and $\delta$.} 
For setting 1, we plot in Figure \ref{fig: setting 1} the empirical coverage and CI length over $\delta \in \{0,0.1,0.5,1,2\}.$ Our proposed CIs achieve the desired coverage level for $n\geq 200.$ The CIs get shorter with increasing $n$ or $\delta$: the lengths of CIs for $\delta=2$ are around half of those for $\delta=0$. This shows that a positive $\delta$ effectively reduces the CI length in setting 1, where the maximin integration is unstable. 

\vspace{-5mm}
\begin{figure}[htp!]
\centering
\begin{subfigure}[b]{\textwidth}
\centering
 \includegraphics[width=0.8\linewidth]{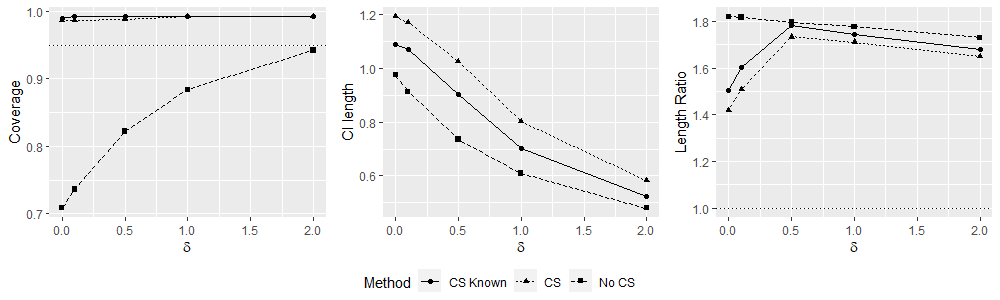}
 \caption{Setting 3(a) with covariate shift}
\end{subfigure}

\begin{subfigure}[b]{\textwidth}
\centering
 \includegraphics[width=0.8\linewidth]{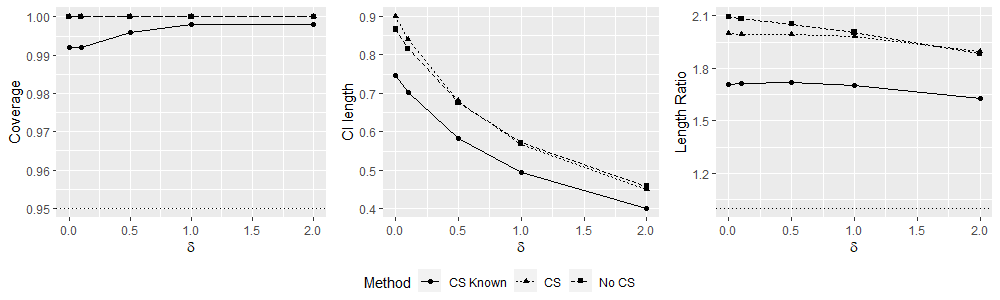}
 \caption{Setting 3(b) with no covariate shift}
\end{subfigure}
\caption{\small Comparison of covariate shift and no covariate shift algorithms ($n=500$). {``CS Known", ``CS" and ``No CS" represent Algorithm \ref{algo: SAR} with known $\Sigma^{\T}$, Algorithm \ref{algo: SAR}, and Algorithm \ref{algo: SAR} with no covariate shift, respectively.} ``Coverage" and ``CI Length" stand for the empirical coverage and the average length of our proposed CI, respectively; ``Length Ratio" represents the ratio of the average length of our proposed CI to the normality CI.}
\label{fig: method comparison}
\end{figure}
{\noindent \bf Covariate shift.} We modify Algorithm \ref{algo: SAR} to two extra scenarios: (1) $\Sigma^{\T}$ is known; (2) no covariate shift between the target and source populations. In both scenarios, we present how to construct a debiased estimator $\widehat{\Gamma}^{\T}$ in Section \ref{sec: high-dim Gamma} in the supplement. We then implement Algorithm \ref{algo: SAR} with the modified $\widehat{\Gamma}^{\T}$. We shall refer to the corresponding methods as Algorithm \ref{algo: SAR} with known $\Sigma^{\T}$ and Algorithm \ref{algo: SAR} with no covariate shift. 

The top of Figure \ref{fig: method comparison} corresponds to the simulation settings with covariate shift and $n=500$. The no covariate shift algorithm does not achieve 95\% coverage due to the bias of assuming no covariate shift. In contrast, the covariate shift algorithms (with or without knowing $\Sigma^{\T}$) achieve the 95\% coverage level, and the CI constructed with known $\Sigma^{\T}$ is shorter as it does not need to quantify the uncertainty of estimating $\Sigma^{\T}.$ 
The bottom of Figure \ref{fig: method comparison} corresponds to the setting with no covariate shift. All algorithms achieve the desired coverage level. The results for $n=200$ are reported in Figure \ref{fig: method comparison n=200} in the supplement.



\section{Real Data Applications}
\label{sec: real data}
We analyze a genome-wide association study \citep{bloom2013finding} on yeast colony growth based on $n=1008$ saccharomyces cerevisiae segregants crossbred from a laboratory and a wine strain. \citet{bloom2013finding} selected $4410$ Single Nucleotide Polymorphisms (SNPs) for the data analysis. We further apply LD screening \citep{calus2018snprune} to remove SNPs with absolute correlation above $0.85$ and end up with $p=513$ SNPs for the regression analysis. The outcome variables are the end-point colony sizes under different growth media. These outcome variables are normalized to have a variance of $1$. We consider four source growth media: ``Ethanol", ``Lactose", ``5-Fluorouracil", and ``Xylose". The model \eqref{eq: multi-group model} is applied here with $L=4$, and each $1\leq l\leq 4$ corresponds to the data for one growth medium.

We start with the preliminary analysis of whether the regression vectors in \eqref{eq: multi-group model} are heterogeneous across the four growth media. 
For $1\leq l\leq 4$, we construct the debiased lasso estimator $\widehat{b}^{(l)}_{j}$ as in \eqref{eq: linear separate} with $\xnew=e_j$ for $1\leq j\leq p,$ and obtain the corresponding covariance matrix as ${\rm Cov}(\widehat{b}^{(l)})\in \R^{p\times p}.$ For $1\leq l_1<l_2\leq L$, we test $H_0: {b}^{(l_1)}={b}^{(l_2)}$ by extending the bootstrap methods in \citet{dezeure2017high}. We generate the bootstrap samples $Z^{(1)},\cdots, Z^{(1000)}$ following 
$Z^{(k)}\sim N\left({\bf 0}, {\rm Cov}(\widehat{b}^{(l_1)})+{\rm Cov}(\widehat{b}^{(l_2)})\right)$ for $1\leq k\leq 1000.$ We compute the maximum statistics $T^{(l_1,l_2)}_{\rm obs}=\max_{1\leq j\leq p}|\widehat{b}^{(l_1)}_{j}-\widehat{b}^{(l_2)}_{j}|$ and calculate the p-value as 
$\frac{1}{1000}\sum_{k=1}^{1000}{\bf 1}(\|Z^{(k)}\|_{\infty}\geq T^{(l_1,l_2)}_{\rm obs})$. As reported in Table \ref{tab: pvalue}, the small p-values indicate the data heterogeneity across the four media.

\begin{table}[H]
\centering
\resizebox{0.6\linewidth}{!}{
\begin{tabular}[t]{|l|c|c|c|c|}
\hline
 & Ethanol & Lactose & 5-Fluorouracil & Xylose\\
\hline
Ethanol & - & 0.001 & 0.059 & 0.146\\
\hline
Lactose & - & - & 0.001 & 0.003\\
\hline
5-Fluorouracil & - & - & - & 0.027\\
\hline
\end{tabular}}
\caption{\small p-values for homogeneity test of regression vectors in \eqref{eq: multi-group model}. For example, $0.003$ stands for the p-value of testing whether the regression vectors for media ``Lactose" and ``Xylose" are the same.}
\label{tab: pvalue}
\end{table}

\subsection{Maximin effects: summary of stable associations across source media}
In Figure \ref{fig: real-data}, we demonstrate that the maximin effect summarizes the stable associations across the four source media. In particular, an SNP with a significant maximin effect tends to have consistent effects across source media. Due to space constraints, we report the inference results for a representative subset $\mathcal{S}$ of SNPs in Figure \ref{fig: real-data} rather than reporting results for all $513$ SNPs used for the regression analysis. Figure \ref{fig: real-data} illustrates that SNPs with indexes $420, 443, 437, 423, 245, 424$ have homogeneous effects across the source media, and the corresponding maximin effect is significant. The gene KRE33 containing SNP $420$ is an essential gene for yeast \citep{cherry2012saccharomyces}, which is a gene absolutely required to maintain life provided that all nutrients are available \citep{zhang2009deg}.

\begin{figure}[htp!]
\centering
\includegraphics[width=0.8\linewidth]{selected_snp.pdf}
\vspace{-5mm}
\caption{The top panel plots debiased estimators and corresponding CIs for $\{b^{(l)}_j\}_{1\leq l\leq 4, j\in \mathcal{S}}$ with the index set $\mathcal{S}=\{420, 443, 437, 423, 245, 424, 364, 229, 6, 177, 63, 84\}$. The bottom panel plots CIs for $\{[\bp_{\delta}]_j\}_{j\in \mathcal{S}}$ in the no covariate shift setting with $\delta\in \{0,0.2,0.5\}$.}
\label{fig: real-data}
\end{figure}
 Figure \ref{fig: real-data} also illustrates the SNPs with insignificant maximin effects, which correspond to the following three types of heterogeneous regression effects: (1) the SNPs (e.g., with indexes 364, 229) have opposite effects across different media; (2) the SNPs (e.g., with indexes 6, 177) only have a significant effect on one medium; (3) the SNPs (e.g., with indexes 63, 84) do not have any significant effect across different growth media. Figure \ref{fig: real-data} shows that the CIs for the ridge-type maximin effect get shorter with a larger penalty level $\delta$, which is coherent with the simulation results reported in Figures \ref{fig: setting 1} and \ref{fig: method comparison}.

\subsection{Generalizability of maximin effects to test media}
We demonstrate the generalizability of the maximin effect by examining seven test media: ``Lactate", ``SDS", ``Trehalose", ``6-Azauracil", ``YNB", ``YPD", and ``YPD.4C". In this application, the target distribution $\mathbb{Q}$ represents the joint distribution of $p=513$ SNPs and the colony growth size under a specific test medium. There is no covariate shift  since the covariates observations are the same across different growth media, and the only difference is the outcome variable (colony size). Due to the difference in growth media, the conditional outcome distribution $\mathbb{Q}_{Y|X}$ in the test media will likely differ from $\{\mathbb{P}^{(l)}_{Y|X}\}_{1\leq l\leq 4}$ in the source media. The outcome observations for the seven test media are only used to validate the maximin effect's generalizability instead of constructing the maximin effect. 

In the following, we examine whether the stable associations captured by the maximin effect can be generalized to the test media. Mainly, we investigate whether the SNPs with significant maximin effects also have significant effects in the test media. We use each test medium's own SNP and outcome data and conduct multiple testing to choose SNPs with significant effects. We report the results in Table \ref{tab: SNP index}, where we adjust for the multiplicity by applying the BH procedure \citep{benjamini1995controlling} and controlling FDR below $0.1$.

\begin{table}[htp!]
\centering
\resizebox{\linewidth}{!}{
\begin{tabular}{|c|c|c|c|}
\hline
&Media name & Number & Indexes of significant SNPs \\
\hline
\multirow{4}{*}{Source Media}&Ethanol & 6 & 73,186,419,420,423,443\\
&Lactose & 5 & 323,420,442,443,451\\
&5-Fluorouracil & 16 & 16,126,130,282,330,364,366,396,399,420,423,424,442,458,462,497\\
&Xylose & 12 & 73,80,207,245,356,364,420,423,437,443,459,496\\
\hline
\multirow{7}{*}{Test Media}&Lactate & 6 & 1,53,324,420,437,443\\
&SDS & 5 & 256,257,364,420,459\\
&Trehalose & 9 & 1,79,324,349,364,420,437,443,496\\
&6-Azauracil & 6 & 73,420,424,437,442,459\\
&YNB & 9 & 27,207,208,254,282,420,423,442,499\\
&YPD & 12 & 24,73,207,231,359,420,423,424,437,442,443,459\\
&YPD.4C & 6 & 73,342,364,420,423,459\\
\hline
\end{tabular}}
\caption{\small Significant SNPs for each growth medium after controlling the FDR below $0.1$; for example, for the ``Ethanol" medium, there are 6 significant SNPs with indexes $73,186,419,420,423,443.$}
\label{tab: SNP index}
\end{table}

\begin{figure}[H]
\centering
\includegraphics[width=0.95\linewidth]{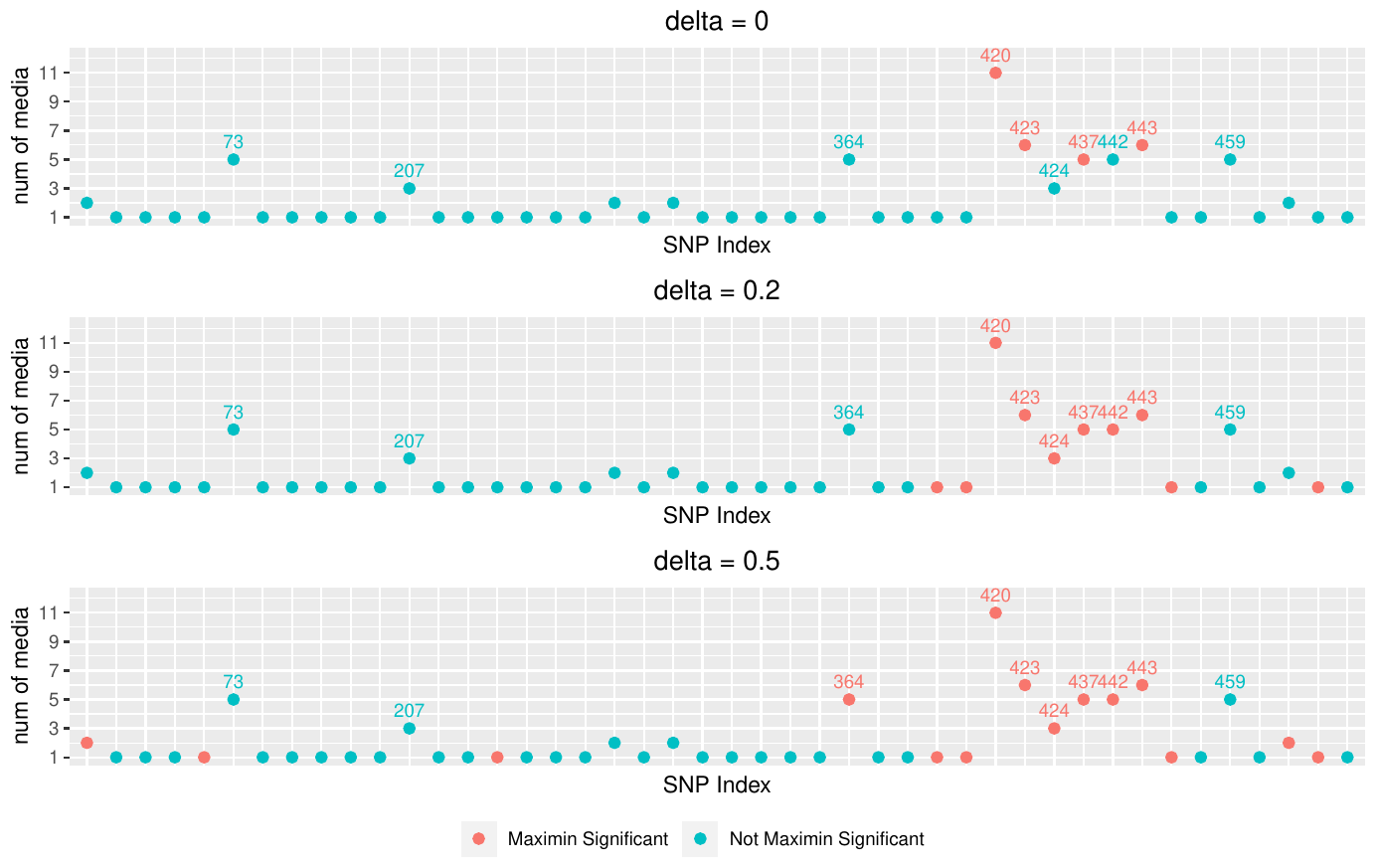}
\vspace{-3mm}
\caption{\small The y-axis represents the number of media, and the x-axis represents the SNP indexes. As an example, the SNP 420 is significant over 11 media. The red color indicates the maximin significance. The top, middle, and bottom panels correspond to the maximin effects with no ridge penalty, penalty $\delta=0.2$, and $\delta=0.5$, respectively.}
\label{fig: snps-withoutCS}
\end{figure}

We conduct the maximin significance test by applying the BH procedure to the p-values defined in \eqref{eq: p-val} and controlling the false discovery rate (FDR) below $0.1$. After adjusting for multiplicity, we obtain the maximin significant SNPs as $\{420, 423, 437, 443\}.$ 
In Figure \ref{fig: snps-withoutCS}, we plot a subset of SNPs that are maximin significant or significant in at least one source or test media. For every SNP, we report the number of growth media on which it has significant effects. Of all eleven media, SNP 420 is significant in all, SNPs 423 and 443 are significant in six, and SNP 437 is significant in five. The maximin significant SNPs $\{420, 423, 437, 443\}$ are shown to have generalizable effects for the test media. As reported in Table \ref{tab: SNP index}, the SNP 420 is significant in all seven test media, the SNP 437 is significant in four test media, and the SNPs 423 and 443 are significant in three test media. Figure \ref{fig: snps-withoutCS} also demonstrates that a larger group of maximin significant SNPs can be identified after increasing the ridge penalty $\delta$, where the identified SNPs 442 and 424 are significant in three and two test media, respectively. We report the names of the genes containing the maximin significant SNPs in Table \ref{table: gene names} in the supplement.

\vspace{-5mm}

\vspace{-3mm}
Figure \ref{fig: snps-withoutCS} demonstrates that the maximin significant SNPs are likely replicable for source and test media. We compare our maximin integration to the empirical risk minimization (ERM), which pools over the data from four source media and implements the standard debiased estimators and the following-up FDR control on this combined data. The ERM method identifies significant SNPs $\{420, 423, 437, 246, 357, 419\}$ with the SNPs $\{420, 423, 437\}$ being also identified as maximin significant. The important SNP 443 is maximin significant but not identified using ERM. Moreover, the ERM method selects some SNPs without generalizable effects: the SNP 419 is only identified as significant over a single source medium but not any test medium; the SNPs $246, 357$ are insignificant over any of the eleven growth media. This comparison illustrates that the maximin integration identifies SNPs with more generalizable effects across different environments than the ERM method. The stable associations summarized by maximin effects are easier to generalize to the target populations, which might have potential distribution shifts resulting from the different growth media used.

\section{Conclusion and Discussions}
\label{sec: discussion}
This paper advocates integrating multi-source data with maximin effects, a new data-fusion tool extracting generalizable information from heterogeneous data. The stable associations summarized by the maximin effects are more likely to generalize to a range of target populations that may have distributional shifts from the source populations. The maximin integration contrasts with other multi-source learning algorithms, including the meta-analysis and the regression analysis based on the merged data, which do not accommodate the distributional shifts between the source and target populations. Our proposed sampling approach addresses inference challenges arising in the maximin integration and helps address other non-standard inference problems. Interesting directions include inference for maximin effects when the linear models in \eqref{eq: multi-group model} are misspecified \citep{wasserman2014discussion,buhlmann2015high} and construction of distributionally robust models with machine learning prediction models. Both questions are left for future research.

\section*{Supplement}
The supplement contains all proofs and additional methods, theories, and numerical results.

 \setstretch{0.9}
\putbib
\end{bibunit}


\newpage 
\begin{bibunit}
\setstretch{1.32}


\appendix
\renewcommand{\thefigure}{S\arabic{figure}}
\setcounter{figure}{0}
\renewcommand{\thetable}{S\arabic{table}}
\setcounter{table}{0}
\setcounter{page}{1}


\begin{center}
\large Supplement to ``Inference for Maximin Effects: \\Identifying Stable Associations across Multiple Studies"
\end{center}
The supplementary materials are organized as follows,
\begin{enumerate}
\item In Section \ref{sec: additional discussion}, we provide additional discussions, methods, and theories.
\item In Section \ref{sec: non-regularity challenge}, we further discuss the non-regularity and instability challenges for confidence interval construction with bootstrap and subsampling methods.
\item We present the proofs of Theorems \ref{thm: sampling accuracy} and \ref{thm: inference for linear} in Sections \ref{sec: sampling proof} and \ref{sec: inference results}, respectively. In Section \ref{sec: prop proof}, we present the proofs of Propositions \ref{prop: identification} to
\ref{prop: general variance}. In Section \ref{sec: additional proof}, we provide the proofs of extra lemmas. 
\item In Section \ref{sec: additional num}, we present additional numerical studies. 
\end{enumerate}
\paragraph{Additional notations}. For positive sequences $a_n$ and $b_n$, $a_n \lesssim b_n$ means that $\exists C > 0$ such that $a_n \leq C b_n$ for all $n$;
$a_n \asymp b_n $ if $a_n \lesssim b_n$ and $b_n \lesssim a_n$, and $a_n \ll b_n$ if $\limsup_{n\rightarrow\infty} {a_n}/{b_n}=0$. For a set $S$, $\left|S\right|$ denotes its cardinality and $S^{c}$ denotes its complement. For a vector $x\in \R^{p}$ and a subset $S\subset[p]$, $x_{S}$ is the sub-vector of $x$ with indices in $S$ and $x_{-S}$ is the sub-vector with indices in $S^{c}$. The $\ell_q$ norm of a vector $x$ is defined as $\|x\|_{q}=\left(\sum_{l=1}^{p}|x_l|^q\right)^{\frac{1}{q}}$ for $q \geq 0$ with $\|x\|_0=\left|\{1\leq l\leq p: x_l \neq 0\}\right|$ and $\|x\|_{\infty}=\max_{1\leq l \leq p}|x_l|$. For a matrix $A$, we use $\lambda_{j}(A)$, $\|A\|_{F}$, $\|A\|_2$ and $\|A\|_{\infty}$ to denote its $j$-th largest singular value, Frobenius norm, spectral norm, and element-wise maximum norm, respectively.  For a matrix $X$, $X_{i}$ and $X_{\cdot, j}$ are used to denote its $i$-th row and $j$-th column, respectively; for index sets $S_1$ and $S_2$, $X_{S_1,S_2}$ denotes the sub-matrix of $X$ with row and column indices belonging to $S_1$ and $S_2,$ respectively; $X_{S_1}$ denotes the sub-matrix of $X$ with row indices belonging to $S_1.$ For random objects $X_1$ and $X_2$, we use $X_1\stackrel{d}{=}X_2$ to denote that they are equal in distribution. 
For a sequence of random variables $X_n$ indexed by $n$, we use $X_n \cip X$ and $X_{n} \cid X$ to represent that $X_n$ converges to $X$ in probability and in distribution, respectively.

\section{Additional Discussions}
\label{sec: additional discussion}
 In Section \ref{sec: minimax fairness}, we discuss the connection to minimax group fairness.
In Section \ref{sec: maximin proj}, we formulate the maximin projection as a form of the covariate-shift maximin effect.
In Section \ref{sec: high-dim Gamma}, we provide the details on inference for $\Gamma^{\T}$ in high dimensions.
\subsection{Minimax Group Fairness and Rawlsian Max-min Principle}
\label{sec: minimax fairness}
Fairness is an important consideration for designing the machine learning algorithm. In particular, the algorithm trained to maximize average performance on the training data set might under-serve or even cause harm to a sub-population of individuals \citep{dwork2012fairness}. The goal of the {\it group fairness} is to build a model satisfying a certain fairness notation (e.g. statistical parity) across predefined sub-populations. However, such fairness notation can be typically achieved by downgrading the performance on the benefitted groups without improving the disadvantaged ones \citep{martinez2020minimax,diana2021minimax}.  To address this, \cite{martinez2020minimax,diana2021minimax} proposed the \text{minimax group fairness} algorithm which ensures certain fairness principle and also maximizes the utility for each sub-population. We assume that we have access to the i.i.d data $\{Y_i, X_i, A_i\}_{1\leq i\leq n}$, where for the $i$-th observation, $Y_i$ and $X_i\in \R^{p}$ denote the outcome and the covariates, respectively, and $A_i$ denotes the sensitive variable (e.g. age or sex). The training data can be separated into different sub-groups depending on the value of the sensitive variable $A_i.$ For a discrete $A_i$, we use $\mathcal{A}$ to denote the set of all possible values that $A_i$ can take. Then the minimax group fairness can be defined as 
\begin{equation*}
\beta^{\rm mm-fair}:=\argmin_{\beta} \max_{a\in \mathcal{A}} \E_{Y_i, X_i\mid A_i=a}\ell(Y_i, X_i^{\intercal}\beta)=\argmax_{\beta} \min_{a\in \mathcal{A}} \E_{Y_i, X_i\mid A_i=a}[-\ell(Y_i, X_i^{\intercal}\beta)]
\end{equation*} 
where $\ell(Y_i, X_i^{\intercal}\beta)$ denotes a loss function and $-\ell(Y_i, X_i^{\intercal}\beta)$ can be viewed as a reward/utility function. In terms of utility maximization, the idea of minimax group fairness estimator dates at least back to Rawlsian max-min fairness \citep{rawls2001justice}. When the distribution of $X_i$ does not change with the value of $A_i$, then minimax group fairness estimator $\beta^{\rm mm-fair}$ is equivalent to the minimax or the maximin estimator, where the group label is determined by the value of $A_i.$

\subsection{Individualized Treatment Effect: Maximin Projection}
\label{sec: maximin proj}
\citet{shi2018maximin} proposed the maximin projection algorithm to construct the optimal treatment regime for new patients by leveraging training data from different groups with heterogeneity in optimal treatment decision. As explained in \citet{shi2018maximin}, the heterogeneity in optimal treatment decision might come from patients' different enrollment periods and/or the treatment quality from different healthcare centers. 
Particularly, \cite{shi2018maximin} considered that the data is collected from $L$ heterogeneous groups. For the $l$-th group with   $1\leq l\leq L$, let $Y^{(l)}_i\in \R$, $A^{(l)}_{i}$ and $X^{(l)}_{i}\in \R^{p}$ denote the outcome, the treatment and the baseline covariates, respectively. \cite{shi2018maximin} considered the following model for the data in group $l$, 
$$
Y^{(l)}_i=h_{l}(X^{(l)}_{i})+A^{(l)}_{i} \cdot [(b^{(l)})^{\intercal}X^{(l)}_{i}+c]+e^{(l)}_i \quad \text{with}\quad \E(e^{(l)}_i\mid X^{(l)}_{i}, A^{(l)}_{i})=0,
$$
where $h_l: \R^{p}\rightarrow \R$ denotes the unknown baseline function for the group $l$ and the vector $b^{(l)}\in \R^{p}$ describes the individualized treatment effect. 
To address the heterogeneity in optimal treatment regimes, \cite{shi2018maximin} has proposed the maximum projection 
\begin{equation}
\bpt=\argmax_{\|\beta\|_2\leq 1} \min_{1\leq l\leq L} \beta^{\intercal}b^{(l)}.
\label{eq: maximin proj}
\end{equation}
After identifying $\bpt,$ we may construct the treatment regime for a new patient with covariates $\xnew$ by testing
\begin{equation}
H_0: \xnew^{\intercal}\bpt+c< 0.
\label{eq: DTR maximin test}
\end{equation}
The following Proposition \ref{prop: maximin proj} identifies the maximin projection $\bpt.$ Through comparing it with Proposition \ref{prop: identification}, we note that the maximin projection is proportional to the general maximin effect defined in \eqref{eq: general maximin} with $\Sigma^{\mathbb{Q}}={\rm I}$ and hence the identification of $\bp(\mathbb{I})$ is instrumental in identifying $\bpt$, which provides the strong motivation for statistical inference for  $\xnew^{\intercal}\bp(\mathbb{I}).$

\begin{Proposition}
The maximum projection $\bpt$ in \eqref{eq: maximin proj} satisfies
$
\bpt= \frac{1}{\|\bp({\mathbb I})\|_2} \bp({\mathbb I})$ with  $\bp({\mathbb I})=\sum_{l=1}^{L}\gamma^{*}_{l} b^{(l)}$ where 
$\gamma^{*}=\argmin_{\gamma \in \Delta^{L}} \gamma^{\intercal} \Gamma^{\mathbb I}\gamma$ and $\Gamma_{lk}^{\mathbb I}=(b^{(l)})^{\intercal}b^{(k)}$ for $1\leq l,k \leq L.$
\label{prop: maximin proj}
\end{Proposition}
We refer to \cite{shi2018maximin} for more details on the maximin projection in the low-dimensional setting.
Our proposed sampling method in Section \ref{sec: method} is useful in devising statistical inference methods for $\bpt$ in high dimensions.

\subsection{Debiased Estimators of $\Gamma^{\T}$}
\label{sec: high-dim Gamma}

We present the details about constructing the debiased estimator ${\covest}^{\T}$ in \eqref{eq: cov est general}.  We estimate $\{b^{(l)}\}_{1\leq l\leq L}$ by applying Lasso \citep{tibshirani1996regression} to the sub-sample with the index set $A_l$: 
\begin{equation}
\widehat{b}_{init}^{(l)}=\argmin_{b \in \RR^{p}}\frac{\|Y^{(l)}_{A_l}-X^{(l)}_{A_l}b\|_2^2}{2|A_l|}+ {\lambda_l} \sum_{j=1}^{p} \frac{\|X^{(l)}_{A_l, j}\|_2}{\sqrt{|A_l|}} |b_j|, \; \text{with}\; \lambda_l=\sqrt{\frac{(2+c)\log p}{|A_l|}}\sigma_{l} 
\label{eq: Lasso splitting}
\end{equation}
for some constant $c>0.$ The Lasso estimators $\{\widehat{b}^{(l)}_{init}\}_{1\leq l\leq L}$ are implemented by the R-package \texttt{glmnet} \citep{friedman2010regularization} with tuning parameters $\{\lambda_{l}\}_{1\leq l\leq L}$ chosen by cross validation. We may also construct the initial estimator $\widehat{b}_{init}^{(l)}$ by tuning-free penalized estimators \citep{sun2012scaled, belloni2011square}. Define $$\widehat{\Sigma}^{\T}=\frac{1}{|B|}\sum_{i\in B} X^{\T}_{i} (X^{\T}_{i})^{\intercal},\quad \widehat{\Sigma}^{(l)}=\frac{1}{|B_l|}\sum_{i\in B_l} X^{(l)}_{i} [X^{(l)}_{i}]^{\intercal}, \quad \widetilde{\Sigma}^{\T}=\frac{1}{|A|}\sum_{i\in A} X^{\T}_{i}
(X^{\T}_{i})^{\intercal}.$$
For $1\leq l,k\leq L$, the plug-in estimator $[\widehat{b}_{init}^{(l)}]^{\intercal}\widehat{\Sigma}^{\T}\widehat{b}_{init}^{(k)}$ has the error decomposition:
\begin{equation*}
\begin{aligned}
(\widehat{b}_{init}^{(l)})^{\intercal} \widehat{\Sigma}^{\T}\widehat{b}_{init}^{(k)}-(b^{(l)})^{\intercal}\Sigma^{\T} b^{(k)}&=(\widehat{b}_{init}^{(k)})^{\intercal}\widehat{\Sigma}^{\T}(\widehat{b}_{init}^{(l)}-b^{(l)})+(\widehat{b}_{init}^{(l)})^{\intercal}\widehat{\Sigma}^{\T}(\widehat{b}_{init}^{(k)}-b^{(k)})\\
&-(\widehat{b}_{init}^{(l)}-b^{(l)})^{\intercal}\widehat{\Sigma}^{\T}(\widehat{b}_{init}^{(k)}-b^{(k)})+(b^{(l)})^{\intercal}(\widehat{\Sigma}^{\T}-\Sigma^{\T})b^{(k)}.
\end{aligned}
\end{equation*}

The debiased estimator in \eqref{eq: cov est general} is to correct the plug-in estimator $[\widehat{b}_{init}^{(l)}]^{\intercal}\widehat{\Sigma}^{\T}\widehat{b}_{init}^{(k)}$ by approximating the estimation error $(\widehat{b}_{init}^{(k)})^{\intercal}\widehat{\Sigma}^{\T}(\widehat{b}_{init}^{(l)}-b^{(l)})$ with 
$-\frac{1}{|B_l|}[\widehat{u}^{(l,k)}]^{\intercal}[X^{(l)}_{B_l}]^{\intercal}(Y^{(l)}_{B_l}-X^{(l)}_{B_l}\widehat{b}_{init}^{(l)}),$ where the projection direction $\widehat{u}^{(l,k)}$ is constructed as follows,
\begin{align}
\widehat{u}^{(l,k)}=\;\argmin_{u\in \R^{p}} u^{\intercal} \widehat{\Sigma}^{(l)} u \quad \text{subject to}&\; \|\widehat{\Sigma}^{(l)} u-\omega^{(k)}\|_{\infty} \leq  \|\omega^{(k)}\|_{2} \mu_{l} \label{eq: constraint 1}\\
&\; \left|[\omega^{(k)}]^{\intercal}\widehat{\Sigma}^{(l)} u-\|\omega^{(k)}\|_2^2\right|\leq \|\omega^{(k)}\|_2^2\mu_{l} \label{eq: constraint 2}\\
&\; \left\|X_{B_l} u\right\|_{\infty}\leq \|\omega^{(k)}\|_2 \tau_l \label{eq: constraint 3}
\end{align}
with $\mu_{l}\asymp \sqrt{\log p/|B_l|}$, $\tau_l\asymp \sqrt{\log n_l},$ and 
\begin{equation}
\omega^{(k)}=\widetilde{\Sigma}^{\T} \widehat{b}_{init}^{(k)}\in \R^{p}.
\label{eq: omega k}
\end{equation}
Similarly, we approximate the bias $(\widehat{b}_{init}^{(l)})^{\intercal}\widehat{\Sigma}^{\T}(\widehat{b}_{init}^{(k)}-b^{(k)})$  by 
$-\frac{1}{|B_k|}[\widehat{u}^{(k,l)}]^{\intercal}[X^{(k)}_{B_k}]^{\intercal}(Y^{(k)}_{B_k}-X^{(k)}_{B_k}\widehat{b}_{init}^{(k)}).$

We decompose the error of approximating $(\widehat{b}_{init}^{(k)})^{\intercal}\widehat{\Sigma}^{\T}(\widehat{b}_{init}^{(l)}-b^{(l)})$  by $-\frac{1}{|B_l|}[\widehat{u}^{(l,k)}]^{\intercal}[X^{(l)}_{B_l}]^{\intercal}(Y^{(l)}_{B_l}-X^{(l)}_{B_l}\widehat{b}_{init}^{(l)})$ as \begin{equation}
-\frac{1}{|B_l|}[\widehat{u}^{(l,k)}]^{\intercal}[X^{(l)}_{B_l}]^{\intercal}\epsilon^{(l)}_{B_l}+ [\widehat{\Sigma}^{(l)}\widehat{u}^{(l,k)}-\widehat{\Sigma}^{\T}\widehat{b}_{init}^{(k)}]^{\intercal}(\widehat{b}_{init}^{(l)}-b^{(l)}).
\label{eq: appro illustration}
\end{equation}
We now provide intuitions on why the projection direction $\widehat{u}^{(l,k)}$ proposed in \eqref{eq: constraint 1}, \eqref{eq: constraint 2} and \eqref{eq: constraint 3} ensures a small approximation error in \eqref{eq: appro illustration}. The objective $u^{\intercal} \widehat{\Sigma}^{(l)} u$ in \eqref{eq: constraint 1} is proportional to the variance of the first term in \eqref{eq: appro illustration}. The constraint set in \eqref{eq: constraint 1} implies $
\widehat{\Sigma}^{(l)}\widehat{u}^{(l,k)}-\widehat{\Sigma}^{\T}\widehat{b}_{init}^{(k)}\approx \widehat{\Sigma}^{(l)}\widehat{u}^{(l,k)}-\omega^{(k)}\approx {\bf 0},$ which guarantees the second term of \eqref{eq: appro illustration} to be small. 
The additional constraint \eqref{eq: constraint 2} is seemingly useless to control the approximation error in \eqref{eq: appro illustration}. However, this additional constraint ensures that the first term in \eqref{eq: appro illustration} dominates the second term in \eqref{eq: appro illustration}, which is critical in constructing an asymptotically normal estimator of $\Gamma^{\T}_{l,k}.$ The additional constraint \eqref{eq: constraint 2} is particularly useful in the covariate shift setting, that is, $\Sigma^{(l)}\neq \Sigma^{\T}$ for some $1\leq l\leq L.$ The last constraint \eqref{eq: constraint 3} is useful in establishing the asymptotic normality of the debiased estimator for non-Gaussian errors. We believe that the constraint \eqref{eq: constraint 3} is only needed for technical reasons.

To construct $\widehat{u}^{(l,k)}$ in \eqref{eq: cov est general}, we solve the dual problem of \eqref{eq: constraint 1} and \eqref{eq: constraint 2}, 
\begin{equation}
\widehat{h}=\argmin_{h \in \R^{p+1}} h^{\intercal} H^{\intercal}\widehat{\Sigma}^{(l)}H h/4+(\omega^{(k)})^{\intercal} H h/{\|\omega^{(k)}\|_2}+\lambda \|h\|_1\;\text{with}\; H=\left[\begin{matrix} {\omega^{(k)}}/{\|\omega^{(k)}\|_2}  , \mathbf{I}_{p\times p} \end{matrix} \right], 
\label{eq: dual problem}
\end{equation}
where we adopt the notation $0/0=0$. The objective value of this dual problem is unbounded from below when $H^{\intercal}\widehat{\Sigma}^{(l)}H$ is singular and $\lambda$ is near zero. We choose the smallest $\lambda>0$ such that the dual problem is bounded from below and construct $\widehat{u}^{(l,k)}=-\tfrac{1}{2}(\widehat{h}_{-1}+\widehat{h}_1 {\omega^{(k)}}/{\|\omega^{(k)}\|_2}).$

We estimate the covariance between $\widehat{\Gamma}^{\T}_{l_1,k_1}-{\Gamma}^{\T}_{l_1,k_1}$ and $\widehat{\Gamma}^{\T}_{l_2,k_2}-{\Gamma}^{\T}_{l_2,k_2}$ by $$\widehat{\bf V}_{\pi(l_1,k_1),\pi(l_2,k_2)}=\widehat{\bf V}^{(a)}_{\pi(l_1,k_1),\pi(l_2,k_2)}+\widehat{\bf V}^{(b)}_{\pi(l_1,k_1),\pi(l_2,k_2)},$$ where
{\small
\begin{equation}
\begin{aligned}
\widehat{\bf V}^{(a)}_{\pi(l_1,k_1),\pi(l_2,k_2)}=&\frac{\widehat{\sigma}_{l_1}^2}{|B_{l_1}|}(\widehat{u}^{(l_1,k_1)})^{\intercal}\widehat{\Sigma}^{(l_1)} \left[\widehat{u}^{(l_2,k_2)} {\bf 1}(l_2=l_1)+\widehat{u}^{(k_2,l_2)} {\bf 1}(k_2=l_1)\right]+\\
&\frac{\widehat{\sigma}_{k_1}^2}{|B_{k_1}|}(\widehat{u}^{(k_1,l_1)})^{\intercal}\widehat{\Sigma}^{(k_1)} \left[\widehat{u}^{(l_2,k_2)} {\bf 1}(l_2=k_1)+\widehat{u}^{(k_2,l_2)} {\bf 1}(k_2=k_1)\right],
\end{aligned}
\label{eq: var 1st}
\end{equation}}
and
{\small
\begin{equation}
\widehat{\bf V}^{(b)}_{\pi(l_1,k_1),\pi(l_2,k_2)}={\frac{1}{|B|N_{\T}}\sum_{i=1}^{N_{\T}} \left((\widehat{b}_{init}^{(l_1)})^{\intercal} X_{i}^{\T} (\widehat{b}_{init}^{(k_1)})^{\intercal}X_{i}^{\T}(\widehat{b}_{init}^{(l_2)})^{\intercal} X_{i}^{\T} (\widehat{b}_{init}^{(k_2)})^{\intercal}X_{i}^{\T} -(\widehat{b}_{init}^{(l_1)})^{\intercal}\bar{\Sigma}^{\T}\widehat{b}_{init}^{(k_1)}(\widehat{b}_{init}^{(l_2)})^{\intercal}\bar{\Sigma}^{\T}\widehat{b}_{init}^{(k_2)}\right)}
\label{eq: var 2nd}
\end{equation}}
with $\widehat{\sigma}_{l}^2$ defined in \eqref{eq: asymp var}, $\bar{\Sigma}^{\T}=\frac{1}{N_{\T}}\sum_{i=1}^{N_{\T}}X_{i}^{\T}(X_{i}^{\T})^{\intercal}$, and $\widehat{\Sigma}^{(l)}=\frac{1}{|B_l|}\sum_{i\in B_l} X^{(l)}_{i} [X^{(l)}_{i}]^{\intercal}$ for $1\leq l\leq L,$

\begin{Remark}\rm
If $\Sigma^{\T}$ is known, we modify ${\covest}^{\T}$ in \eqref{eq: cov est general} by replacing $\widehat{\Sigma}^{\T}$ by $\Sigma^{\T}$  and $\omega^{(k)}$ in \eqref{eq: omega k} by $\omega^{(k)}=\Sigma^{\T}\widehat{b}_{init}^{(k)}.$ The covariance matrix of the estimator (with known $\Sigma^{\T}$) will be $\widehat{\V}^{(a)}$ defined in \eqref{eq: cov def a} since $\widehat{\V}^{(b)}$ is used to quantify the uncertainty of estimating $\Sigma^{\T}$ but there is no uncertainty of estimating $\Sigma^{\T}.$ The estimator constructed with the knowledge of $\Sigma^{\T}$ typically has a smaller variance than the estimator ${\covest}^{\T}$ in \eqref{eq: cov est general} since there is no uncertainty of estimating $\Sigma^{\T}$; see Figure \ref{fig: method comparison} in the main paper for numerical comparisons. 
\end{Remark}
\subsubsection{Theoretical Justification}
In the following, we provide the theoretical guarantee of our proposed estimator $\widehat{\Gamma}^{\T}_{l,k}.$
Define \begin{equation}
\V_{\pi(l_1,k_1),\pi(l_2,k_2)}=\V^{(a)}_{\pi(l_1,k_1),\pi(l_2,k_2)}+\V^{(b)}_{\pi(l_1,k_1),\pi(l_2,k_2)},
\label{eq: cov def}
\end{equation}
with 
\begin{equation}
\begin{aligned}
\V^{(a)}_{\pi(l_1,k_1),\pi(l_2,k_2)}&=\frac{\sigma_{l_1}^2}{|B_{l_1}|}(\widehat{u}^{(l_1,k_1)})^{\intercal}\widehat{\Sigma}^{(l_1)} \left[\widehat{u}^{(l_2,k_2)} {\bf 1}(l_2=l_1)+\widehat{u}^{(k_2,l_2)} {\bf 1}(k_2=l_1)\right]\\
&+\frac{\sigma_{k_1}^2}{|B_{k_1}|}(\widehat{u}^{(k_1,l_1)})^{\intercal}\widehat{\Sigma}^{(k_1)} \left[\widehat{u}^{(l_2,k_2)} {\bf 1}(l_2=k_1)+\widehat{u}^{(k_2,l_2)} {\bf 1}(k_2=k_1)\right],
\end{aligned}
\label{eq: cov def a}
\end{equation}
and
\begin{equation}
\V^{(b)}_{\pi(l_1,k_1),\pi(l_2,k_2)}=\frac{1}{|B|}(\E [b^{(l_1)}]^{\intercal}X_{i}^{\T}[b^{(k_1)}]^{\intercal}X_{i}^{\T}[b^{(l_2)}]^{\intercal}X_{i}^{\T}[b^{(k_2)}]^{\intercal}X_{i}^{\T}-(b^{(l_1)})^{\intercal}\Sigma^{\T} b^{(k_1)}(b^{(l_2)})^{\intercal}\Sigma^{\T} b^{(k_2)}).
\label{eq: cov def b}
\end{equation}

Our analysis relies on the asymptotic normality of the point estimator $\widehat{\Gamma}^{\T}$. However, we shall emphasize that our analysis only requires the following Proposition \ref{prop: verification of A3}, which establishes that the marginal distribution of $\widehat{\Gamma}_{l,k}^{\T}-\Gamma_{l,k}^{\T}$ is approximately normal. This marginal limiting distribution is a weaker requirement than the joint asymptotic normality of $\{\widehat{\Gamma}_{l,k}^{\T}-\Gamma_{l,k}^{\T}\}_{1\leq k\leq l\leq L}$. However, the results established in the following proposition is already sufficient for our use of proving Theorem \ref{thm: sampling accuracy}. We present its proof at Section
\ref{sec: covariate shift}.
\begin{Proposition} Consider the model \eqref{eq: multi-group model}. Suppose that Conditions {\rm (A1)} and {\rm (A2)} hold, then 
the estimator $\widehat{\Gamma}^{\T}$ in \eqref{eq: cov est general} satisfies \eqref{eq: condition for sampling 2}
\begin{equation}
\liminf_{n,p\rightarrow \infty}\PP\left(\max_{1\leq l,k\leq L}\frac{\left|\widehat{\Gamma}_{l,k}^{\T}-\Gamma_{l,k}^{\T}\right|}{\sqrt{\widehat{\V}_{\pi(l,k),\pi(l,k)}+d_0/n}}\leq 1.05\cdot z_{{\alpha_0}/[L(L+1)]}\right)\geq 1-\alpha_0,
\label{eq: condition for sampling 2}
\end{equation}
for any $\alpha_0\in(0,0.01].$
\label{prop: verification of A3} 
\end{Proposition}

The proof of Proposition \ref{prop: verification of A3} relies on the following Propositions
\ref{prop: limiting Gamma univariate}
and \ref{prop: general variance}. The proofs of  Propositions \ref{prop: limiting Gamma univariate} and \ref{prop: general variance} can be found in Sections \ref{sec: covariate shift proof 1} and \ref{sec: variance bound}, respectively. 

\begin{Proposition} Consider the model \eqref{eq: multi-group model}. Suppose Condition {\rm (A1)} holds, $\tfrac{s \log p}{\min\{n,N_{\T}\}} \rightarrow 0$ with $n=\min_{1\leq l\leq L} n_{l}$ and $s=\max_{1\leq l\leq L}\|b^{(l)}\|_0$. Then the proposed estimator $\widehat{\Gamma}^{\T}_{l,k}\in \R^{L\times L}$ in \eqref{eq: cov est general} in the main paper satisfies 
$
\widehat{\Gamma}^{\T}_{l,k}-\Gamma^{\T}_{l,k}=D_{l,k}+\Rem_{l,k}, 
$ where
\begin{equation}
\frac{D_{l,k}}{\sqrt{\V_{\pi(l,k),\pi(l,k)}}}\cid \mathcal{N}(0,1),
\label{eq: limiting univariate}
\end{equation} 
with $\V$ defined in \eqref{eq: cov def}; for $1\leq l,k\leq L,$ with probability larger than $1-\min\{n,p\}^{-c}$ for a constant $c>0,$  the reminder term $\Rem_{l,k}$ satisfies \begin{equation}
\left|\Rem_{l,k}\right|\lesssim (1+\|\omega^{(k)}\|_2+\|\omega^{(l)}\|_2)\frac{s\log p}{n} +(\|b^{(k)}\|_2+\|b^{(l)}\|_2)\sqrt{\frac{s (\log p)^2}{n N_{\T}}},
\label{eq: bias}
\end{equation}
where  $c>0$ is a positive constant and $\omega^{(k)}$ and $\omega^{(l)}$ are defined in \eqref{eq: omega k}.  
\label{prop: limiting Gamma univariate}
\end{Proposition}

\begin{Proposition}
Suppose that the assumptions of Proposition \ref{prop: limiting Gamma univariate} hold. Then with probability larger than $1-\min\{n,p\}^{-c}$, the diagonal element ${\V}_{\pi(l,k),\pi(l,k)}$ in \eqref{eq: cov def} for $(l,k)\in \mathcal{I}_{L}$ satisfies,
\begin{equation}
\frac{\|\omega^{(l)}\|^2_2}{n_k} +\frac{\|\omega^{(k)}\|^2_2}{\nl} \lesssim{\V}^{(a)}_{\pi(l,k),\pi(l,k)}\lesssim \frac{\|\omega^{(l)}\|^2_2}{n_k} +\frac{\|\omega^{(k)}\|^2_2}{\nl},\quad {\V}^{(b)}_{\pi(l,k),\pi(l,k)}\lesssim {\frac{\|b^{(l)}\|^2_2\|b^{(k)}\|^2_2}{{N_{\T}}}},
\label{eq: diagonal order}
\end{equation}
where $c>0$ is a positive constant and $\omega^{(l)}$ and $\omega^{(k)}$ are defined in \eqref{eq: omega k}.
\begin{itemize} 
\item If $\Sigma^{\T}$ is known, then with probability larger than $1-\min\{n,p\}^{-c},$
$$
n\cdot {\V}_{\pi(l,k),\pi(l,k)} \lesssim \|b^{(k)}\|_2^2+\|b^{(l)}\|_2^2+s\log p/n.
$$
\item If $\Sigma^{\T}$ is unknown, then with probability larger than $1-\min\{n,p\}^{-c},$ {\small
\begin{equation}
n\cdot {\V}_{\pi(l,k),\pi(l,k)}\lesssim \left(1+{\frac{p}{N_{\T}}}\right)^2\left(\|b^{(k)}\|_2^2+\|b^{(l)}\|_2^2+s\frac{\log p}{n}\right)+\frac{n}{N_{\T}}\|b^{(l)}\|^2_2\|b^{(k)}\|^2_2.
\label{eq: variance 1}
\end{equation}}
\end{itemize}
\label{prop: general variance}
\end{Proposition}

\subsubsection{Special settings:  known $\Sigma^{\T}$ and no covariate shift}
\label{sec: high-dim Gamma noshift}

We consider the no covariate shift setting and will simplify the procedure of estimating $\Gamma^{\T}$. For $1\leq l\leq L$, we estimate $b^{(l)}$ by applying Lasso to the whole data set $(X^{(l)},Y^{(l)})$: 
\begin{equation}
\widehat{b}^{(l)}=\argmin_{b \in \RR^{p}} \|Y^{(l)}-X^{(l)}b\|_2^2/(2n_l)+\lambda_{l}\sum_{j=1}^{p} {\|X^{(l)}_{\cdot,j}\|_2}/{\sqrt{n_l}} \cdot |b_j|
\label{eq: Lasso}
\end{equation}
with $\lambda=\sqrt{(2+c)\log p/n_l} \sigma_l$ for some constant $c>0.$ 
Since $\Sigma^{(l)}=\Sigma^{\T}$ for $1\leq l\leq L$, we define $$
\widehat{\Sigma}=\frac{1}{\sum_{l=1}^{L} \nl+N_{\T}}\left(\sum_{l=1}^{L}
\sum_{i=1}^{\nl} X^{(l)}_{i}[X^{(l)}_{i}]^{\intercal} +\sum_{i=1}^{N_{\T}} X^{(l)}_{i}[X^{(l)}_{i}]^{\intercal}\right)
$$
and estimate $\Gamma_{l,k}$ by \begin{equation}
{\covest}^{\T}=(\widehat{b}^{(l)})^{\intercal} \widehat{\Sigma}\widehat{b}^{(k)}+(\widehat{b}^{(l)})^{\intercal}\frac{1}{n_k}[X^{(k)}]^{\intercal}(Y^{(k)}-X^{(k)}\widehat{b}^{(k)})+(\widehat{b}^{(k)})^{\intercal}\frac{1}{n_l}[X^{(l)}]^{\intercal}(Y^{(l)}-X^{(l)}\widehat{b}^{(l)}).
\label{eq: inner est}
\end{equation} 
This estimator can be viewed as a special case of \eqref{eq: cov est general} by taking $\widehat{u}^{(l,k)}$ and $\widehat{u}^{(k,l)}$ as $\widehat{b}^{(k)}$ and $\widehat{b}^{(l)}$, respectively. Neither the optimization in \eqref{eq: constraint 1} and \eqref{eq: constraint 2} nor the sample splitting is needed for constructing the debiased estimator in the no covariate shift setting.

We estimate the covariance between $\widehat{\Gamma}^{\T}_{l_1,k_1}-{\Gamma}^{\T}_{l_1,k_1}$ and $\widehat{\Gamma}^{\T}_{l_2,k_2}-{\Gamma}^{\T}_{l_2,k_2}$ by 
\begin{equation}
\widehat{\V}_{\pi(l_1,k_1),\pi(l_2,k_2)}=\widehat{\V}^{(1)}_{\pi(l_1,k_1),\pi(l_2,k_2)}+\widehat{\V}^{(2)}_{\pi(l_1,k_1),\pi(l_2,k_2)}
\label{eq: cov def no shift estimation}
\end{equation} 
where {\begin{equation*}
\begin{aligned}
\widehat{\V}^{(1)}_{\pi(l_1,k_1),\pi(l_2,k_2)}&=\frac{\widehat{\sigma}_{l_1}^2}{n_{l_1}^2}[b^{(l_1)}]^{\intercal}[X^{(l_1)}]^{\intercal} X^{(l_1)} \left[b^{(l_2)} {\bf 1}(l_2=l_1)+b^{(k_2)} {\bf 1}(k_2=l_1)\right]\\
&+\frac{\widehat{\sigma}_{k_1}^2}{n_{k_1}^2}[b^{(k_1)}]^{\intercal}[X^{(k_1)}]^{\intercal} X^{(k_1)}\left[b^{(l_2)} {\bf 1}(l_2=k_1)+b^{(k_2)} {\bf 1}(k_2=k_1)\right]
\end{aligned}
\label{eq: cov def 1 est}
\end{equation*}
{\small
\begin{equation*}
\begin{aligned}
&\widehat{\V}^{(2)}_{\pi(l_1,k_1),\pi(l_2,k_2)}={\frac{\sum_{i=1}^{N_{\T}} \left((\widehat{b}^{(l_1)})^{\intercal} X_{i}^{\T} (\widehat{b}^{(k_1)})^{\intercal}X_{i}^{\T}(\widehat{b}^{(l_2)})^{\intercal} X_{i}^{\T} (\widehat{b}^{(k_2)})^{\intercal}X_{i}^{\T} -(\widehat{b}^{(l_1)})^{\intercal}\widehat{\Sigma}\widehat{b}^{(k_1)}(\widehat{b}^{(l_2)})^{\intercal}\widehat{\Sigma}\widehat{b}^{(k_2)}\right)}{(\sum_{l=1}^{L}n_{l}+N_{\T})^2}}\\
&+{\frac{\sum_{l=1}^{L}\sum_{i=1}^{n_l} \left((\widehat{b}^{(l_1)})^{\intercal} X^{(l)}_{i} (\widehat{b}^{(k_1)})^{\intercal}X^{(l)}_{i}(\widehat{b}^{(l_2)})^{\intercal} X^{(l)}_{i} (\widehat{b}^{(k_2)})^{\intercal}X^{(l)}_{i} -(\widehat{b}^{(l_1)})^{\intercal}\widehat{\Sigma}\widehat{b}^{(k_1)}(\widehat{b}^{(l_2)})^{\intercal}\widehat{\Sigma}\widehat{b}^{(k_2)}\right)}{(\sum_{l=1}^{L}n_{l}+N_{\T})^2}}
\end{aligned}
\label{eq: cov def 2 est}
\end{equation*}}
}
\section{Inference Challenges with Bootstrap and Subsampling}
\label{sec: non-regularity challenge}
We demonstrate the challenges of confidence interval construction for the maximin effects with bootstrap and subsampling methods. 
\subsection{Simulation Settings (I-1) to (I-10)} 
\label{sec: I settings}
We focus on the no covariate shift setting with $\Sigma^{(l)} = {\bf I}_p$ for $1\leq l\leq L$ and $\Sigma^{\T}={\bf I}_p$. In the following, we describe how to generate the settings (I-1) to (I-6) with non-regularity and instability. We set $L=4$. For $1\leq l \leq L$, we generate $b^{(l)}$ as $b^{(l)}_j = j/20+ \kappa^{(l)}_j$  for $1\leq j \leq 5$  with  $\{\kappa^{(l)}_j\}_{1\leq j\leq 5} \stackrel{i.i.d.}{\sim} \mathcal{N}(0, \sigma_{\rm irr}^2)$, $b^{(l)}_j = j/20$ for $6\leq j \leq 10$, and $b^{(l)}_j=0$ for $11\leq j\leq p$. Set $\xnew_{j}=1$ for $1\leq j\leq 5$ and zero otherwise.  
We consider the setting (I-0) as the special setting with $\sigma_{\rm irr}=0$, that is, $b^{(1)}=\cdots=b^{(L)}.$
We choose the following six combinations of $\sigma_{\rm irr}$ and the random seed for generating $\kappa^{(l)}_j,$
\begin{enumerate}
\item[] (I-1) $\sigma_{\rm irr}=0.05$, ${\rm seed}=42$; (I-2) $\sigma_{\rm irr}=0.05$, ${\rm seed}=20$; (I-3) $\sigma_{\rm irr}=0.10$, ${\rm seed}=36$;
\item[](I-4) $\sigma_{\rm irr}=0.15$, ${\rm seed}=17$; (I-5) $\sigma_{\rm irr}=0.20$, ${\rm seed}=12$; (I-6) $\sigma_{\rm irr}=0.25$, ${\rm seed}=31$.
\end{enumerate}
In addition, we generate the following non-regular settings:  \begin{enumerate}
\item[(I-7)] $L=2$; $b^{(1)}_1=2$, $b^{(1)}_j=j/40$ for $2\leq j \leq 10$ and $b^{(1)}_j=0$ otherwise; $b^{(2)}_1=-0.03$, $b^{(2)}_j=j/40$ for $2\leq j\leq 10$ and $b^{(1)}_j=0$ otherwise; ${\xnew}=e_1.$
\item[(I-8)] Same as (I-7) except for $b^{(l)}_j=(10-j)/40$ for $11\leq j\leq 20$ and $l=1,2$;
\item [(I-9)] Same as (I-7) except for $b^{(l)}_j=1 $ for $2\leq j\leq 30$ and $l=1,2$.
\end{enumerate}
Finally, we generate (I-10) as a favorable setting without non-regularity or instability.
\begin{enumerate}
\item[(I-10)] $L=2$; $b^{(1)}_j=j/20$ for $1\leq j\leq 10$, $b^{(2)}_j=-j/20$ for $1\leq j \leq10$; $[{\xnew}]_{j}=j/5$ for $1\leq j\leq 5$ and $[{\xnew}]_{j}=0$ otherwise.
\end{enumerate}
\subsection{Challenges for Bootstrap and Subsampling: Numerical Evidence}
\label{sec: low-dim method}

In Section \ref{sec: sim}, we have reported the under-coverage of the normality CIs in high dimensions.  In the following, we explore a low dimensional setting with $p=30$ and $n_1=\cdots=n_{L}=n=1000$. We shall compare our proposed CI, the CI assuming asymptotic normality \citep{rothenhausler2016confidence}, and CIs by the subsampling or bootstrap methods. We describe these methods in the following. 

%

\vspace{-5mm}
\paragraph{Magging estimator and CI assuming asymptotic normality.}
In low-dimensional setting, the Magging estimator has been proposed in \citet{buhlmann2015magging} to estimate the maximin effect. 
In low dimensions, the regression vector $b^{(l)}$ is estimated by the ordinary least square estimator $\widehat{b}^{(l)}_{\rm OLS}$ for $1\leq l\leq L$ and the covariance matrix $\Sigma$ is estimated by the sample covariance matrix 
$\widehat{\Sigma}=\frac{1}{\sum_{l=1}^{L}n_l} \sum_{l=1}^{L}\sum_{i=1}^{n} X^{(l)}_i [X^{(l)}_i]^{\intercal}.$ 
Then the Magging estimator in low dimension is of the form,
\begin{equation}
\widehat{\beta}^{\rm magging}=\sum_{l=1}^{L}\widehat{\gamma}_{l} \widehat{b}_{\rm OLS}^{(l)} \quad \text{with}\quad \widehat{\gamma}\coloneqq \argmin_{\gamma \in \Delta^{L}} \gamma^{\intercal}\widehat{\Gamma}\gamma
\label{eq: optimal weight low}
\end{equation}
where $\widehat{\Gamma}_{lk}=(\widehat{b}_{\rm OLS}^{(l)})^{\intercal}\widehat{\Sigma} \widehat{b}_{\rm OLS}^{(k)}$ for $1\leq l,k \leq L$ and $\Delta^{L}=\{\gamma\in \RR^{L}: \gamma_j\geq0,\; \sum_{j=1}^{L}\gamma_j=1\}$ is the simplex over $\R^{L}$.  \citet{rothenhausler2016confidence} have established the asymptotic normality of the magging estimator under certain conditions, which essentially ruled out the non-regularity and instability settings. In the following, we shall show that the CI assuming asymptotic normality fails to provide valid inference for the low-dimensional maximin effects in the presence of non-regularity or instability. In particular, we construct a normality CI of the form 
\begin{equation}
(\xnew^{\intercal}\widehat{\beta}^{\rm magging}-1.96\cdot \widehat{\rm SE},\xnew^{\intercal}\widehat{\beta}^{\rm magging}+1.96\cdot \widehat{\rm SE}),
\label{eq: normality CI low}
\end{equation}
 where  $\widehat{\rm SE}$ denotes the sample standard deviation of $\xnew^{\intercal}\widehat{\beta}^{\rm magging}$ calculated based on $500$ simulations. Since $\widehat{\rm SE}$ is calculated in an oracle way, this normality CI is a favorable implementation of the CI construction in \citet{rothenhausler2016confidence}. 

\paragraph{Bootstrap and subsampling.} We briefly describe the implementation of the bootstrap and subsampling methods. We compute the point estimator for the original data as in \eqref{eq: optimal weight low}, denoted as $\hat{\theta}$. For $1\leq l\leq L,$ we randomly sample $m$ observations (with/without replacement) from $(X^{(l)},Y^{(l)})$ and use this generated sample to compute the point estimator $\hat{\theta}^{m,j}$ as in \eqref{eq: optimal weight low}. We conduct the random sampling $500$ times to obtain $\{\hat{\theta}^{m,j}\}_{1\leq j\leq 500}$ and define the empirical CDF as,
$L_n(t) = \frac{1}{500} \sum_{j=1}^{500} {\bf 1}\left(\sqrt{m}(\hat{\theta}^{m,j} - \hat{\theta}) \leq t \right),$
where ${\bf 1}$ denotes the indicator function. Define $\hat{t}_{\alpha/2}$ as the minimum $t$ value such that $L_n(t)\geq \alpha/2$ and $\hat{t}_{1-\alpha/2}$ as the minimum $t$ value such that $L_n(t)\geq 1-\alpha/2.$ 
We construct the bootstrap/subsampling confidence interval as
$
\left[ \hat{\theta} - \frac{\hat{t}_{1-\alpha/2}}{\sqrt{n}}, \hat{\theta} - \frac{\hat{t}_{\alpha/2}}{\sqrt{n}}\right].
$


\begin{table}[H]
\centering
\resizebox{\linewidth}{!}{
\begin{tabular}[t]{|c|c|c|c|c|c|c|c|c|c|c|c|c|}
\hline
\multicolumn{1}{|c|}{Setting} & \multicolumn{1}{c|}{ } & \multicolumn{4}{c|}{m-out-of-n subsampling} & \multicolumn{5}{c|}{m-out-of-n bootstrap} &
\multicolumn{2}{c|}{Proposed}\\
\cline{3-6} \cline{7-11} \cline{12-13}
$p=30$ & normality & $m=200$  & $m=300$ & $m=400$ & $m=500$ & $m=200$ & $m=300$ & $m=400$ & $m=500$ & $m=1000$ & Cov & L-ratio\\
\hline
(I-0) & 0.976 & 0.824  & 0.822 & 0.802 & 0.812 & 0.852 & 0.828 & 0.876 & 0.894 & 0.916 & 1.000 & 1.344\\
\hline
(I-1) & 0.686 & 0.380  & 0.390 & 0.380 & 0.408 & 0.432 & 0.460 & 0.502 & 0.490 & 0.536 & 0.956 & 1.678\\
\hline
(I-2) & 0.808 & 0.418 & 0.456 & 0.474 & 0.446 & 0.456 & 0.510 & 0.532 & 0.546 & 0.602 & 0.990 & 1.723\\
\hline
(I-3) & 0.770 & 0.392  & 0.494 & 0.464 & 0.440 & 0.482 & 0.480 & 0.516 & 0.544 & 0.634 & 0.984 & 1.766\\
\hline
(I-4) & 0.816 & 0.620 & 0.668 & 0.670 & 0.668 & 0.672 & 0.694 & 0.710 & 0.700 & 0.794 & 0.990 & 1.686\\
\hline
(I-5) & 0.790 & 0.612 & 0.626 & 0.626 & 0.594 & 0.628 & 0.664 & 0.702 & 0.732 & 0.732 & 1.000 & 1.843\\
\hline
(I-6) & 0.806 & 0.590 & 0.636 & 0.654 & 0.632 & 0.626 & 0.682 & 0.698 & 0.712 & 0.760 & 0.994 & 1.833\\
\hline
(I-7) & 0.824 & 0.914 & 0.932 & 0.908 & 0.888 & 0.890 & 0.950 & 0.934 & 0.950 & 0.952 & 0.996 & 5.078\\
\hline
(I-8) & 0.888 & 0.912 & 0.934 & 0.884 & 0.856 & 0.914 & 0.916 & 0.932 & 0.932 & 0.958 & 0.996 & 4.017\\
\hline
(I-9) & 0.900 & 0.834 & 0.822 & 0.788 & 0.778 & 0.914 & 0.914 & 0.916 & 0.862 & 0.914 & 0.996 & 2.061\\
\hline
(I-10) & 0.954 & 0.956 & 0.904 & 0.866 & 0.836 & 0.956 & 0.948 & 0.962 & 0.964 & 0.942 & 1.000 & 1.725\\
\hline
\end{tabular}}
\caption{Empirical coverage of the normality CI in \eqref{eq: normality CI low}, the CI by subsampling, the CI by $m$ out of $n$ bootstrap and our proposed CI, where the column indexed with L-ratio denoting the ratio of the average length of our proposed CI to that of the normality CI.}
\label{tab: subsampling}
\end{table}

In Table \ref{tab: subsampling}, we report the empirical coverage of the normality CI in \eqref{eq: normality CI low}, the CI by subsampling, the CI by $m$ out of $n$ bootstrap,  and our proposed CI. The normality CI in \eqref{eq: normality CI low} and the CIs by subsampling and bootstrap methods are in general under-coverage for settings (I-1) to (I-9). Our proposed CI achieves the desired coverage level at the expense of a wider interval. For the favorable setting (I-10), bootstrap methods achieve the desired coverage level while subsampling methods only work for a small $m,$ which is an important requirement for the validity of subsampling methods \citep{politis1999subsampling}.

In Section \ref{sec: challenge subsampling}, we discuss why subsampling methods fail to provide valid inference for the maximin effect in the presence of non-regularity.

\subsection{Challenges for Bootstrap and Subsampling Methods: A Theoretical View}
\label{sec: challenge subsampling}
We illustrate the challenge of bootstrap and subsampling methods for the maximin effects in non-regular settings. As a remark, the following argument is not a rigorous proof but of a similar style to the discussion of \cite{andrews2000inconsistency}, explaining why the subsampling methods do not completely solve the non-regular inference problems. The main difficulty appears in a  near boundary setting with  
\begin{equation}
\gamma_1=\frac{{\mu}_1}{\sqrt{n}} \quad \text{for a positive constant}\quad \mu_1>0,
\label{eq: setting}
\end{equation}
where $\gamma_1$ denotes the weight of the first group.

To illustrate the problem, we consider the special setting $L=2$, $n_1=n_2=n$ and $b^{(1)}$ and $b^{(2)}$ are known. The first coefficient of the maximin effect $\bp$ can be expressed as
$
\bp_1=b^{(1)}_1\cdot\gamma_1+b^{(2)}_1\cdot(1-\gamma_1)=(b^{(1)}_1-b^{(2)}_1)\cdot \gamma_1+b^{(2)}_1.
$
For this special scenario, the only uncertainty is from estimating $\gamma_1$ since $\Sigma^{\T}$ is unknown. 
We estimate $\gamma_1$ by $$\widehat{\gamma}_1=\max\{\bar{\gamma}_1,0\} \quad\text{with}\quad \bar{\gamma}_1=\frac{\widehat{\Gamma}_{22}-\widehat{\Gamma}_{12}}{\widehat{\Gamma}_{11}+\widehat{\Gamma}_{22}-2\widehat{\Gamma}_{12}},$$
where $\widehat{\Gamma}_{12}=[b^{(1)}]^{\intercal}\widehat{\Gamma}^{\T}b^{(2)},$ $\widehat{\Gamma}_{11}=[b^{(1)}]^{\intercal}\widehat{\Gamma}^{\T}b^{(1)},$ and $\widehat{\Gamma}_{22}=[b^{(2)}]^{\intercal}\widehat{\Gamma}^{\T}b^{(2)}.$ In the definition of $\widehat{\gamma}_1$, we do not restrict it to be smaller than 1 as this happens with a high probability under our current setting \eqref{eq: setting}.
We then estimate $\bp_1$ by 
$
\widehat{\beta}_1=(b^{(1)}_1-b^{(2)}_1)\cdot \widehat{\gamma}_1+b^{(2)}_1. 
$

We separately subsample $\{X^{(1)}_{i},Y^{(1)}_{i}\}_{1\leq i\leq n}$ and $\{X^{(2)}_{i},Y^{(2)}_{i}\}_{1\leq i\leq n}$ and use $m$ to denote the subsample size. For $1\leq t\leq T$ with a positive integer $T>0$, denote the $t$-th subsampled data as $\{X^{(*,t,1)}_{i},Y^{(*,t,1)}_{i}\}_{1\leq i\leq m}$ and $\{X^{(*,t,2)}_{i},Y^{(*,t,2)}_{i}\}_{1\leq i\leq m}$. We apply these subsampled data sets to compute the sample covariance matrix  $\widehat{\Sigma}^{(*,t)}$. Then we compute $\bar{\gamma}^{(*,t)}_1$ as 
$$\widehat{\gamma}^{(*,t)}_1=\max\{\bar{\gamma}^{(*,t)}_1,0\} \quad\text{with}\quad \bar{\gamma}^{(*,t)}_1=\frac{\widehat{\Gamma}^{(*,t)}_{22}-\widehat{\Gamma}^{(*,t)}_{12}}{\widehat{\Gamma}^{(*,t)}_{11}+\widehat{\Gamma}^{(*,t)}_{22}-2\widehat{\Gamma}^{(*,t)}_{12}},$$
with $\widehat{\Gamma}^{(*,t)}_{12}=[b^{(1)}]^{\intercal}\widehat{\Sigma}^{(*,t)}b^{(2)},$ $\widehat{\Gamma}^{(*,t)}_{11}=[b^{(1)}]^{\intercal}\widehat{\Sigma}^{(*,t)} b^{(1)},$ and $\widehat{\Gamma}^{(*,t)}_{22}=[b^{(2)}]^{\intercal}\widehat{\Sigma}^{(*,t)} b^{(2)}.$ Then we construct the subsampling estimator 
$\widehat{\beta}^{(*,t)}_1=(b^{(1)}_1-b^{(2)}_1)\cdot \widehat{\gamma}^{(*,t)}_1+b^{(2)}_1.$

We assume $\sqrt{n}(\bar{\gamma}_1-\gamma_1)$ and $\sqrt{n}(\bar{\gamma}^{(*,t)}_1-\bar{\gamma}_1)$ share the same limiting normal distribution. Specifically, we assume $\sqrt{n}(\bar{\gamma}_1-\gamma_1)\cid Z$ with $Z\sim N(0,V_{\gamma})$ and conditioning on the observed data, $\sqrt{n}(\bar{\gamma}^{(*,t)}_1-\bar{\gamma}_1)\cid Z.$ Then for the setting \eqref{eq: setting}, we have 
\begin{equation}
\sqrt{n}\left(\widehat{\gamma}_1-\gamma_1\right)\cid \max\{Z,-\mu_1\}. 
\label{eq: limiting dist}
\end{equation}

In the following, we will show that $\sqrt{m}(\widehat{\gamma}^{(*,t)}_1-\widehat{\gamma}_1)$ does not approximate the limiting distribution of $\sqrt{n}(\widehat{\gamma}_1-\gamma_1)$ in \eqref{eq: limiting dist} if $\gamma_1=\frac{{\mu}_1}{\sqrt{n}}$. Note that
\begin{equation*}
\begin{aligned}
\sqrt{m}(\widehat{\gamma}^{(*,t)}_1-\widehat{\gamma}_1)=&\max\{\sqrt{m}[\bar{\gamma}^{(*,t)}_1-\widehat{\gamma}_1],-\sqrt{m}\widehat{\gamma}_1\}\\
=&\max\{\sqrt{m}[\bar{\gamma}^{(*,t)}_1-\bar{\gamma}_1]+\sqrt{m}[\bar{\gamma}_1-\gamma_1],-\sqrt{m}{\gamma}_1\}-\sqrt{m}(\widehat{\gamma}_1-\gamma_1).\\
\end{aligned}
\end{equation*}
When $m\ll n$ and $\gamma_1\asymp 1/\sqrt{n},$ then the following event happens with a probability larger than $1-n^{-c}$ for a small positive constant $c>0$, 
$$
\mathcal{A}_0=\left\{\max\{\sqrt{m}[\bar{\gamma}_1-\gamma_1],\sqrt{m}{\gamma}_1,\sqrt{m}(\widehat{\gamma}_1-\gamma_1)\}\lesssim \sqrt{\frac{m\log n}{n}}\right\}.
$$
Then conditioning on the event $\mathcal{A}_0,$
$
\sqrt{m}(\widehat{\gamma}^{(*,t)}_1-\widehat{\gamma}_1) \cid \max\{Z,0\},
$
which is different from the limiting distribution of $\sqrt{n}(\widehat{\gamma}_1-\gamma_1)$  in \eqref{eq: limiting dist}.

\section{Proofs of Propositions \ref{prop: identification}, \ref{prop: DRO ridge}, \ref{prop: maximin proj}, \ref{prop: limiting Gamma univariate}, \ref{prop: general variance} and \ref{prop: verification of A3}}
\label{sec: prop proof}
\subsection{Proof of Proposition \ref{prop: identification}}
We now supply a proof of Proposition \ref{prop: identification}, which follows from \eqref{eq: general maximin} and the proof of Theorem 1 in   \cite{meinshausen2015maximin}. 
We start with the proof of \eqref{eq: general maximin} in the main paper. For any $\Tar \in \mathcal{C}(\T_{X}),$ we express its conditional outcome model as $\Tar_{Y|X}=\sum_{l=1}^{L} q_l \cdot \mathbb{P}_{Y|X}^{(l)}$ for some weight vector $q\in \Delta^{L}.$  Then we have \begin{equation}
\begin{aligned}
\E_{X_i, Y_i\sim \Tar} \left[Y_i^2-(Y_i-X_i^{\intercal}\beta)^2\right]=&\E_{X_i, Y_i\sim\Tar} \left[2Y_i X_i^{\intercal}\beta-\beta^{\intercal}X_i X_i^{\intercal}\beta \right]\\
=&\E_{X_i\sim \T_{X}}  \left[\sum_{l=1}^{L} q_l \cdot \E_{Y_i\mid X_i\sim \mathbb{P}^{(l)}_{Y|X}}\left[2Y_i X_i^{\intercal}\beta-\beta^{\intercal}X_i X_i^{\intercal}\beta \right]\right]\\
=&\sum_{l=1}^{L} q_l \cdot \E_{X_i\sim \T_{X}}  \E_{Y_i\mid X_i\sim \mathbb{P}^{(l)}_{Y|X}}\left[2Y_i X_i^{\intercal}\beta-\beta^{\intercal}X_i X_i^{\intercal}\beta \right].
\end{aligned}
\label{eq: expression mixture}
\end{equation}
By the outcome model \eqref{eq: multi-group model}, we have 
$$\E_{X_i\sim \T_{X}} \E_{Y_i\mid X_i\sim \mathcal{P}^{(l)}_{Y|X}}\left[2Y_i X_i^{\intercal}\beta-\beta^{\intercal}X_i X_i^{\intercal}\beta \right]=2 b^{(l)}\Sigma^{\T}\beta-\beta^{\intercal}\Sigma^{\T}\beta,$$
where $\Sigma^{\T}=\E X_{1}^{\T} (X_{1}^{\T})^{\intercal}.$ 
Together with \eqref{eq: expression mixture}, we have 
\begin{equation*}
\begin{aligned}
R_{\T}(\beta)
=\min_{q\in \Delta^{L}}\sum_{l=1}^{L} q_l \cdot \left[2 b^{(l)}\Sigma^{\T}\beta-\beta^{\intercal}\Sigma^{\T}\beta\right]=\min_{b\in \mathbb{B}} \left[2 b^{\intercal}\Sigma^{\T}\beta- \beta^{\intercal}\Sigma^{\T} \beta\right],
\end{aligned}
\end{equation*}
where $\mathbb{B}=\{b\in \R^{p}: b=\sum_{l=1}^{L}q_l\cdot b^{(l)} \; \text{with}\; q\in \Delta^{L}\}.$
The above equation establishes $
\bp=\argmax_{\beta\in \R^{p}} \min_{b\in \mathbb{B}}  \left[2 b^{\intercal}\Sigma^{\T}\beta- \beta^{\intercal}\Sigma^{\T} \beta\right],
$ which is \eqref{eq: general maximin} in the main paper. We decompose $\Sigma^{\T}=C^{\intercal} C$ such that $C$ is invertible. Define $\widetilde{\mathbb{B}}=C^{-1} \mathbb{B}.$ Then we have  $\bp=C^{-1}\xi^*$ with 
\begin{equation}
\xi^*=\argmax_{\xi \in \R^{p}} \min_{u\in \widetilde{\mathbb{B}}}  \left[2 u^{\intercal}\xi-\xi^{\intercal}\xi\right].
\label{eq: original}
\end{equation}
If we interchange $\min$ and $\max$ in the above equation, then we have 
\begin{equation}
\xi^{*}=\argmin_{\xi \in \widetilde{\mathbb{B}}} \xi^{\intercal}\xi
\label{eq: inter-change}
\end{equation}
We will justify this inter-change by showing that the solution $\xi^{*}$ defined in \eqref{eq: inter-change} is the solution to \eqref{eq: original}. For any $\nu\in [0,1]$ and $\mu\in \widetilde{\mathbb{B}}$, we use the fact $\xi^{*}+\nu(\mu-\xi^{*})\in \widetilde{\mathbb{B}}$ and obtain
$
\|\xi^{*}+\nu(\mu-\xi^{*})\|_2^2\geq \|\xi^{*}\|_2^2.
$
This leads to 
$
(\xi^{*})^{\intercal} \mu-(\xi^{*})^{\intercal}\xi^{*}\geq 0, 
$
and hence 
$
2(\xi^{*})^{\intercal} \mu-(\xi^{*})^{\intercal}\xi^{*}\geq (\xi^{*})^{\intercal}\xi^{*}$ {for any}  $\mu\in \widetilde{\mathbb{B}}.
$
By taking $\xi$ as $\xi^{*}$ in the optimization problem \eqref{eq: original}, we have 
\begin{equation*}
\max_{\xi \in \R^{p}} \min_{u\in \widetilde{\mathbb{B}}}  \left[2 u^{\intercal}\xi-\xi^{\intercal}\xi\right]\geq \min_{u\in \widetilde{\mathbb{B}}}  \left[2 u^{\intercal}\xi^{*}-[\xi^{*}]^{\intercal}\xi^{*}\right]\geq  (\xi^{*})^{\intercal}\xi^{*}.
\end{equation*}
In \eqref{eq: original}, if we take $u=\xi^{*}$, we have 
\begin{equation*}
\max_{\xi \in \R^{p}} \min_{u\in \widetilde{\mathbb{B}}}  \left[2 u^{\intercal}\xi-\xi^{\intercal}\xi\right]\leq \max_{\xi \in \R^{p}} \left[2 [\xi^{*}]^{\intercal}\xi-\xi^{\intercal}\xi\right]= (\xi^{*})^{\intercal}\xi^{*}.
\end{equation*}
By matching the above two bounds, $\xi^{*}$ is the optimal solution to \eqref{eq: original} and 
$$\max_{\xi \in \R^{p}} \min_{u\in \widetilde{\mathbb{B}}}  \left[2 u^{\intercal}\xi-\xi^{\intercal}\xi\right]=[\xi^{*}]^{\intercal}\xi^{*}.
$$
Since $\bp=C^{-1}\xi^*$ and $\Sigma^{\T}=C^{\intercal} C$, we have 
\begin{equation}
\bp=\argmin_{\beta \in {\mathbb{B}}} \beta^{\intercal}\Sigma^{\T}\beta
\label{eq: direct conclusion}
\end{equation}
and 
$
\max_{\beta \in \R^{p}} \min_{b\in {\mathbb{B}}}  \left[2 b^{\intercal}\Sigma^{\T}\beta-\beta^{\intercal}\Sigma^{\T}\beta\right]= [\bp]^{\intercal}\Sigma^{\T}\bp.
$ Note that ${\mathbf B}=\begin{pmatrix}b^{(1)},\ldots,b^{(L)}\end{pmatrix}\in \R^{p\times L}.$
We establish \eqref{eq: optimal weight} by combining \eqref{eq: direct conclusion} and the fact that $\beta \in {\mathbb{B}}$ can be expressed as $\beta=\B\gamma$ for $\gamma\in \Delta^{L}.$

\subsection{Proof of Proposition \ref{prop: DRO ridge}}
For $\delta>0$, we have $\Gamma+\delta\cdot {\rm I}$ to be positive definite. We apply Lemma \ref{lemma: gamma accuracy} in the supplement to establish 
the uniqueness of $\bp_{\delta}(\T).$ Define $\Delta=\delta\cdot U \Lambda^{-2} U^{\intercal}\in \R^{p\times p}$ and recall $\B=U\Lambda V^{\intercal}.$
 When $\B$ has the rank $L$, we have 
\begin{equation}
\B^{\intercal} \Delta \B=\delta\cdot {\bf I}_{L\times L}.
\label{eq: key equation}
\end{equation}
\noindent \underline{\bf Proof of \eqref{eq: maximin general shift ridge}}.
By applying Proposition \ref{prop: identification}, we show that $\bp_{\delta}(\T)$ defined in \eqref{eq: maximin general shift ridge} can be expressed as,
\begin{equation}
\bp_{\delta}(\T)=\sum_{l=1}^{L}[\gamma_{\delta}^{*}(\T)]_{l} b^{(l)} \quad \text{with}\quad \gamma_{\delta}^{*}(\T)=\argmin_{\gamma \in \Delta^{L}} \gamma^{\intercal}\Gamma^{\T^{\delta}} \gamma,
\label{eq: optimal weight ridge explicit}
\end{equation}
where $\Gamma^{\T^{\delta}}=[{\mathbf B}]^{\intercal}\E(X_i+W_i)(X_i+W_i)^{\intercal}{\mathbf B}.$ Since $W_i\in \R^{p}$ is generated as $W_i=\sqrt{\delta} \cdot U W_i^{0}$ with $W_i^{0}\sim N({\bf 0}, \Lambda^{-2})$ and $W_i^{0}\in \R^{L}$ being independent of $X_i$, we further have 
\begin{equation*}
\begin{aligned}
\Gamma^{\T^{\delta}}=[{\mathbf B}]^{\intercal}\E(X_i+W_i)(X_i+W_i)^{\intercal}{\mathbf B}=[{\mathbf B}]^{\intercal} \Sigma^{\T}{\mathbf B}+\delta\cdot [{\mathbf B}]^{\intercal}U\Lambda^{-2} U^{\intercal}{\mathbf B}=[{\mathbf B}]^{\intercal} \Sigma^{\T}{\mathbf B}+\delta \cdot {\rm I},
\end{aligned}
\end{equation*}
where the last equality follows from \eqref{eq: key equation}. Combined with \eqref{eq: optimal weight ridge explicit}, we establish that the definition of $\bp_{\delta}(\T)$ in \eqref{eq: maximin general shift ridge} is the same as that in \eqref{eq: optimal weight ridge} in the main paper.


\noindent \underline{\bf Proof of \eqref{eq: reward reduction}}. It follows from Proposition \ref{prop: identification}, the definition of $\bp_{\delta}$ in \eqref{eq: optimal weight ridge}, and \eqref{eq: key equation} that
\begin{equation*}
\bp_{\delta}=\max_{\beta\in \R^{p}}\min_{b\in \mathbb{B}}  \left[2 b^{\intercal}(\Sigma^{\T}+\Delta)\beta- \beta^{\intercal}(\Sigma^{\T}+\Delta)\beta\right] 
\end{equation*}
and
\begin{equation}
\min_{b\in \mathbb{B}}  \left[2 b^{\intercal}(\Sigma^{\T}+\Delta)\bp_{\delta}- [\bp_{\delta}]^{\intercal}(\Sigma^{\T}+\Delta)\bp_{\delta}\right]=[\bp_{\delta}]^{\intercal}(\Sigma^{\T}+\Delta)\bp_{\delta}=[\gamma^{*}_{\delta}]^{\intercal}\left(\Gamma^{\T}+\delta\cdot{\rm I}\right)\gamma^{*}_{\delta}
\label{eq: delta optimality}
\end{equation}
Now we compute the lower bound for $R_{\T}(\bp_{\delta}) =\min_{b\in \mathbb{B}}  \left[2 b^{\intercal}\Sigma^{\T}\bp_{\delta}- [\bp_{\delta}]^{\intercal}\Sigma^{\T} \bp_{\delta}\right].$

With $\bp_{\delta}=\B \gamma^{*}_{\delta}$, we have $[\bp_{\delta}]^{\intercal}\Delta\bp_{\delta}=\delta \|\gamma^{*}_{\delta}\|_2^2$ and further establish 
\begin{equation}
\begin{aligned}
&\min_{b\in \mathbb{B}}  \left[2 b^{\intercal}(\Sigma^{\T}+\Delta)\bp_{\delta}- [\bp_{\delta}]^{\intercal}(\Sigma^{\T}+\Delta)\bp_{\delta}\right]\\
&=\min_{b\in \mathbb{B}}  \left[2 b^{\intercal}(\Sigma^{\T}+\Delta)\bp_{\delta}- [\bp_{\delta}]^{\intercal}\Sigma^{\T}\bp_{\delta}\right]-\delta \|\gamma^{*}_{\delta}\|_2^2\\
&\leq \min_{b\in \mathbb{B}}  \left[2 b^{\intercal}\Sigma^{\T}\bp_{\delta}- [\bp_{\delta}]^{\intercal}\Sigma^{\T}\bp_{\delta}\right]+2 \max_{b\in \mathbb{B}} b^{\intercal}\Delta\bp_{\delta}-\delta \|\gamma^{*}_{\delta}\|_2^2\\
&=R_{\T}(\bp_{\delta})+2\delta \max_{\gamma\in \Delta^{L}} \gamma^{\intercal}\gamma^{*}_{\delta}-\delta \|\gamma^{*}_{\delta}\|_2^2
\end{aligned}
\label{eq: key upper}
\end{equation}
where the last equality follows from \eqref{eq: key equation}. 
We combine \eqref{eq: delta optimality} and \eqref{eq: key upper} and establish 
\begin{equation}
\begin{aligned}
R_{\T}(\bp_{\delta})&\geq [\gamma^{*}_{\delta}]^{\intercal}\left(\Gamma^{\T}+\delta\cdot{\rm I}\right)\gamma^{*}_{\delta}-2\delta \max_{\gamma\in \Delta^{L}} \gamma^{\intercal}\gamma^{*}_{\delta}+\delta \|\gamma^{*}_{\delta}\|_2^2\\
&\geq [\gamma^{*}]^{\intercal}\Gamma^{\T}\gamma^{*}+2\delta \|\gamma^{*}_{\delta}\|_2^2-2\delta \max_{\gamma\in \Delta^{L}} \gamma^{\intercal}\gamma^{*}_{\delta}\\
&=R_{\T}(\bp)-2\delta \left(\max_{\gamma\in \Delta^{L}} \gamma^{\intercal}\gamma^{*}_{\delta}-\|\gamma^{*}_{\delta}\|_2^2\right)
\end{aligned}
\label{eq: important lower}
\end{equation}
where the second inequality follows from the definition of $\gamma^{*}.$
Note that $\max_{\gamma\in \Delta^{L}} \gamma^{\intercal}\gamma^{*}_{\delta}-\|\gamma^{*}_{\delta}\|_2^2\geq 0$ and 
$\max_{\gamma\in \Delta^{L}} \gamma^{\intercal}\gamma^{*}_{\delta}-\|\gamma^{*}_{\delta}\|_2^2=\|\gamma^{*}_{\delta}\|_{\infty}-\|\gamma^{*}_{\delta}\|_{2}^2,$ which establishes the first inequality in  \eqref{eq: reward reduction}.

We use $j^{*}\in[L]$ to denote the index such that $[\gamma^{*}_{\delta}]_{j^{*}}=\|\gamma^{*}_{\delta}\|_{\infty}.$ Then we have 
\begin{equation}
\begin{aligned}
\|\gamma^{*}_{\delta}\|_{\infty}-\|\gamma^{*}_{\delta}\|_{2}^2=[\gamma^{*}_{\delta}]_{j^{*}}-[\gamma^{*}_{\delta}]_{j^{*}}^2-\sum_{l\neq j^{*}}[\gamma^{*}_{\delta}]_{l}^2
&\leq [\gamma^{*}_{\delta}]_{j^{*}}-[\gamma^{*}_{\delta}]_{j^{*}}^2-\frac{1}{L-1}\left(\sum_{l\neq j^{*}}[\gamma^{*}_{\delta}]_{l}\right)^2\\
&= [\gamma^{*}_{\delta}]_{j^{*}}-[\gamma^{*}_{\delta}]_{j^{*}}^2-\frac{1}{L-1}(1-[\gamma^{*}_{\delta}]_{j^{*}})^2
\end{aligned}
\label{eq: careful upper}
\end{equation}
We take the maximum value of the right hand side with respect to $[\gamma^{*}_{\delta}]_{j^{*}}$ over the domain $[1/L,1].$ Then we obtain $\max_{\frac{1}{L}\leq [\gamma^{*}_{\delta}]_{j^{*}}\leq 1}\left([\gamma^{*}_{\delta}]_{j^{*}}-[\gamma^{*}_{\delta}]_{j^{*}}^2-\frac{1}{L-1}(1-[\gamma^{*}_{\delta}]_{j^{*}})^2\right)=\frac{1}{4}\left(1-\frac{1}{L}\right),$
where the maximum value is achieved at $[\gamma^{*}_{\delta}]_{j^{*}}=\frac{1+\frac{1}{L}}{2}.$
Combined with \eqref{eq: important lower} and \eqref{eq: careful upper}, we establish the second inequality in  \eqref{eq: reward reduction}.
\subsection{Proof of Proposition \ref{prop: maximin proj}}
We can write the maximin definition in the following form
\begin{equation}
\bpt=\argmax_{\|\beta\|_2\leq 1} \min_{b \in \B} \beta^{\intercal}b
\label{eq: maximin proj another}
\end{equation}
where $\B=\{b^{(1)},\ldots,b^{(L)}\}.$
Since $b^{\intercal}\beta$ is linear in $b$, we can replace $\B$ with its convex hull $\CB$ and have
$
\bpt=\argmax_{\|\beta\|_2\leq 1} \min_{b \in \CB} b^{\intercal}\beta
$
We exchange the max and min in the above equation and have 
$
\min_{b \in \CB}\max_{\|\beta\|_2\leq 1}  b^{\intercal}\beta=\min_{b\in \CB} \|b\|_2.
$ 
We define 
$\xi=\argmin_{b\in \CB} \|b\|_2.$
We claim that $\xi^{*}=\xi/\|\xi\|$ is the optimal solution of \eqref{eq: maximin proj another}. For any $\mu\in \CB$, we have $\xi+\nu(\mu-\xi)\in \CB$ for $\nu\in[0,1]$ and have 
$
\|\xi+\nu(\mu-\xi)\|_2^2\geq \|\xi\|_2^2$ for any  $\nu\in [0,1].$
By taking $\nu\rightarrow 0,$ we have 
$
\mu^{\intercal}\xi-\|\xi\|_2^2\geq 0.
$
By dividing both sides by $\|\xi\|_2$, we have 
\begin{equation}
\mu^{\intercal}\xi^{*} \geq \|\xi\|_2 \quad \text{for any} \quad \mu\in \mathbb{B}.
\label{eq: key lower bound} 
\end{equation}
In the definition of \eqref{eq: maximin proj another}, we take $\beta=\xi^{*}$ and have 
\begin{equation}
\max_{\|\beta\|_2\leq 1} \min_{b\in \CB} b^{\intercal}\beta \geq  \min_{b\in \CB} b^{\intercal}\xi^{*}\geq \|\xi\|_2
\label{eq: lower}
\end{equation}
where the last inequality follows from \eqref{eq: key lower bound} .
Additionally, we take $b=\xi$ in the definition of \eqref{eq: maximin proj another} and have 
$
\max_{\|\beta\|_2\leq 1} \min_{b\in \CB} b^{\intercal}\beta\leq \max_{\|\beta\|_2\leq 1} \xi^{\intercal}\beta=\|\xi\|_2
$
Combined with \eqref{eq: lower}, we have shown that $\xi^{*}=\argmax_{\|\beta\|_2\leq 1} \min_{b\in \CB} b^{\intercal}\beta$ that is, $\bpt=\xi^{*}.$

\subsection{High probability events}
We introduce the following events to facilitate the proofs of Propositions \ref{prop: limiting Gamma univariate} and \ref{prop: general variance}.
\begin{equation}
\begin{aligned}
\mathcal{G}_0&=\left\{\left\|\frac{1}{n_l}[X^{(l)}]^{\intercal}\epsilon^{(l)}\right\|_{\infty}\lesssim \sqrt{\frac{\log p}{n_l}}\quad \text{for} \; 1\leq l\leq L\right\},\\
\mathcal{G}_1&=\left\{\max\left\{\|\widehat{b}^{(l)}_{init}-b^{(l)}\|_2,\frac{1}{\sqrt{n_{l}}}\|X^{(l)}(\widehat{b}^{(l)}_{init}-b^{(l)})\|_2\right\}\lesssim \sqrt{\|{b}^{(l)}\|_0\frac{\log p}{\nl}}\sigma_{l} \quad \text{for} \; 1\leq l\leq L\right\},\\
\mathcal{G}_2&=\left\{ \|\widehat{b}^{(l)}_{init}-b^{(l)}\|_1\lesssim \|{b}^{(l)}\|_0\sqrt{\frac{\log p}{\nl}}\sigma_{l}, \|[\widehat{b}_{init}^{(l)}-b^{(l)}]_{\mathcal{S}_{l}^c}\|_1\leq C \|[\widehat{b}_{init}^{(l)}-b^{(l)}]_{\mathcal{S}_{l}}\|_1\quad \text{for} \; 1\leq l\leq L\right\},\\
\mathcal{G}_3&=\left\{|\widehat{\sigma}_{l}^2-\sigma_{l}^2|\lesssim {\|{b}^{(l)}\|_0\frac{\log p}{\nl}}+\sqrt{\frac{\log p}{{n_l}}} \quad \text{for} \; 1\leq l\leq L\right\},
\end{aligned}
\label{eq: high prob event 1}
\end{equation}
where $\mathcal{S}_l\subset[p]$ denotes the support of $b^{(l)}$ for $1\leq l\leq L$ and $C>0$ is a positive constant. Recall that, for $1\leq l\leq L$, $\widehat{b}_{init}^{(l)}$ is the Lasso estimator defined in \eqref{eq: Lasso splitting} with  $\lambda_l=\sqrt{{(2+c)\log p}/{|A_l|}}\sigma_{l}$ for some constant $c>0$; $\widehat{\sigma}_l^2=\|Y^{(l)}-X^{(l)}\widehat{b}^{(l)}\|_2^2/n_l$ for $1\leq l\leq L$ with $\widehat{b}^{(l)}$ denoting  the Lasso estimator based on the non-split data. 

We further define the following events,
\begin{equation}
\begin{aligned}
\mathcal{G}_4&=\left\{\|\widetilde{\Sigma}^{\T}-{\Sigma}^{\T}\|_2\lesssim \sqrt{\frac{p}{N_{\T}}}+\frac{p}{N_{\T}}\right\},\\
\mathcal{G}_5&=\left\{ \max_{\mathcal{S}\subset[p], |\mathcal{S}|\leq s}\max_{\|w_{\mathcal{S}^c}\|_1\leq C\|w_{\mathcal{S}}\|_1}\left|\frac{w^{\intercal}\left(\frac{1}{N_{\T}}\sum_{i=1}^{N_{\T}}X^{\T}_{i}[X^{\T}_{i}]^{\intercal}\right)w}{w^{\intercal}E(X^{\T}_{i}[X^{\T}_{i}])w}-1\right| \lesssim {\frac{s\log p}{N_{\T}}}\right\},\\
\mathcal{G}_6(w,v,t)&=\left\{\left|{w^{\intercal}\left(\widehat{\Sigma}^{\T}-\Sigma^{\T}\right)v}\right|+\left|{w^{\intercal}\left(\widetilde{\Sigma}^{\T}-\Sigma^{\T}\right)v}\right|\lesssim t\frac{\|(\Sigma^{\T})^{1/2} w\|_2\|(\Sigma^{\T})^{1/2} v\|_2}{\sqrt{N_{\T}}}\right\},
\label{eq: high prob event 2}
\end{aligned}
\end{equation}
where $\widetilde{\Sigma}^{\T}=\frac{1}{|A|}\sum_{i\in A} X^{\T}_{i, \cdot}
(X^{\T}_{i, \cdot})^{\intercal}$, $\widehat{\Sigma}^{\T}=\frac{1}{|B|}\sum_{i\in B} X^{\T}_{i, \cdot}
(X^{\T}_{i, \cdot})^{\intercal}$ and  $t>0$ is any positive constant and $w,v\in \R^{p}$ are pre-specified vectors. 

\begin{Lemma}
Suppose that Condition {\rm (A1)} holds and $s\lesssim n/\log p$, then 
\begin{equation}
\PP\left(\cap_{j=0}^{3}\mathcal{G}_j\right)\geq 1-\min\{n,p\}^{-c},
\label{eq: control of event 1}
\end{equation} 
\begin{equation}
\PP\left(\mathcal{G}_4\cap\mathcal{G}_5\right)\geq 1-p^{-c},
\label{eq: control of event 2}
\end{equation}
\begin{equation}
\PP\left(\mathcal{G}_6(w,v,t)\right)\geq 1-2\exp(-ct^2), 
\label{eq: control of event 3}
\end{equation}
for some positive constant $c>0.$
\label{lem: high prob literature}
\end{Lemma}
The above high-probability statement \eqref{eq: control of event 1} follows from the existing literature results on the analysis of Lasso estimators and we shall point to the exact literature results. Specifically, the control of the probability of $\mathcal{G}_0$ follows from Lemma 6.2 of \cite{buhlmann2011statistics}. Regarding the events $\mathcal{G}_1$ and $\mathcal{G}_2$, the control of  $\|\widehat{b}^{(l)}_{init}-b^{(l)}\|_1,$ $\|\widehat{b}^{(l)}_{init}-b^{(l)}\|_2$ and $\frac{1}{\sqrt{n_{l}}}\|X^{(l)}(\widehat{b}^{(l)}_{init}-b^{(l)})\|_2$ can be found in Theorem 3 of \cite{ye2010rate}, Theorem 7.2 of \cite{bickel2009simultaneous} or Theorem 6.1 of \cite{buhlmann2011statistics}; the control of  $\|[\widehat{b}_{init}^{(l)}-b^{(l)}]_{\mathcal{S}_{l}^c}\|_1\leq C \|[\widehat{b}_{init}^{(l)}-b^{(l)}]_{\mathcal{S}_{l}}\|_1$ can be found in Corollary B.2 of \cite{bickel2009simultaneous} or Lemma 6.3 of \cite{buhlmann2011statistics}. For the event $\mathcal{G}_3$, its probability can be controlled as Theorem 2 or (20) in \cite{sun2012scaled}.

If $X^{\T}_{i}$ is sub-gaussian, it follows from equation (5.26) of \cite{vershynin2010introduction} that the event $\mathcal{G}_4$ holds with a probability larger than $1-\exp(-cp)$ for some positive constant $c>0.$; it follows from Theorem 1.6 of \cite{zhou2009restricted} that the event $\mathcal{G}_5$ holds with a probability larger than $1-p^{-c}$ for some positive constant $c>0$. The proof of \eqref{eq: control of event 3} follows from Lemma 10 in the supplement of \cite{cai2020semisupervised}.

%
%

\subsection{Proof of Proposition \ref{prop: general variance}}
\label{sec: variance bound}
We have the expression for the diagonal element of $\V$ as
\begin{equation*}
\begin{aligned}
\V_{\pi(l,k),\pi(l,k)}&=\frac{\sigma_{l}^2}{|B_{l}|}(\widehat{u}^{(l,k)})^{\intercal}\widehat{\Sigma}^{(l)} \left[\widehat{u}^{(l,k)}+\widehat{u}^{(k,l)} {\bf 1}(k=l)\right]+\frac{\sigma_{k}^2}{|B_{k}|}(\widehat{u}^{(k,l)})^{\intercal}\widehat{\Sigma}^{(k)} \left[\widehat{u}^{(l,k)} {\bf 1}(l=k)+\widehat{u}^{(k,l)}\right]\\
&+\frac{1}{|B|}(\E [b^{(l)}]^{\intercal}X_{i}^{\T}[b^{(k)}]^{\intercal}X_{i}^{\T}[b^{(l)}]^{\intercal}X_{i}^{\T}[b^{(k)}]^{\intercal}X_{i}^{\T}-(b^{(l)})^{\intercal}\Sigma^{\T} b^{(k)}(b^{(l)})^{\intercal}\Sigma^{\T} b^{(k)}).
\end{aligned}
\end{equation*}
We introduce the following lemma, which restates Lemma 1 of \cite{cai2019individualized} in the current paper's terminology.
\begin{Lemma} Suppose that Condition {\rm (A1)} holds, then with probability larger than $1-p^{-c}$,
\begin{equation*}
c \frac{\|\omega^{(k)}\|^2_2}{\nl} \leq \frac{1}{|B_l|}(\widehat{u}^{(l,k)})^{\intercal}\widehat{\Sigma}^{(l)} \widehat{u}^{(l,k)}\leq C \frac{\|\omega^{(k)}\|^2_2}{\nl}, \quad \text{for} \quad 1\leq l,k\leq L,
\end{equation*}
\begin{equation*}
c \frac{\|\omega^{(l)}\|^2_2}{n_k} \leq  \frac{1}{|B_k|}(\widehat{u}^{(k,l)})^{\intercal}\widehat{\Sigma}^{(k)} \widehat{u}^{(l,k)}\leq C \frac{\|\omega^{(l)}\|^2_2}{n_k},  \quad \text{for} \quad 1\leq l,k\leq L,
\end{equation*}
for some positive constants $C>c>0$.
\label{lem: enhanced variance}
\end{Lemma}
The bounds for $\V^{(a)}_{\pi(l,k),\pi(l,k)}$ in \eqref{eq: diagonal order} follow from Lemma \ref{lem: enhanced variance}. 
Since $X_{i}^{\T}$ is sub-gaussian, we have
\begin{equation}
\left|\E [b^{(l_1)}]^{\intercal}X_{i}^{\T}[b^{(k_1)}]^{\intercal}X_{i}^{\T}[b^{(l_2)}]^{\intercal}X_{i}^{\T}[b^{(k_2)}]^{\intercal}X_{i}^{\T}\right|\lesssim \|b^{(l_1)}\|_2 \|b^{(k_1)}\|_2\|b^{(l_2)}\|_2\|b^{(k_2)}\|_2,
\label{eq: var 2a}
\end{equation}
and
\begin{equation}
(b^{(l_1)})^{\intercal}\Sigma^{\T} b^{(k_1)}(b^{(l_2)})^{\intercal}\Sigma^{\T} b^{(k_2)}\lesssim\|b^{(l_1)}\|_2 \|b^{(k_1)}\|_2\|b^{(l_2)}\|_2\|b^{(k_2)}\|_2.
\label{eq: var 2b}
\end{equation}
We establish the upper bound for $\V^{(b)}_{\pi(l,k),\pi(l,k)}$ in \eqref{eq: diagonal order} by taking $l_1=l_2=l$ and $k_1=k_2=k.$

For the setting of known ${\Sigma}^{\T}$, on the event $\mathcal{G}_2$ defined in \eqref{eq: high prob event 1}, we establish
$$\|\omega^{(k)}\|_2=\|{\Sigma}^{\T} \widehat{b}_{init}^{(k)}\|_2\lesssim\lambda_{\max}(\Sigma^{\T})\|\widehat{b}_{init}^{(k)}\|_2\lesssim \lambda_{\max}(\Sigma^{\T})\left(\|{b}^{(k)}\|_2+\sqrt{s\log p/n}\right).$$

To establish \eqref{eq: variance 1}, we control $\|\omega^{(k)}\|_2=\|\widetilde{\Sigma}^{\T} \widehat{b}_{init}^{(k)}\|_2$ as follows, 
\begin{equation*}
\begin{aligned}
\|\widetilde{\Sigma}^{\T} \widehat{b}_{init}^{(k)}\|_2\leq \|{\Sigma}^{\T} \widehat{b}_{init}^{(k)}\|_2+\|(\widetilde{\Sigma}^{\T}-{\Sigma}^{\T}) \widehat{b}_{init}^{(k)}\|_2\leq \lambda_{\max}(\Sigma^{\T})\|\widehat{b}_{init}^{(k)}\|_2+\|\widetilde{\Sigma}^{\T}-{\Sigma}^{\T}\|_2\|\widehat{b}_{init}^{(k)}\|_2.
\end{aligned}
\end{equation*}
On the event $\mathcal{G}_4$, we establish
$
\|\omega^{(k)}\|_2\lesssim \lambda_{\max}(\Sigma^{\T})\left(1+\sqrt{\frac{p}{N_{\T}}}+\frac{p}{N_{\T}}\right)\left(\|{b}^{(k)}\|_2+\sqrt{s\log p/n}\right).
$
With a similar bound for $\|\omega^{(l)}\|_2$, we establish \eqref{eq: variance 1}.

 \subsection{Proof of Proposition \ref{prop: limiting Gamma univariate}}
 \label{sec: covariate shift proof 1}
We decompose the error ${\covest}^{\T}-\cov^{\T}$ as
 \begin{equation}
\begin{aligned}
{\covest}^{\T}-\cov^{\T}&=\frac{1}{|B_l|}(\widehat{u}^{(l,k)})^{\intercal}[X^{(l)}_{B_{l}}]^{\intercal}\epsilon^{(l)}_{B_l}+\frac{1}{|B_k|}(\widehat{u}^{(k,l)})^{\intercal}[X^{(k)}_{B_{k}}]^{\intercal}\epsilon^{(k)}_{B_k}\\
&+(b^{(l)})^{\intercal}(\widehat{\Sigma}^{\T}-\Sigma^{\T})b^{(k)}-(\widehat{b}_{init}^{(l)}-b^{(l)})^{\intercal}\widehat{\Sigma}^{\T}(\widehat{b}_{init}^{(k)}-b^{(k)})\\
&+(\widehat{\Sigma}^{\T}\widehat{b}_{init}^{(k)}-\widehat{\Sigma}^{(l)} \widehat{u}^{(l,k)})^{\intercal}(\widehat{b}_{init}^{(l)}-b^{(l)})+(\widehat{\Sigma}^{\T}\widehat{b}_{init}^{(l)}-\widehat{\Sigma}^{(l)} \widehat{u}^{(k,l)})^{\intercal}(\widehat{b}_{init}^{(k)}-b^{(k)}).
\end{aligned}
\label{eq: decomposition general}
\end{equation}

We define $D_{l,k}=D^{(a)}_{l,k}+D^{(b)}_{l,k}$ with 
$D^{(a)}_{l,k}=\frac{1}{|B_l|}(\widehat{u}^{(l,k)})^{\intercal}[X^{(l)}_{B_{l}}]^{\intercal}\epsilon^{(l)}_{B_l}+\frac{1}{|B_k|}(\widehat{u}^{(k,l)})^{\intercal}[X^{(k)}_{B_{k}}]^{\intercal}\epsilon^{(k)}_{B_k},$
and 
$D^{(b)}_{l,k}=(b^{(l)})^{\intercal}(\widehat{\Sigma}^{\T}-\Sigma^{\T})b^{(k)}.$
Define $\Rem_{l,k}$ as
\begin{equation*}
\begin{aligned}
\Rem_{l,k}=&-(\widehat{b}_{init}^{(l)}-b^{(l)})^{\intercal}\widehat{\Sigma}^{\T}(\widehat{b}_{init}^{(k)}-b^{(k)})+(\widehat{\Sigma}^{\T}\widehat{b}_{init}^{(k)}-\widehat{\Sigma}^{(l)} \widehat{u}^{(l,k)})^{\intercal}(\widehat{b}_{init}^{(l)}-b^{(l)})\\
&+(\widehat{\Sigma}^{\T}\widehat{b}_{init}^{(l)}-\widehat{\Sigma}^{(l)} \widehat{u}^{(k,l)})^{\intercal}(\widehat{b}_{init}^{(k)}-b^{(k)}).
\end{aligned}
\label{eq: reminder term}
\end{equation*}
By \eqref{eq: decomposition general}, we have ${\covest}^{\T}-\cov^{\T}=D_{l,k}+\Rem_{l,k}.$
Note that $D^{(a)}_{l,k}$ is a function of $X^{\T}_{A}$, $\{X^{(l)}\}_{1\leq l\leq L}$ and $\{\epsilon^{(k)}\}_{1\leq l\leq L}$ and $D^{(b)}_{l,k}$ is a function of $X^{\T}_{B}$ and hence $D^{(a)}_{l,k}$ is independent of $D^{(b)}_{l,k}.$

\paragraph{Limiting distribution of $D_{l,k}.$}
We check the Lindeberg's condition and establish the asymptotic normality of $D_{l,k}$. 
In the following, we focus on the setting $l\neq k$ and the proof can be extended to the setting $l=k.$
We write \begin{align*}
\frac{D^{(a)}_{l,k}}{{\sqrt{\V_{\pi(l,k),\pi(l,k)}}} }&=\frac{1}{\sqrt{\V_{\pi(l,k),\pi(l,k)}}} \left((\widehat{u}^{(l,k)})^{\intercal} \frac{1}{|B_l|}\sum_{i\in B_l} {X^{(l)}_{i}}\epsilon^{(l)}_{i}+(\widehat{u}^{(k,l)})^{\intercal} \frac{1}{|B_k|}\sum_{i\in B_k} {X^{(k)}_{i}}\epsilon^{(k)}_{i}\right),\\
\frac{D^{(b)}_{l,k}}{{\sqrt{\V_{\pi(l,k),\pi(l,k)}}} }&=\frac{1}{{\sqrt{\V_{\pi(l,k),\pi(l,k)}}}}\frac{1}{|B|}\sum_{i\in B}[b^{(l)}]^{\intercal} \left(X^{\T}_{i} (X^{\T}_{i})^{\intercal}-\Sigma^{\T}\right)b^{(k)}.
\end{align*} 
Define 
\begin{equation*}
W_{l,i}=\frac{1}{|B_l|\sqrt{\V_{\pi(l,k),\pi(l,k)}}}(\widehat{u}^{(l,k)})^{\intercal}{X^{(l)}_{i}}\epsilon^{(l)}_{i} \quad \text{for}\quad i\in B_l, 
\end{equation*}
\begin{equation*}
W_{k,i}=\frac{1}{|B_k|\sqrt{\V_{\pi(l,k),\pi(l,k)}}}(\widehat{u}^{(k,l)})^{\intercal}{X^{(k)}_{i}}\epsilon^{(k)}_{i} \quad \text{for}\quad i\in B_k,
\end{equation*}
and 
$$W_{\T,i}=\frac{1}{|B|{\sqrt{\V_{\pi(l,k),\pi(l,k)}}}}[b^{(l)}]^{\intercal} \left(X^{\T}_{i} (X^{\T}_{i})^{\intercal}-\Sigma^{\T}\right)b^{(k)} \quad \text{for}\quad i \in B.
$$
Then we have 
$$\frac{D_{l,k}}{{\sqrt{\V_{\pi(l,k),\pi(l,k)}}}}=\sum_{i\in B_l} W_{l,i}+\sum_{i\in B_k} W_{k,i}+\sum_{i\in B} W_{\T,i}.$$


We use $\mathcal{O}_1=\{X^{\T}_{i}\}_{i\in A}\cup\{X^{(l)},Y^{(l)}_{A_l}\}_{1\leq l\leq L}$  to denote the subset of data used for computing $\widehat{u}^{(l,k)}$ and $\widehat{u}^{(k,l)}$. 
Conditioning on $\mathcal{O}_1,$  then $\{W_{l,i}\}_{i\in B_l}$, $\{W_{k,i}\}_{i\in B_k}$ and $\{W_{\T,i}\}_{i\in B}$ are independent random variables. 
Note that $\E(W_{l,i}\mid \mathcal{O}_1)=0$ for $i\in B_l$, $\E(W_{k,i}\mid \mathcal{O}_1)=0$ for $i\in B_k$ and $\E(W_{\T,i}\mid \mathcal{O}_1)=0$ for $i\in B.$ Furthermore, we have 
\begin{equation}
\sum_{i\in B_l}\E(W^2_{l,i}\mid \mathcal{O}_1)+\sum_{i\in B_k}\E(W^2_{k,i}\mid \mathcal{O}_1)+\sum_{i\in B}\E(W^2_{\T,i}\mid \mathcal{O}_1)=1.
\label{eq: var constraint}
\end{equation}
Define the event $\mathcal{G}_7=\left\{\frac{\|\omega^{(l)}\|^2_2}{n_k} +\frac{\|\omega^{(k)}\|^2_2}{\nl} \lesssim{\V}^{(a)}_{\pi(l,k),\pi(l,k)} \right\},$
and it follows from Proposition \ref{prop: general variance} that 
\begin{equation}
\PP\left(\mathcal{O}_1\in \mathcal{G}_7\right)\geq 1-\min\{n,p\}^{-c}.
\label{eq: inter prob control}
\end{equation}
Let ${o}_1$ denote one element of the event $\mathcal{G}_7.$
To establish the asymptotic normality, it is sufficient to check the following Lindeberg's condition: for any constant $c>0$,
\begin{equation}
\begin{aligned}
&\sum_{i\in B_l}\E\left(W_{l,i}^2 \mathbf{1}{\left\{\left|W_{l,i}\right|\geq c\right\}}\mid \mathcal{O}_1={o}_1\right)+\sum_{i\in B_k}\E\left(W_{k,i}^2 \mathbf{1}{\left\{\left|W_{k,i}\right|\geq c\right\}}\mid \mathcal{O}_1={o}_1\right)\\
&+\sum_{i\in B}\E\left(W_{\T,i}^2 \mathbf{1}{\left\{\left|W_{\T,i}\right|\geq c\right\}}\mid \mathcal{O}_1={o}_1\right)\rightarrow 0.
\end{aligned}
\label{eq: lin condition}
\end{equation}
We apply the optimization constraint 
in \eqref{eq: constraint 3} and establish
\begin{equation*}
\begin{aligned}
&\sum_{i\in B_l}\E\left(W_{l,i}^2 \mathbf{1}{\left\{\left|W_{l,i}\right|\geq c\right\}}\mid \mathcal{O}_1\right)+\sum_{i\in B_k}\E\left(W_{k,i}^2 \mathbf{1}{\left\{\left|W_{k,i}\right|\geq c\right\}}\mid \mathcal{O}_1\right)\\
&\leq \sum_{i\in B_l}\frac{[\sigma_l^2(\widehat{u}^{(l,k)})^{\intercal}X^{(l)}_{i}]^2}{|B_l|^2{\V_{\pi(l,k),\pi(l,k)}}}\E\left(\frac{(\epsilon^{(l)}_{i})^2}{\sigma_l^2} \mathbf{1}{\left\{\left|\epsilon^{(l)}_{i}\right|\geq \frac{c |B_l|\sqrt{\V_{\pi(l,k),\pi(l,k)}}}{\|\omega^{(k)}\|_2\sqrt{\log |B_l|}}\right\}}\mid \mathcal{O}_1 \right)\\
&+\sum_{i\in B_k}\frac{[\sigma_k^2(\widehat{u}^{(k,l)})^{\intercal}X^{(k)}_{i}]^2}{|B_k|^2{\V_{\pi(l,k),\pi(l,k)}}}\E\left(\frac{(\epsilon^{(k)}_{i})^2}{\sigma_k^2} \mathbf{1}{\left\{\left|\epsilon^{(k)}_{i}\right|\geq \frac{c |B_k|\sqrt{\V_{\pi(l,k),\pi(l,k)}}}{\|\omega^{(l)}\|_2\sqrt{\log |B_k|}}\right\}}\mid \mathcal{O}_1 \right)\\
&\leq \sum_{i\in B_l}\frac{[\sigma_l^2(\widehat{u}^{(l,k)})^{\intercal}X^{(l)}_{i}]^2}{|B_l|^2{\V_{\pi(l,k),\pi(l,k)}}}\E\left(\frac{(\epsilon^{(l)}_{i})^2}{\sigma_l^2} \mathbf{1}{\left\{\left|\epsilon^{(l)}_{i}\right|\geq \frac{c |B_l|\sqrt{\V^{(a)}_{\pi(l,k),\pi(l,k)}}}{\|\omega^{(k)}\|_2\sqrt{\log |B_l|}}\right\}}\mid \mathcal{O}_1 \right)\\
&+\sum_{i\in B_k}\frac{[\sigma_k^2(\widehat{u}^{(k,l)})^{\intercal}X^{(k)}_{i}]^2}{|B_k|^2{\V_{\pi(l,k),\pi(l,k)}}}\E\left(\frac{(\epsilon^{(k)}_{i})^2}{\sigma_k^2} \mathbf{1}{\left\{\left|\epsilon^{(k)}_{i}\right|\geq \frac{c |B_k|\sqrt{\V^{(a)}_{\pi(l,k),\pi(l,k)}}}{\|\omega^{(l)}\|_2\sqrt{\log |B_k|}}\right\}}\mid \mathcal{O}_1 \right)
\end{aligned}
\end{equation*}
where the last inequality follows from $\V_{\pi(l,k),\pi(l,k)}\geq \V^{(a)}_{\pi(l,k),\pi(l,k)}.$

By the definition of $\mathcal{G}_7$, we condition on $\mathcal{O}_1={o}_1$ with ${o}_1\in \mathcal{G}_7$ and further upper bound  the right hand side of the above inequality by 
\begin{equation}
\begin{aligned}
&\max_{1\leq i\leq n_l}\E\left(\frac{(\epsilon^{(l)}_{i})^2}{\sigma_l^2} \mathbf{1}{\left\{\left|\epsilon^{(l)}_{i}\right|\geq \frac{c |B_l|\sqrt{\V^{(a)}_{\pi(l,k),\pi(l,k)}}}{\|\omega^{(k)}\|_2\sqrt{\log |B_l|}}\right\}}\mid \mathcal{O}_1={o}_1\right)\\
&+\max_{1\leq i\leq n_k}\E\left(\frac{(\epsilon^{(k)}_{i})^2}{\sigma_l^2} \mathbf{1}{\left\{\left|\epsilon^{(k)}_{i}\right|\geq \frac{c |B_k|\sqrt{\V^{(a)}_{\pi(l,k),\pi(l,k)}}}{\|\omega^{(l)}\|_2\sqrt{\log |B_k|}}\right\}}\mid \mathcal{O}_1={o}_1 \right)\\ 
&\leq \max_{1\leq i\leq n_l}\E\left(\frac{(\epsilon^{(l)}_{i})^2}{\sigma_l^2} \mathbf{1}{\left\{\left|\epsilon^{(l)}_{i}\right|\geq \frac{c |B_l|\sqrt{\frac{\|\omega^{(l)}\|^2_2}{n_k} +\frac{\|\omega^{(k)}\|^2_2}{\nl}}}{\|\omega^{(k)}\|_2\sqrt{\log |B_l|}}\right\}}\mid \mathcal{O}_1={o}_1 \right)\\
&+\max_{1\leq i\leq n_k}\E\left(\frac{(\epsilon^{(k)}_{i})^2}{\sigma_l^2} \mathbf{1}{\left\{\left|\epsilon^{(k)}_{i}\right|\geq \frac{c |B_k|\sqrt{\frac{\|\omega^{(l)}\|^2_2}{n_k} +\frac{\|\omega^{(k)}\|^2_2}{\nl}}}{\|\omega^{(l)}\|_2\sqrt{\log |B_k|}}\right\}}\mid \mathcal{O}_1={o}_1 \right)
&
\lesssim \left(\frac{\log n}{n}\right)^{\frac{c}{2}},
\end{aligned}
\label{eq: lin condition 2}
\end{equation}
where the last inequality follows from $\E([\epsilon^{(l)}_i]^{2+c}_i\mid X^{(l)}_{i})\leq C$ in Condition (A1). 

Define $J_i=[b^{(l)}]^{\intercal} \left(X^{\T}_{i} (X^{\T}_{i})^{\intercal}-\Sigma^{\T}\right)b^{(k)},$ and 
\begin{align*}
\bar{W}_{\T,i}=\frac{1}{|B|{\sqrt{\V^{(b)}_{\pi(l,k),\pi(l,k)}}}}[b^{(l)}]^{\intercal} \left(X^{\T}_{i} (X^{\T}_{i})^{\intercal}-\Sigma^{\T}\right)b^{(k)} =\frac{J_i}{{\sqrt{|B| {\rm Var}(J_i)}}}\quad \text{for}\quad i \in B.
\end{align*} 
Note that $|{W}_{\T,i}|\leq |\bar{W}_{\T,i}|,$ and then we have 
\begin{equation*}
\begin{aligned}
\sum_{i\in B}\E\left(W_{\T,i}^2 \mathbf{1}{\left\{\left|W_{\T,i}\right|\geq c\right\}}\mid \mathcal{O}_1\right)&=\sum_{i\in B}\E\left(W_{\T,i}^2 \mathbf{1}{\left\{\left|W_{\T,i}\right|\geq c\right\}}\right)\\
 &\leq \sum_{i\in B}\E\left(\bar{W}_{\T,i}^2 \mathbf{1}{\left\{\left|\bar{W}_{\T,i}\right|\geq c\right\}}\right)\\
 &\leq \E \left({J_i^2}/{{\rm Var}(J_i)}\right) \cdot \mathbf{1}\left({|J_i|}/\sqrt{{\rm Var}(J_i)}\geq c\sqrt{|B|}\right).
\end{aligned}
\end{equation*}
Together with the dominated convergence theorem, we have  $\sum_{i\in B}\E\left(W_{\T,i}^2 \mathbf{1}{\left\{\left|W_{\T,i}\right|\geq c\right\}}\mid \mathcal{O}_1\right)\rightarrow 0.$
Combined with \eqref{eq: lin condition 2}, we establish \eqref{eq: lin condition}. Hence, for ${o}_1\in \mathcal{G}_7$, we establish $$\frac{D_{l,k}}{{\sqrt{\V_{\pi(l,k),\pi(l,k)}}}}\mid \mathcal{O}_1={o}_1 \cid \mathcal{N}(0,1).$$ We calculate its characteristic function
\begin{equation*}
\begin{aligned}
&\E \exp\left(it \cdot \frac{D_{l,k}}{{\sqrt{\V_{\pi(l,k),\pi(l,k)}}}}\right)-e^{-t^2/2}\\
&=\int \E \left(\left[\exp\left(it \cdot \frac{D_{l,k}}{{\sqrt{\V_{\pi(l,k),\pi(l,k)}}}}\right)\mid \mathcal{O}_1={o}_1\right]-e^{-t^2/2}\right)\cdot {\bf 1}_{o_1\in \mathcal{G}_7} \cdot \mu(o_1)\\
&+\int \E \left[\exp\left(it \cdot \frac{D_{l,k}}{{\sqrt{\V_{\pi(l,k),\pi(l,k)}}}}\right)\mid \mathcal{O}_1={o}_1\right]\cdot {\bf 1}_{o_1\not \in \mathcal{G}_7}\cdot \mu(o_1)-e^{-t^2/2}\cdot \PP(\mathcal{E}^c_3).
\end{aligned}
\end{equation*}
Combined with \eqref{eq: inter prob control} and the bounded convergence theorem, we establish 
$\frac{D_{l,k}}{{\sqrt{\V_{\pi(l,k),\pi(l,k)}}}}\overset{d}{\rightarrow} \mathcal{N}(0,1).$

\paragraph{Control of $\Rem_{l,k}$ in \eqref{eq: bias}.}
We introduce the following lemma, whose proof is presented in Section \ref{sec: bias general}.
{
\begin{Lemma} Suppose that Condition {\rm (A1)} holds, then with probability larger than $1-\min\{n,p\}^{-c},$ we have
\begin{equation}
\small
\left|(\widehat{b}_{init}^{(l)}-b^{(l)})^{\intercal}\widehat{\Sigma}^{\T}(\widehat{b}_{init}^{(k)}-b^{(k)})\right| \lesssim \sqrt{\frac{\|b^{(l)}\|_0 \|b^{(k)}\|_0(\log p)^2}{{\nl n_k}}},
\label{eq: error bound 1}
\end{equation}
\begin{equation}
\small
\left|(\widehat{\Sigma}^{\T}\widehat{b}_{init}^{(k)}-\widehat{\Sigma}^{(l)} \widehat{u}^{(l,k)})^{\intercal}(\widehat{b}_{init}^{(l)}-b^{(l)})\right| \lesssim \|\omega^{(k)}\|_2 {\frac{\|b^{(l)}\|_0 \log p}{\nl}}+ \|\widehat{b}^{(k)}_{init}\|_2\sqrt{\frac{\|b^{(l)}\|_0 (\log p)^2}{\nl N_{\T}}},
\label{eq: error bound 2}
\end{equation}
\begin{equation}
\small
\left|(\widehat{\Sigma}^{\T}\widehat{b}_{init}^{(l)}-\widehat{\Sigma}^{(l)} \widehat{u}^{(k,l)})^{\intercal}\widehat{\Sigma}^{\T}(\widehat{b}_{init}^{(k)}-b^{(k)})\right| \lesssim \|\omega^{(l)}\|_2 {\frac{\|b^{(k)}\|_0 \log p}{n_k}}+ \|\widehat{b}^{(l)}_{init}\|_2\sqrt{\frac{\|b^{(k)}\|_0 (\log p)^2}{n_k N_{\T}}}.
\label{eq: error bound 3}
\end{equation}
\label{lem: bias general}	
\end{Lemma}
}
On the event $\mathcal{G}_1,$ we have $\|\widehat{b}^{(l)}_{init}\|_2\leq \|{b}^{(l)}\|_2+\sqrt{{\frac{\|b^{(k)}\|_0 \log p}{n_k}}}.$ Combining this inequality with Lemma \ref{lem: bias general}, we establish the upper bound for $\Rem_{l,k}$ in \eqref{eq: bias}.

\subsection{Proof of Proposition \ref{prop: verification of A3}}
\label{sec: covariate shift}
We start with the proof of Proposition \ref{prop: verification of A3} for the covariate shift setting, which relies on Propositions \ref{prop: limiting Gamma univariate} and \ref{prop: general variance}.  On the event $\mathcal{E}_1$ defined in \eqref{eq: event}, we apply the definitions in \eqref{eq: covariance rescaled} and have 
\begin{equation}
\|\widehat{\V}-\V\|_2\leq \frac{d_0}{3n}.
\label{eq: variance bound}
\end{equation}
Note that 
\begin{equation}
\begin{aligned}
\frac{\left|\widehat{\Gamma}_{l,k}^{\T}-\Gamma_{l,k}^{\T}\right|}{\sqrt{\widehat{\V}_{\pi(l,k),\pi(l,k)}+d_0/n}}&\leq \frac{\left|\widehat{\Gamma}_{l,k}^{\T}-\Gamma_{l,k}^{\T}\right|}{\sqrt{{\V}_{\pi(l,k),\pi(l,k)}+2d_0/3n}}\\
&\leq \frac{\left|D_{l,k}\right|}{\sqrt{{\V}_{\pi(l,k),\pi(l,k)}+2d_0/3n}}+\frac{\left|\Rem_{l,k}\right|}{\sqrt{{\V}_{\pi(l,k),\pi(l,k)}+2d_0/3n}},
\end{aligned}
\label{eq: decomposition}
\end{equation}
where $D_{l,k}$ and $\Rem_{l,k}$ are defined in Proposition \ref{prop: limiting Gamma univariate}. 

It follows from \eqref{eq: limiting univariate} that
\begin{equation}
\liminf_{n,p\rightarrow \infty}\PP\left(\max_{1\leq l,k\leq L}\frac{\left|D_{l,k}\right|}{\sqrt{{\V}_{\pi(l,k),\pi(l,k)}+(2d_0/3n)}}\leq  z_{{\alpha_0}/[L(L+1)]}\right)\geq 1-\alpha_0.
\label{eq: bound A3-1}
\end{equation}
Combining \eqref{eq: bias} and \eqref{eq: diagonal order}, we apply the boundedness on $\max_{1\leq l\leq L}\|b^{(l)}\|_2$ and establish that, with probability larger than $1-\min\{n,p\}^{-c},$
$$\frac{\left|\Rem_{l,k}\right|}{\sqrt{{\V}_{\pi(l,k),\pi(l,k)}+2d_0/(3n)}}\lesssim \frac{s\log p}{\sqrt{n}}+\sqrt{\frac{s (\log p)^2}{N_{\T}}}.$$
By the condition $\sqrt{n}\gg s\log p$ and $N_{\T}\gg s (\log p)^2,$ we then establish that, with probability larger than $1-\min\{n,p\}^{-c},$ 
$$\max_{1\leq l,k\leq L}\frac{\left|\Rem_{l,k}\right|}{\sqrt{{\V}_{\pi(l,k),\pi(l,k)}+3d_0/(3n)}}\leq  0.05\cdot z_{{\alpha_0}/[L(L+1)]}$$
Combined with \eqref{eq: decomposition} and \eqref{eq: bound A3-1}, we establish \eqref{eq: condition for sampling 2}.

\section{Proof of Theorem \ref{thm: sampling accuracy}} 
\label{sec: sampling proof}

We first introduce some notations. For $L>0$ and $\alpha_0\in(0,0.01],$ define 
\begin{equation}
C^{*}(\alpha_0)=c^{*}(\alpha_0)\cdot{\rm Vol}\left[\frac{L(L+1)}{2}\right]  \quad \text{with}\quad c^{*}(\alpha_0)=\frac{\exp\left(-\frac{L(L+1)}{3}\frac{z^2_{{\alpha_0}/[L(L+1)]}(n\cdot \lambda_{\max}(\V)+\frac{4}{3}d_0)}{n\cdot \lambda_{\min}(\V)+\frac{2}{3}d_0}\right)}{\sqrt{2\pi}\prod_{i=1}^{\frac{L(L+1)}{2}} \left[n \cdot \lambda_i({\bf V})+{4d_0}/{3}\right]^{1/2}},
\label{eq: sampling constants}
\end{equation} 
where ${\rm Vol}\left[\frac{L(L+1)}{2}\right]$ denotes the volume of a unit ball in $L(L+1)/2$ dimensions. 

We prove the following lemma in Section \ref{sec: lower bound constant}, which shows that $C^{*}(\alpha_0)$ and $c^{*}(\alpha_0)$ are lower bounded by a positive constant with a high probability. 
\begin{Lemma} Consider the model \eqref{eq: multi-group model}. Suppose Condition {\rm (A1)} holds, $\tfrac{s \log p}{\min\{n,N_{\T}\}} \rightarrow 0$ with $n=\min_{1\leq l\leq L} n_{l}$ and $s=\max_{1\leq l\leq L}\|b^{(l)}\|_0$. 
If $N_{\T}\gtrsim\max\{n,p\}$, then with probability larger than $1-\min\{n,p\}^{-c}$ for some positive constant $c>0,$ then $\min\{C^{*}(\alpha_0),c^*(\alpha_0)\}\geq c'$ for a positive constant $c'>0.$
\label{lem: lower bounded constant}
\end{Lemma}

The following theorem is a generalization of Theorem \ref{thm: sampling accuracy} in the main paper, which implies Theorem \ref{thm: sampling accuracy} by setting $\delta=0.$ 
\begin{Theorem} Consider the model \eqref{eq: multi-group model}. Suppose Conditions {\rm (A1)} and {\rm (A2)} hold.  If $\err_n(M)$ defined in \eqref{eq: sampling ratio} satisfies $\err_n(M) \ll \min\{1,c^{*}(\alpha_0), \lambda_{\min}(\Gamma^{\T})+\delta\}$ where $c^{*}(\alpha_0)$ is defined in \eqref{eq: sampling constants}, then 
 \begin{equation}
\liminf_{n,p\rightarrow \infty} \liminf_{M\rightarrow \infty}\PP\left(\min_{m\in \mathbb{M}}\|\widehat{\gamma}^{\m}_{\delta}-\gamma^{*}_{\delta}\|_2
\leq \frac{\sqrt{2} \err_n(M)}{\lambda_{\min}(\Gamma^{\T})+\delta} \cdot \frac{1}{\sqrt{n}}\right)\geq 1-\alpha_0,
 \label{eq: resampling weight accuracy}
 \end{equation}
where $\alpha_0\in(0,0.01]$ is the pre-specified constant used in the construction of $\mathbb{M}$ in \eqref{eq: generating condition}.
\label{thm: sampling accuracy ridge}
\end{Theorem}


%

The main idea of the proof is follows: we ultilize the property  of $\widehat{\Gamma}^{\T}$ established in Proposition \ref{prop: verification of A3}. For these $\widehat{\Gamma}^{\T}$, we establish the following result in Section \ref{sec: resampling approximation},
\begin{equation}
\liminf_{n,p\rightarrow \infty} \liminf_{M\rightarrow \infty}\PP\left(\min_{m\in \MM}\|\widehat{\Gamma}^{\m}-{\Gamma}^{\T}\|_F\leq \sqrt{2} \err_n(M)/{\sqrt{n}}\right)\geq 1-\alpha_0,
 \label{eq: resampling approximation}
\end{equation}
where $\mathbb{M}$ is defined in \eqref{eq: generating condition}. In Section \ref{sec: resampling weight accuracy}, we apply \eqref{eq: resampling approximation} to establish Theorem \ref{thm: sampling accuracy ridge}.

\subsection{Proof of \eqref{eq: resampling approximation}}
\label{sec: resampling approximation}  

Denote the data by $\Data$, that is, $\Data=\{X^{(l)},Y^{(l)}\}_{1\leq l\leq L}\cup \{X^{\T}\}.$
Recall $n=\min_{1\leq l\leq L} \nl$ and write $\widehat{\Gamma}=\widehat{\Gamma}^{\T}.$  Define $\widehat{Z}=\sqrt{n}\left({\rm vecl}(\widehat{\Gamma})-{\rm vecl}(\Gamma)\right),$ and $Z^{\m}=\sqrt{n}[{\rm vecl}(\widehat{\Gamma})-{\rm vecl}(\widehat{\Gamma}^{\m})]$ for $1\leq m \leq M.$ For the given data $\Data$, $\widehat{Z}$ is fixed. We have the expression  
\begin{equation}
\widehat{Z}-\Zm=\sqrt{n}[{\rm vecl}(\widehat{\Gamma}^{\m})-{\rm vecl}(\Gamma)].
\label{eq: rescaling expression}
\end{equation}
We define the rescaled covariance matrices as
\begin{equation}
\Cov=n {\bf V}\quad \text{and} \quad \widehat{\Cov}=n \widehat{\bf V},
\label{eq: covariance rescaled}
\end{equation}
 with ${\bf V}$ and $\widehat{\bf V}$ defined in \eqref{eq: cov def} and \eqref{eq: cov est}, respectively.  
 The density of the rescaled $Z^{\m}$ is
\begin{equation*}
f(Z^{\m}=Z\mid  {\Data})=\frac{\exp\left(-\frac{1}{2}Z^{\intercal}(\widehat{\Cov}+d_0 {\bf I})^{-1}{Z}\right) }{\sqrt{2\pi {\rm det}(\widehat{\Cov}+d_0 {\bf I}) }}.
\end{equation*}
We define the following function to facilitate the proof,
\begin{equation}
g({Z})=\frac{1}{\sqrt{2\pi {\rm det}(\Cov+\frac{4}{3} d_0 {\bf I}) }} \exp\left(-\frac{1}{2}{Z}^{\intercal}({\Cov}+\frac{2}{3}d_0 {\bf I})^{-1}{Z}\right).
\label{eq: expression density}
\end{equation}
We define the following events for the data $\Data$,
\begin{equation}
\begin{aligned}
\mathcal{E}_1&=\left\{\|\widehat{\Cov}-\Cov\|_2<d_0/3\right\},\\
\mathcal{E}_2&=\left\{\max_{1\leq l,k\leq L}\frac{\left|\widehat{Z}_{\pi(l,k)}\right|}{\sqrt{\widehat{\Cov}_{\pi(l,k),\pi(l,k)}+d_0}}\leq 1.05\cdot z_{{\alpha_0}/[L(L+1)]}\right\},
%
\end{aligned}
\label{eq: event}
\end{equation}
where $\|\widehat{\Cov}-\Cov\|_2$ denotes the spectral norm of the matrix  $\widehat{\Cov}-\Cov.$ 
The following lemma shows that the event $\mathcal{E}_1$ holds with a high probability, whose proof is presented in Section \ref{sec: event lemma 1}.

\begin{Lemma} Suppose that the conditions of Theorem \ref{thm: sampling accuracy} hold, then we have
\begin{equation}
\PP\left(\mathcal{E}_1\right)\geq 1-\min\{N_{\T},n,p\}^{-c}
\label{eq: high prob general}
\end{equation}
\label{lem: event lemma 1}
for some positive constant $c>0.$
\end{Lemma}

Together with \eqref{eq: condition for sampling 2}, we establish 
\begin{equation}
\liminf_{n,p\rightarrow\infty}\PP\left(\mathcal{E}_1\cap \mathcal{E}_2\right)\geq 1-\alpha_0.
\label{eq: high event for sampling}
\end{equation}
Let $A\succ B$ denote that the matrix $A-B$ is positive definite, respectively.  On the event $\Data\in \mathcal{E}_1$, we have $\Cov+\frac{4}{3} d_0 {\bf I}\succ \widehat{\Cov}+d_0 {\bf I}
\succ \Cov+\frac{2}{3} d_0 {\bf I},$
which implies
\begin{equation}
f(\Zm=Z\mid {\Data}) \cdot {\bf 1}_{\{\Data\in \mathcal{E}_1\}}  \geq g(Z) \cdot{\bf 1}_{ \{\Data\in \mathcal{E}_1\}}.
\label{eq: simplified lower bound}
\end{equation} 
for any $Z\in \R^{L(L+1)/2}.$ Furthermore, on the event $\mathcal{E}_1\cap \mathcal{E}_2,$ we have 
\begin{equation*}
\begin{aligned}
\frac{1}{2}\widehat{Z}^{\intercal}(\Cov+\frac{2}{3} d_0 {\bf I})^{-1}\widehat{Z}&\leq \frac{L(L+1)}{4}\frac{\max_{1\leq l,k\leq L}(\widehat{\Cov}_{\pi(l,k),\pi(l,k)}+d_0)\cdot (1.05\cdot z_{{\alpha_0}/[L(L+1)]})^2}{\lambda_{\min}(\Cov)+\frac{2}{3}d_0}\\
&\leq  \frac{L(L+1)}{3}\frac{z^2_{{\alpha_0}/[L(L+1)]}(\lambda_{\max}(\Cov)+\frac{4}{3}d_0)}{\lambda_{\min}(\Cov)+\frac{2}{3}d_0}.
\end{aligned}
\end{equation*}
Hence, on the event $\mathcal{E}_1\cap \mathcal{E}_2,$ we have   
\begin{equation}
g(\widehat{Z})\geq c^{*}(\alpha_0).
\label{eq: key constant 2}
\end{equation}

We further lower bound the targeted probability in \eqref{eq: resampling approximation} as
\begin{equation}
\begin{aligned}
&\PP\left(\min_{m\in \MM}\|\widehat{\Gamma}^{\m}-{\Gamma}^{\T}\|_F\leq \sqrt{2} \err_n(M)/{\sqrt{n}}\right)
\geq \PP\left(\min_{m\in \MM}\|\Zm-\widehat{Z}\|_2 \leq \err_n(M)\right)\\
&\geq \E_{\Data}\left[\PP\left(\min_{m\in \MM}\|\Zm-\widehat{Z}\|_2 \leq \err_n(M)\mid \Data\right)\cdot {\bf 1}_{ \Data\in \mathcal{E}_1\cap \mathcal{E}_2}\right],
\end{aligned}
\label{eq: lower 1}
\end{equation}
where $\PP(\cdot \mid \Data)$ denotes the conditional probability given the observed data $\Data$ and $\E_{\Data}$ denotes the expectation taken with respect to the observed data $\Data.$ 

For $m\not \in \MM,$ the definition of $\MM$ implies that there exists $1\leq k_0\leq l_0\leq L$ such that 
$$
\frac{\left|{Z}^{\m}_{\pi(l_0,k_0)}\right|}{\sqrt{\widehat{\Cov}_{\pi(l_0,k_0),\pi(l_0,k_0)}+d_0}}\geq 1.1\cdot z_{{\alpha_0}/[L(L+1)]}.$$
Hence, on the event $\mathcal{E}_1\cap \mathcal{E}_2,$
\begin{equation*}
\|\Zm-\widehat{Z}\|_2\geq \left|{Z}^{\m}_{\pi(l_0,k_0)}-\widehat{Z}_{\pi(l_0,k_0)}\right|\geq \sqrt{\frac{2d_0}{3}}\cdot 0.05\cdot z_{{\alpha_0}/[L(L+1)]}.
\end{equation*}
There exists a positive integer $M_0>0$ such that for $M\geq M_0$, $\err_n(M)\leq \sqrt{\frac{2d_0}{3}}\cdot 0.05\cdot z_{{\alpha_0}/[L(L+1)]}$ and 
$\min_{m\not\in \MM}\|\Zm-\widehat{Z}\|_2\geq \err_n(M).$ In the following analysis, we consider the sampling size $M$ that is larger than $M_0.$ As a consequence, for $\Data\in \mathcal{E}_1\cap \mathcal{E}_2,$
\begin{equation}
\PP\left(\min_{m\in \MM}\|\Zm-\widehat{Z}\|_2 \leq \err_n(M)\mid \Data\right)=\PP\left(\min_{1\leq m\leq M}\|\Zm-\widehat{Z}\|_2 \leq \err_n(M)\mid \Data\right).
\label{eq: key equivalience}
\end{equation}
Together with \eqref{eq: lower 1}, we have
\begin{equation}
\begin{aligned}
&\PP\left(\min_{m\in \MM}\|\widehat{\Gamma}^{\m}-{\Gamma}^{\T}\|_F\leq \sqrt{2} \err_n(M)/{\sqrt{n}}\right)\\
&\geq \E_{\Data}\left[\PP\left(\min_{1\leq m\leq M}\|\Zm-\widehat{Z}\|_2 \leq \err_n(M)\mid \Data\right)\cdot {\bf 1}_{ \Data\in \mathcal{E}_1\cap \mathcal{E}_2}\right].
\end{aligned}
\label{eq: lower 2}
\end{equation}

Note that
\begin{equation*}
\begin{aligned}
\PP\left(\min_{1\leq m\leq M}\|\Zm-\widehat{Z}\|_2 \leq \err_n(M)\mid \Data\right)
&=1-\PP\left(\min_{1\leq m\leq M}\|\Zm-\widehat{Z}\|_2 \geq \err_n(M)\mid \Data\right)\\
&=1-\prod_{1\leq m\leq M}\left[1-\PP\left(\|\Zm-\widehat{Z}\|_2 \leq \err_n(M)\mid \Data\right)\right]\\
\end{aligned}
\label{eq: exp lower}
\end{equation*}
where the second equality follows from the conditional independence of $\{\Zm\}_{1\leq m\leq M}$ given the data $\mathcal{O}.$  Since $1-x\leq e^{-x}$, we further lower bound the above expression as 
\begin{equation*}
\begin{aligned}
\PP\left(\min_{1\leq m\leq M}\|\Zm-\widehat{Z}\|_2 \leq \err_n(M)\mid \Data\right)
&\geq 1-\prod_{1\leq m\leq M}\exp\left[-\PP\left(\|\Zm-\widehat{Z}\|_2 \leq \err_n(M)\mid \Data\right)\right]\\
&= 1-\exp\left[-M\cdot \PP\left(\|\Zm-\widehat{Z}\|_2 \leq \err_n(M)\mid \Data\right)\right].
\end{aligned}
\end{equation*} 
Hence, we have 
\begin{equation}
\begin{aligned}
&\PP\left(\min_{1\leq m\leq M}\|\Zm-\widehat{Z}\|_2 \leq \err_n(M)\mid \Data\right)\cdot {\bf 1}_{ \Data\in \mathcal{E}_1\cap \mathcal{E}_2}\\
&\geq \left(1-\exp\left[-M\cdot \PP\left(\|\Zm-\widehat{Z}\|_2 \leq \err_n(M)\mid \Data\right)\right]\right)\cdot {\bf 1}_{ \Data\in \mathcal{E}_1\cap \mathcal{E}_2}\\
&=1-\exp\left[-M\cdot\PP\left(\|\Zm-\widehat{Z}\|_2 \leq \err_n(M)\mid \Data\right)\cdot {\bf 1}_{ \Data\in \mathcal{E}_1\cap \mathcal{E}_2}\right].
\end{aligned}
\label{eq: connection}
\end{equation}

We apply the inequality \eqref{eq: simplified lower bound} and establish the following lower bound,
\begin{equation}
\begin{aligned}
&\PP\left(\|\Zm-\widehat{Z}\|_2 \leq \err_n(M)\mid \Data\right)\cdot {\bf 1}_{ \Data\in \mathcal{E}_1\cap \mathcal{E}_2}
\\
=& \int f(Z^{\m}=Z\mid \Data) \cdot {\bf 1}_{\left\{\|Z-\widehat{Z}\|_2 \leq \err_n(M)\right\}}d Z \cdot {\bf 1}_{ \Data\in \mathcal{E}_1\cap \mathcal{E}_2}\\
\geq &\left[\int g(Z) \cdot{\bf 1}_{\left\{\|Z-\widehat{Z}\|_2 \leq \err_n(M)\right\}} d Z \right]\cdot {\bf 1}_{ \Data\in \mathcal{E}_1\cap \mathcal{E}_2}\\
=& \left[\int g(\widehat{Z}) \cdot{\bf 1}_{\left\{\|Z-\widehat{Z}\|_2 \leq \err_n(M)\right\}} d Z \right] \cdot {\bf 1}_{ \Data\in \mathcal{E}_1\cap \mathcal{E}_2}\\
&+\left[\int [g(Z)-g(\widehat{Z})] \cdot{\bf 1}_{\left\{\|Z-\widehat{Z}\|_2 \leq \err_n(M)\right\}} d Z \right]\cdot {\bf 1}_{ \Data\in \mathcal{E}_1\cap \mathcal{E}_2}.\end{aligned}
\label{eq: decomposition density}
\end{equation}
By \eqref{eq: key constant 2}, we have 
\begin{equation}
\begin{aligned}
&\int g(\widehat{Z}) \cdot {\bf 1}_{\left\{\|Z-\widehat{Z}\|_2 \leq \err_n(M)\right\}}d Z \cdot {\bf 1}_{ \Data\in \mathcal{E}_1\cap \mathcal{E}_2}
\geq c^{*}(\alpha_0) \cdot \int {\bf 1}_{\left\{\|Z-\widehat{Z}\|_2 \leq \err_n(M)\right\}}d Z \cdot {\bf 1}_{ \Data\in \mathcal{E}_1\cap \mathcal{E}_2}\\
&\geq c^{*}(\alpha_0)\cdot {\rm Vol}(L(L+1)/2)\cdot [\err_n(M)]^{L(L+1)/2} \cdot {\bf 1}_{ \Data\in \mathcal{E}_1\cap \mathcal{E}_2},
\end{aligned}
\label{eq: main density}
\end{equation}
where ${\rm Vol}(L(L+1)/2)$ denotes the volume of the unit ball in $\frac{L(L+1)}{2}$-dimension.

Note that there exists $t\in (0,1)$ such that
$$g(Z)-g(\widehat{Z})=[\triangledown g(\widehat{Z}+t(Z-\widehat{Z}))]^{\intercal} (Z-\widehat{Z}),$$
with $\triangledown g(w)=\frac{\exp\left(-\frac{1}{2}{w}^{\intercal}({\Cov}+\frac{2}{3}d_0 {\bf I})^{-1}{w}\right)}{\sqrt{2\pi {\rm det}(\Cov+\frac{4}{3} d_0 {\bf I}) }} w^{\intercal}({\Cov}+\frac{2}{3}d_0 {\bf I})^{-1}{w}.$ Since $\lambda_{\min}({\Cov}+\frac{2}{3}d_0 {\bf I})\geq \frac{2}{3}d_0$, then $\triangledown g$ is bounded from the above and $\left|g(Z)-g(\widehat{Z})\right|\leq C \|Z-\widehat{Z}\|_2$ for a positive constant $C>0$. Then we establish  
\begin{equation}
\begin{aligned}
&\left|\int [g(Z)-g(\widehat{Z})] \cdot{\bf 1}_{\left\{\|Z-\widehat{Z}\|_2 \leq \err_n(M)\right\}} d Z \cdot {\bf 1}_{ \Data\in \mathcal{E}_1\cap \mathcal{E}_2}
\right|\\
&\leq C \cdot \err_n(M)\cdot \int {\bf 1}_{\left\{\|Z-\widehat{Z}\|_2 \leq \err_n(M)\right\}}d Z \cdot {\bf 1}_{ \Data\in \mathcal{E}_1\cap \mathcal{E}_2}\\
&= C \cdot \err_n(M)\cdot {\rm Vol}(L(L+1)/2)\cdot [\err_n(M)]^{L(L+1)/2} \cdot {\bf 1}_{ \Data\in \mathcal{E}_1\cap \mathcal{E}_2}.
\end{aligned}
\label{eq: approx density}
\end{equation}
Since $\lim_{M\rightarrow \infty} \err_n(M)=0,$ there exists a positive integer $M_1$ such that for $M\geq M_1,$ we have $C \cdot \err_n(M) \leq \frac{1}{2}c^{*}(\alpha_0).$ In the following, we focus on the large integer $M$ that is larger than $\max\{M_1,M_2\}.$ We combine \eqref{eq: decomposition density}, \eqref{eq: main density} and \eqref{eq: approx density} and obtain 
\begin{equation*}
\begin{aligned}
&\PP\left(\|Z^{\m}-\widehat{Z}\|_2 \leq \err_n(M)\mid \Data\right)\cdot {\bf 1}_{ \Data\in \mathcal{E}_1\cap \mathcal{E}_2}\\
&\geq \frac{1}{2}c^{*}(\alpha_0)\cdot{\rm Vol}(L(L+1)/2)\cdot [\err_n(M)]^{L(L+1)/2} \cdot {\bf 1}_{ \Data\in \mathcal{E}_1\cap \mathcal{E}_2}.
\end{aligned}
\end{equation*}
Together with \eqref{eq: connection}, we establish
\begin{equation}
\begin{aligned}
&\PP\left(\min_{1\leq m\leq M}\|Z^{\m}-\widehat{Z}\|_2 \leq \err_n(M)\mid \Data\right)\cdot {\bf 1}_{ \Data\in \mathcal{E}_1\cap \mathcal{E}_2}\\
&\geq 1-\exp\left[-M\cdot\frac{1}{2}c^{*}(\alpha_0)\cdot {\rm Vol}(L(L+1)/2)\cdot[\err_n(M)]^{L(L+1)/2} \cdot {\bf 1}_{\Data\in \mathcal{E}_1\cap \mathcal{E}_2}\right]\\
&=\left(1-\exp\left[-M\cdot\frac{1}{2}c^{*}(\alpha_0)\cdot {\rm Vol}(L(L+1)/2)\cdot[\err_n(M)]^{L(L+1)/2} \right]\right)
\cdot {\bf 1}_{ \Data\in \mathcal{E}_1\cap \mathcal{E}_2}.
\end{aligned}
\label{eq: connection 2}
\end{equation}
Together with \eqref{eq: lower 2}, we have
\begin{equation*}
\begin{aligned}
&\PP\left(\min_{m\in \MM}\|\Zm-\widehat{Z}\|_2 \leq \err_n(M)\right)\\
\geq &\left(1-\exp\left[-M\cdot \frac{1}{2}c^*(\alpha_0)\cdot {\rm Vol}(L(L+1)/2)\cdot[\err_n(M)]^{L(L+1)/2} \right]\right)
\PP\left(\mathcal{E}_1\cap \mathcal{E}_2\right).
\end{aligned}
\end{equation*}
We choose $\err_n(M) =\left[\frac{4 \log n}{C^{*}(\alpha_0) M}\right]^{\frac{2}{L(L+1)}}$ with $C^{*}(\alpha_0)$ defined in \eqref{eq: sampling constants} and establish that, for $m\geq \max\{M_0, M_1\},$
$$\PP\left(\min_{m\in \MM}\|\Zm-\widehat{Z}\|_2 \leq \err_n(M)\right)\geq (1-n^{-1})\cdot \PP\left(\mathcal{E}_1\cap \mathcal{E}_2\right).$$
We apply \eqref{eq: high event for sampling} and establish 
$
\liminf_{n,p\rightarrow \infty} \liminf_{M\rightarrow \infty} \PP\left(\min_{M\in \MM}\|\Zm-\widehat{Z}\|_2 \leq \err_n(M)\right)\geq 1-\alpha_0.
$
By the rescaling in \eqref{eq: rescaling expression}, we establish \eqref{eq: resampling approximation}.

\subsection{Proof of Theorem \ref{thm: sampling accuracy ridge}}
\label{sec: resampling weight accuracy}
The proof of Theorem \ref{thm: sampling accuracy ridge} relies on \eqref{eq: resampling approximation} together with the following two lemmas, whose proofs are presented in Section \ref{sec: gamma accuracy} and Section \ref{sec: positive inequality}.
\begin{Lemma} Define 
\begin{equation}
\widehat{\gamma}=\argmin_{\gamma\in \Delta^{L}}\gamma^{\intercal}\widehat{\Gamma}\gamma\quad \text{and}\quad {\gamma}^*=\argmin_{\gamma\in \Delta^{L}}\gamma^{\intercal}{\Gamma}\gamma.
\label{eq: temp def}
\end{equation}
 If $\lambda_{\min}(\Gamma)>0,$ then \begin{equation}
\|\widehat{\gamma}-\gamma^*\|_2 \leq \frac{\|\widehat{\Gamma}-\Gamma\|_2}{\lambda_{\min}({\Gamma})} \|\widehat{\gamma}\|_2 \leq \frac{\|\widehat{\Gamma}-\Gamma\|_F}{\lambda_{\min}({\Gamma})} .
\label{eq: key conclusion}
\end{equation}
\label{lemma: gamma accuracy}
\end{Lemma}

\begin{Lemma}
Suppose that $\Gamma$ is positive semi-definite, then
$
\|\widehat{\Gamma}_{+}-{\Gamma}\|_{F}\leq \|\widehat{\Gamma}-{\Gamma}\|_{F}.
$
\label{lem: positive inequality}
\end{Lemma}

We use $m^*\in \MM$ to denote one index such that 
$\|\widehat{\Gamma}^{[m^*]}-\Gamma^{\T}\|_{F}=\min_{m\in \MM}\|\widehat{\Gamma}^{\m}-{\Gamma}^{\T}\|_F.$
Then we apply Lemma \ref{lemma: gamma accuracy} with $\widehat{\Gamma}=(\widehat{\Gamma}^{[m^*]}+\delta \cdot {\rm I})_{+}$ and $\Gamma=\Gamma^{\T}+\delta \cdot {\rm I}$ and establish 
\begin{equation}
\|\widehat{\gamma}^{[m^*]}_{\delta}-\gamma^*_{\delta}\|_2 \leq \frac{\|(\widehat{\Gamma}^{[m^*]}+\delta \cdot {\rm I})_{+}-(\Gamma^{\T}+\delta \cdot {\rm I})\|_{F}}{\lambda_{\min}({\Gamma}^{\T})+\delta}\leq \frac{\|\widehat{\Gamma}^{[m^*]}-\Gamma^{\T}\|_{F}}{\lambda_{\min}({\Gamma}^{\T})+\delta},
\label{eq: key connection}
\end{equation} 
where the second inequality follows from Lemma \ref{lem: positive inequality}. 
Together with \eqref{eq: resampling approximation}, we establish \eqref{eq: resampling weight accuracy}.
\subsection{Proof of Lemma \ref{lem: lower bounded constant}}
\label{sec: lower bound constant}
By Proposition \ref{prop: general variance}, under Condition (A1), $s \log p\ll n$ and $N_{\T}\gtrsim\max\{n,p\},$ there exist positive constants $C>0$ and $c>0$ such that $\|n {\bf V}\|_{\infty}\leq C$ and $d_0\leq C$ with probability larger than $1-\min\{n,p\}^{-c}$. Furthermore, since $L$ is finite, $n \lambda_{\max}(V)\lesssim \|n {\bf V}\|_{\infty}.$  Together with $d_0\geq 1$, we establish thatwith probability larger than $1-\min\{n,p\}^{-c}$, $c^*(\alpha_0)\geq c'$ for a positive constant $c'>0.$ The lower bound for $C^{*}(\alpha_0)$ follows from the boundedness on $L$ and the lower bound for $c^*(\alpha_0).$

\subsection{Proof of Lemma \ref{lemma: gamma accuracy}}
\label{sec: gamma accuracy}
By the definition of $\gamma^{*}$ in \eqref{eq: temp def}, we have 
$
({\gamma}^{*})^{\intercal}{\Gamma}{\gamma}^{*} \leq \left[{\gamma}^{*}+t(\widehat{\gamma}-{\gamma}^{*})\right]^{\intercal}{\Gamma} \left[{\gamma}^{*}+t(\widehat{\gamma}-{\gamma}^{*})\right],
$  for any $t\in (0,1)$. This further leads to
$
0 \leq 2t({\gamma}^{*})^{\intercal}{\Gamma} (\widehat{\gamma}-{\gamma}^{*})+t^2(\widehat{\gamma}-{\gamma}^{*})^{\intercal}{\Gamma}(\widehat{\gamma}-{\gamma}^{*}).
$
By taking $t\rightarrow 0+,$ we have
\begin{equation}
({\gamma}^{*})^{\intercal}{\Gamma}(\widehat{\gamma}-\gamma^{*})\geq 0.
\label{eq: key optimal}
\end{equation}
By the definition of $\widehat{\gamma}$ in \eqref{eq: temp def}, we have 
$
\widehat{\gamma}^{\intercal}\widehat{\Gamma} \widehat{\gamma} \leq \left[\widehat{\gamma}+t(\gamma^{*}-\widehat{\gamma})\right]^{\intercal}\widehat{\Gamma} \left[\widehat{\gamma}+t(\gamma^{*}-\widehat{\gamma})\right], 
$ for any $t\in (0,1).$
This gives us 
$
2(\gamma^{*})^{\intercal}\widehat{\Gamma}(\gamma^{*}-\widehat{\gamma})+(t-2)(\gamma^{*}-\widehat{\gamma})^{\intercal}\widehat{\Gamma}(\gamma^{*}-\widehat{\gamma})\geq 0.
$
Since $2-t>0,$ we have 
\begin{equation}
(\gamma^{*}-\widehat{\gamma})^{\intercal}\widehat{\Gamma}(\gamma^{*}-\widehat{\gamma})\leq \frac{2}{2-t}(\gamma^{*})^{\intercal}\widehat{\Gamma}(\gamma^{*}-\widehat{\gamma}).
\label{eq: basic}
\end{equation}
%
It follows from \eqref{eq: key optimal} that
\begin{equation*}
\begin{aligned}
(\gamma^{*})^{\intercal}\widehat{\Gamma}(\gamma^{*}-\widehat{\gamma})=(\gamma^{*})^{\intercal}{\Gamma}(\gamma^{*}-\widehat{\gamma})+(\gamma^{*})^{\intercal}(\widehat{\Gamma}-\Gamma)(\gamma^{*}-\widehat{\gamma})\leq (\gamma^{*})^{\intercal}(\widehat{\Gamma}-\Gamma)(\gamma^{*}-\widehat{\gamma}).
\end{aligned}
\end{equation*}
Combined with \eqref{eq: basic}, we have 
\begin{equation}
\begin{aligned}
(\gamma^{*}-\widehat{\gamma})^{\intercal}\widehat{\Gamma}(\gamma^{*}-\widehat{\gamma})
\leq \frac{2}{2-t} (\gamma^{*})^{\intercal}(\widehat{\Gamma}-\Gamma)(\gamma^{*}-\widehat{\gamma})
\leq \frac{2\|\gamma^{*}\|_2}{2-t} \|\widehat{\Gamma}-\Gamma\|_2\|\gamma^{*}-\widehat{\gamma}\|_2.
\end{aligned}
\label{eq: intermediate result}
\end{equation}

Note that the definitions of $\widehat{\gamma}$ and $\gamma$ are symmetric. Specifically,  $\widehat{\gamma}$ is defined as minimizing a quadratic form of $\widehat{\Gamma}$ while ${\gamma}^*$ is defined with ${\Gamma}.$ We switch the roles of $\{\widehat{\Gamma},\widehat{\gamma}\}$ and $\{\Gamma,\gamma^{*}\}$ in \eqref{eq: intermediate result} and establish 
$
(\gamma^{*}-\widehat{\gamma})^{\intercal}{\Gamma}(\gamma^{*}-\widehat{\gamma}) \leq \frac{2\|\widehat{\gamma}\|_2}{2-t} \|\widehat{\Gamma}-\Gamma\|_2\|\gamma^{*}-\widehat{\gamma}\|_2.
$ If $\lambda_{\min}({\Gamma})>0,$ we apply the above bound by taking $t \rightarrow 0+$ and  establish \eqref{eq: key conclusion}. 
\subsection{Proof of Lemma \ref{lem: positive inequality}}
\label{sec: positive inequality}
Write the eigenvalue decomposition of $\widehat{\Gamma}$ as $\widehat{\Gamma}=\sum_{l=1}^{L}\widehat{A}_{ll}u_l u_l^{\intercal}$. Define 
$$\widehat{\Gamma}_{+}=\sum_{l=1}^{L}\max\{\widehat{A}_{ll},0\}u_l u_l^{\intercal}\quad \text{and}\quad \widehat{\Gamma}_{-}=\sum_{l=1}^{L}-\min\{\widehat{A}_{ll},0\}u_l u_l^{\intercal}.$$
We have $
\widehat{\Gamma}=\widehat{\Gamma}_{+}-\widehat{\Gamma}_{-} \quad \text{with}\quad {\rm Tr}(\widehat{\Gamma}_{+}^{\intercal}\widehat{\Gamma}_{-})=0.$
Since $\Gamma$ is positive semi-definite, we have
$
 {\rm Tr}({\Gamma}^{\intercal}\widehat{\Gamma}_{-})=\sum_{l=1}^{L}-\min\{\widehat{A}_{ll},0\}{\rm Tr}({\Gamma}^{\intercal} u_l u_l^{\intercal})\geq 0.
$
We apply the above two equalities and establish  
\begin{equation*}
\|\widehat{\Gamma}-{\Gamma}\|_{F}^2=\|\widehat{\Gamma}_{+}-{\Gamma}-\widehat{\Gamma}_{-}\|_{F}^2=\|\widehat{\Gamma}_{+}-{\Gamma}\|_{F}^2+\|\widehat{\Gamma}_{-}\|_{F}^2-2 {\rm Tr}(\widehat{\Gamma}_{+}^{\intercal}\widehat{\Gamma}_{-})+2 {\rm Tr}({\Gamma}^{\intercal}\widehat{\Gamma}_{-})
\geq \|\widehat{\Gamma}_{+}-{\Gamma}\|_{F}^2.
\end{equation*}

\section{Proof of Theorem \ref{thm: inference for linear}}
\label{sec: inference results}
We will prove the properties of the CI for the ridge-type maximin effect, which includes Theorem \ref{thm: inference for linear} as a special case.  We first detail the generalized inference procedure in the following, which is a generalization of the inference method in Section \ref{sec: sampling method}. We construct the sampled weight as 
\begin{equation*}
\widehat{\gamma}^{\m}_{\delta}=\argmin_{\gamma\in \Delta^{L}} \gamma^{\intercal}(\widehat{\Gamma}^{\m}+\delta \cdot {\rm I})_{+} \gamma \quad \text{for}\quad \delta\geq 0.
\end{equation*}
For $1\leq m\leq M$, we compute $
\widehat{\xnew^{\intercal}\beta}^{\m}=\sum_{l=1}^{L} [\widehat{\gamma}^{\m}_{\delta}]_{l}\cdot \widehat{\xnew^{\intercal}b^{(l)}},
$ and $
\widehat{\rm se}^{\m}(\xnew)=\sqrt{\sum_{l=1}^{L} [\widehat{\gamma}^{\m}_{\delta}]_{l}^{2}\cdot{\widehat{\rm V}^{(l)}_{\xnew}}}.
$
We construct the sampled interval as, 
\begin{equation}
{\rm Int}_{\alpha}^{\m}(\xnew)=\left(\widehat{\xnew^{\intercal}\beta}^{\m}-z_{\alpha/2} \widehat{{\rm se}}^{\m}(\xnew), \widehat{\xnew^{\intercal}\beta}^{\m}+z_{\alpha/2} \widehat{{\rm se}}^{\m}(\xnew)\right).
\label{eq: sampled interval ridge} 
\end{equation}

We slightly abuse the notation by using ${\rm Int}_{\alpha}^{\m}(\xnew)$ to denote the sampled interval for both $\xnew^{\intercal}\bp_{\delta}$  and $\xnew^{\intercal}\bp.$ 
We construct the CI for $\xnew^{\intercal}\bp_{\delta}$ by aggregating the sampled intervals with $m\in \MM$,
\begin{equation}
{\rm CI}_{\alpha}\left(\xnew^{\intercal}\bp_{\delta}\right)=\cup_{m\in \mathbb{M}} {\rm Int}_{\alpha}^{\m}(\xnew),
\label{eq: aggregation 1 ridge}
\end{equation} 
where $\mathbb{M}$ is defined in \eqref{eq: generating condition} and ${\rm Int}_{\alpha}^{\m}(\xnew)$ is defined in \eqref{eq: sampled interval ridge}.

In the following Theorem \ref{thm: inference for linear ridge}, we establish the coverage and precision properties of ${\rm CI}_{\alpha}\left(\xnew^{\intercal}\bp_{\delta}\right)$ defined in \eqref{eq: aggregation 1 ridge}.  We apply Theorem \ref{thm: inference for linear ridge} with $\delta=0$ and establish Theorem \ref{thm: inference for linear}.  


\begin{Theorem} 
Suppose that the conditions of Theorem \ref{thm: sampling accuracy ridge} hold. Then for any positive constant $\eta_0>0$ used in \eqref{eq: sampled interval ridge}, the confidence interval ${\rm CI}_{\alpha}(\xnew^{\intercal}\bp_{\delta})$ defined in 
\eqref{eq: aggregation 1 ridge} satisfies 
\begin{equation}
\liminf_{n,p\rightarrow \infty}\liminf_{M\rightarrow \infty}\PP\left(\xnew^{\intercal} \bp_{\delta}\in {\rm CI}_{\alpha}\left(\xnew^{\intercal}\bp_{\delta}\right)\right)\geq 1-\alpha-\alpha_0,
\label{eq: coverage ridge}
\end{equation}
where $\alpha\in (0,1/2)$ is the pre-specified significance level and $\alpha_0\in(0,0.01]$ is defined in \eqref{eq: generating condition}. By further assuming $N_{\T}\gtrsim\max\{n,p\}$ and $\lambda_{\min}(\Gamma^{\T})+\delta\gg \sqrt{\log p/\min\{n,N_{\T}\}}$, then there exists some positive constant $C>0$ such that
\begin{equation}
\lim_{n,p\rightarrow \infty} \PP\left(\mathbf{Leng}\left({\rm CI}_{\alpha}\left(\xnew^{\intercal}\bp_{\delta}\right)\right)\leq C\max\left\{ 1,\frac{z_{\alpha_0/[L(L+1)]}}{\lambda_{\min}(\Gamma^{\T})+\delta}\right\}\cdot \frac{\|\xnew\|_2}{\sqrt{n}}\right)=1,
\label{eq: length bound ridge}
\end{equation}
where $\mathbf{Leng}\left({\rm CI}\left(\xnew^{\intercal}\bp_{\delta}\right)\right)$ denotes the interval length and $z_{\alpha_0/[L(L+1)]}$ is the upper $\alpha_0/[L(L+1)]$ quantile of the standard normal distribution.
\label{thm: inference for linear ridge}
\end{Theorem}

Recall the definition
\begin{equation}
{{\rm V}^{(l)}_{\xnew}}=
({{\sigma}_{l}^2}/{\nl^2})[\widehat{v}^{(l)}]^{\intercal}(X^{(l)})^{\intercal}X^{(l)}\widehat{v}^{(l)}  \quad \text{and}\quad
{\widehat{\rm V}^{(l)}_{\xnew}}=
({\widehat{\sigma}_{l}^2}/{\nl^2})[\widehat{v}^{(l)}]^{\intercal}(X^{(l)})^{\intercal}X^{(l)}\widehat{v}^{(l)}.
\end{equation}

In the following, we introduce the definitions of events, which are used to facilitate the proof of Theorem \ref{thm: inference for linear ridge}. 
We shall take $m^*$ as any index such that $\|\widehat{\gamma}^{[m^{*}]}_{\delta}-\gamma^{*}_{\delta}\|_2=\min_{m\in \MM}\|\widehat{\gamma}^{\m}_{\delta}-\gamma^{*}_{\delta}\|_2$.
We introduce the following high-probability events to facilitate the discussion. 
\begin{equation}
\begin{aligned}
\mathcal{E}_3&=\left\{n_l {{\rm V}^{(l)}_{\xnew}}\asymp \|\xnew\|_2 \quad \text{for}\; 1\leq l\leq L\right\},\\
\mathcal{E}_4&=\left\{\max_{1\leq l\leq L}\frac{|\widehat{\xnew^{\intercal}b^{(l)}}-{\xnew^{\intercal}b^{(l)}}|}{\|\xnew\|_2}\lesssim \sqrt{\frac{\log n}{n}}\right\},\\
\mathcal{E}_5&=\left\{\frac{\left|\sum_{l=1}^{L} \left([\widehat{\gamma}_{\delta}^{[m]}]_{l}-[\gamma_{\delta}^{*}]_{l}\right)\widehat{\xnew^{\intercal}b^{(l)}}\right|}{\sqrt{\sum_{l=1}^{L}[{\gamma}^{*}_{\delta}]_{l}^{2}{{\rm V}^{(l)}_{\xnew}}}}\lesssim \sqrt{n}\|\widehat{\gamma}^{[m]}_{\delta}-\gamma^{*}_{\delta}\|_2,\; \text{for}\; 1\leq m\leq M\right\},\\
\mathcal{E}_6&=\left\{\|\widehat{\gamma}^{[m^{*}]}_{\delta}-\gamma^{*}_{\delta}\|_2\leq \frac{\sqrt{2} \err_n(M)}{\lambda_{\min}(\Gamma^{\T})+\delta} \cdot \frac{1}{\sqrt{n}}\right\},\\
\mathcal{E}_7&=\left\{\|\widehat{\Gamma}^{\T}-\Gamma^{\T}\|_2\lesssim \sqrt{\frac{\log p}{\min\{n,N_{\T}\}}}+\frac{p\sqrt{\log p}}{\sqrt{n}N_{\T}}\right\}.
\end{aligned}
\label{eq: high prob inference}
 \end{equation}
We apply Lemma 1 of \cite{cai2019individualized} and establish that
$
\PP\left(\mathcal{E}_3\right)\geq 1-p^{-c},
$ for some positive constant $c>0.$  We introduce the following lemma to justify the asymptotic normality of $\widehat{\xnew^{\intercal}b^{(l)}}$, which follows the same proof as that of Proposition 1 in \cite{cai2019individualized}.
 \begin{Lemma} Consider the model \eqref{eq: multi-group model}. Suppose Conditions {\rm (A1)} and {\rm (A2)} hold, then 
\begin{equation}
\frac{\sum_{l=1}^{L} c_{l} [\widehat{\xnew^{\intercal}b^{(l)}}-{\xnew^{\intercal}b^{(l)}}]}{\sqrt{\sum_{l=1}^{L}c_{l}^{2}{{\rm V}^{(l)}_{\xnew}}}}\cid N(0,1).
\label{eq: limit distribution general}
\end{equation}
where $|c_l|\leq 1$ for any $1\leq l\leq L$, and $\sum_{l=1}^{L}c_l=1.$ 
 \label{lem: linear functional}
\end{Lemma}
We apply Lemma \ref{lem: linear functional} with  $c_l=1$ and $c_j=0$ for $j\neq l$ and show that 
$\PP(\mathcal{E}_4\cap \mathcal{E}_3)\geq 1-\min\{n,p\}^{-c},$ for some positive constant $c>0.$
On the event $\mathcal{E}_3$, we have 
\begin{equation}
\sqrt{\sum_{l=1}^{L}[{\gamma}^{*}_{\delta}]_{l}^{2}{{\rm V}^{(l)}_{\xnew}}}\asymp \frac{\|\gamma^{*}_{\delta}\|_2\|\xnew\|_2}{\sqrt{n}}\asymp \frac{\|\xnew\|_2}{\sqrt{n}},
\label{eq: lower variance}
\end{equation}
where the last asymptotic equivalence holds since $\frac{1}{\sqrt{L}}\leq \|\gamma^{*}_{\delta}\|_2 \leq 1.$ Similarly,  on the event $\mathcal{E}_3,$ 
\begin{equation}
\sqrt{\sum_{l=1}^{L}[\widehat{\gamma}^{\m}_{\delta}]_{l}^{2}{{\rm V}^{(l)}_{\xnew}}}\asymp \frac{\|\widehat{\gamma}^{\m}_{\delta}\|_2\|\xnew\|_2}{\sqrt{n}}\asymp \frac{\|\xnew\|_2}{\sqrt{n}}.
\label{eq: first upper bound}
\end{equation}
Note that 
\begin{equation}
\begin{aligned}
\left|\sum_{l=1}^{L} \left([\widehat{\gamma}_{\delta}^{[m]}]_{l}-[\gamma_{\delta}^{*}]_{l}\right)\widehat{\xnew^{\intercal}b^{(l)}}\right|
&\leq \|\widehat{\gamma}^{[m]}_{\delta}-\gamma^{*}_{\delta}\|_2\sqrt{\sum_{l=1}^{L}[\widehat{\xnew^{\intercal}b^{(l)}}]^2}\\
&\lesssim\|\widehat{\gamma}^{[m]}_{\delta}-\gamma^{*}_{\delta}\|_2\sqrt{\sum_{l=1}^{L}[\widehat{\xnew^{\intercal}b^{(l)}}-{\xnew^{\intercal}b^{(l)}}]^2+\sum_{l=1}^{L}[{\xnew^{\intercal}b^{(l)}}]^2}.
\end{aligned}
\label{eq: error upper}
\end{equation}
By Lemma \ref{lem: linear functional} and \eqref{eq: lower variance}, with probability larger than $1-\min\{n,p\}^{-c}$, 
{\small\begin{equation}
\begin{aligned}
\frac{\left|\sum_{l=1}^{L} \left([\widehat{\gamma}_{\delta}^{[m]}]_{l}-[\gamma_{\delta}^{*}]_{l}\right)\widehat{\xnew^{\intercal}b^{(l)}}\right|}{\sqrt{\sum_{l=1}^{L}[{\gamma}^{*}_{\delta}]_{l}^{2}{{\rm V}^{(l)}_{\xnew}}}}
\lesssim \|\widehat{\gamma}^{[m]}_{\delta}-\gamma^{*}_{\delta}\|_2\cdot\frac{\sqrt{L}\cdot(\log n\cdot\frac{\|\xnew\|_2}{\sqrt{n}}+\max_{l}|\xnew^{\intercal}b^{(l)}|)}{\frac{1}{\sqrt{L}}\cdot \frac{\|\xnew\|_2}{\sqrt{n}}}\lesssim \sqrt{n}\|\widehat{\gamma}^{[m]}_{\delta}-\gamma^{*}_{\delta}\|_2,
 \end{aligned}
 \label{eq: error upper 2}
\end{equation}}
where the last inequality follows from bounded $\|b^{(l)}\|_2$ and finite $L.$ This implies $\PP(\mathcal{E}_5)\geq 1-\min\{n,p\}^{-c}$ for some constant $c>0.$ It follows from \eqref{eq: resampling weight accuracy} that $\liminf_{n,p\rightarrow\infty}\liminf_{M\rightarrow\infty}\PP(\mathcal{E}_6)\geq1-\alpha_0$. It follows from Propositions \ref{prop: limiting Gamma univariate} and \ref{prop: general variance} that $\liminf_{n,p\rightarrow\infty}\PP(\mathcal{E}_7)=1.$  
\subsection{Coverage Property: Proof of \eqref{eq: coverage ridge}} 

By the definition in \eqref{eq: aggregation 1 ridge}, we have 
{\footnotesize
\begin{equation}
\begin{aligned}
\PP\left(\xnew^{\intercal}\bp_{\delta}\not\in {\rm CI}_{\alpha}\left(\xnew^{\intercal}\bp_{\delta}\right)\right)\leq \PP\left(\xnew^{\intercal}\bp_{\delta}\not\in {\rm Int}_{\alpha}^{[m^*]}(\xnew)\right)=\PP\left(\frac{|\widehat{\xnew^{\intercal}\beta}^{[m^*]}_{\delta}-\xnew^{\intercal}\bp_{\delta}|}{\sqrt{\sum_{l=1}^{L}[{\gamma}^{*}_{\delta}]_{l}^{2}{{\rm V}^{(l)}_{\xnew}}}}\geq z_{\alpha/2}\sqrt{\frac{{\sum_{l=1}^{L}[\widehat{\gamma}^{[m^*]}_{\delta}]_{l}^{2}\widehat{\rm V}^{(l)}_{\xnew}}}{{\sum_{l=1}^{L}[{\gamma}^{*}_{\delta}]_{l}^{2}{{\rm V}^{(l)}_{\xnew}}}}}\right).
\end{aligned}
\label{eq: type I bound}
\end{equation}
}
Note that
\begin{equation}
\begin{aligned}
\widehat{\xnew^{\intercal}\beta}^{\m}_{\delta}-\xnew^{\intercal}\bp_{\delta}
=\sum_{l=1}^{L} ([\widehat{\gamma}^{\m}_{\delta}]_{l}-[{\gamma}^{*}_{\delta}]_{l})\cdot\widehat{\xnew^{\intercal}b^{(l)}}+\sum_{l=1}^{L} [{\gamma}^{*}_{\delta}]_{l} \cdot (\widehat{\xnew^{\intercal}b^{(l)}}-{\xnew^{\intercal}b^{(l)}}).
\end{aligned}
\label{eq: type I decomp}
\end{equation}
Note that 
$$\left|\frac{{\sum_{l=1}^{L}[\widehat{\gamma}^{[m^*]}_{\delta}]_{l}^{2}\widehat{\rm V}^{(l)}_{\xnew}}}{{\sum_{l=1}^{L}[{\gamma}^{*}_{\delta}]_{l}^{2}{{\rm V}^{(l)}_{\xnew}}}}-1\right|\leq \left|\frac{{\sum_{l=1}^{L}\left([\widehat{\gamma}^{[m^*]}_{\delta}]_{l}^{2}-[{\gamma}^{*}_{\delta}]_{l}^{2}\right)\widehat{\rm V}^{(l)}_{\xnew}}}{{\sum_{l=1}^{L}[{\gamma}^{*}_{\delta}]_{l}^{2}{{\rm V}^{(l)}_{\xnew}}}}\right|+\left|\frac{{\sum_{l=1}^{L}[{\gamma}^{*}_{\delta}]_{l}^{2}\left(\widehat{\rm V}^{(l)}_{\xnew}-{\rm V}^{(l)}_{\xnew}\right)}}{{\sum_{l=1}^{L}[{\gamma}^{*}_{\delta}]_{l}^{2}{{\rm V}^{(l)}_{\xnew}}}}\right|.$$
On the event $\mathcal{G}_3\cap  \mathcal{E}_3 \cap  \mathcal{E}_6$  with $\mathcal{G}_3$ defined in \eqref{eq: high prob event 1}, 
we apply \eqref{eq: lower variance} and \eqref{eq: first upper bound} to establish $$\left|\sqrt{\frac{{\sum_{l=1}^{L}[\widehat{\gamma}^{[m^*]}_{\delta}]_{l}^{2}\widehat{\rm V}^{(l)}_{\xnew}}}{{\sum_{l=1}^{L}[{\gamma}^{*}_{\delta}]_{l}^{2}{{\rm V}^{(l)}_{\xnew}}}}}-1\right|\leq \kappa_{n,M} \quad \text{with}\quad \kappa_{n,M}=C\frac{1}{\sqrt{n}}+C\frac{k \log p}{n}+\frac{\sqrt{2} \err_n(M)}{\lambda_{\min}(\Gamma^{\T})+\delta} \cdot \frac{1}{\sqrt{n}}.$$ 
On the event $\mathcal{E}_5\cap \mathcal{E}_6,$ we have  
$$\frac{|\sum_{l=1}^{L} (\widehat{\gamma}^{[m^*]}_{\delta}]_{l}-[{\gamma}^{*}_{\delta}]_{l})\cdot\widehat{\xnew^{\intercal}b^{(l)}}|}{\sqrt{\sum_{l=1}^{L}[{\gamma}^{*}_{\delta}]_{l}^{2}{{\rm V}^{(l)}_{\xnew}}}}\lesssim \frac{\sqrt{2} \err_n(M)}{\lambda_{\min}(\Gamma^{\T})+\delta}.$$
We apply the above two inequalities and establish the following result: on the event $\mathcal{G}_3\cap  \mathcal{E}_3\cap \mathcal{E}_5\cap \mathcal{E}_6$, there exists a positive constant $C>0$  such that 
\begin{equation*}
\frac{|\sum_{l=1}^{L} (\widehat{\gamma}^{[m^*]}_{\delta}]_{l}-[{\gamma}^{*}_{\delta}]_{l})\cdot\widehat{\xnew^{\intercal}b^{(l)}}|}{\sqrt{\sum_{l=1}^{L}[{\gamma}^{*}_{\delta}]_{l}^{2}{{\rm V}^{(l)}_{\xnew}}}}\leq \eta_M\cdot z_{\alpha/2}\sqrt{\frac{{\sum_{l=1}^{L}[\widehat{\gamma}^{[m^*]}_{\delta}]_{l}^{2}\widehat{\rm V}^{(l)}_{\xnew}}}{{\sum_{l=1}^{L}[{\gamma}^{*}_{\delta}]_{l}^{2}{{\rm V}^{(l)}_{\xnew}}}}} \; \text{with}\; \eta_M= C \frac{\sqrt{2} \err_n(M)}{\lambda_{\min}(\Gamma^{\T})+\delta}.
\end{equation*}
Define the events 
{\small
\begin{equation*}
\mathcal{E}_9=\left\{\left|\sqrt{\frac{{\sum_{l=1}^{L}[\widehat{\gamma}^{[m^*]}_{\delta}]_{l}^{2}\widehat{\rm V}^{(l)}_{\xnew}}}{{\sum_{l=1}^{L}[{\gamma}^{*}_{\delta}]_{l}^{2}{{\rm V}^{(l)}_{\xnew}}}}}-1\right|\leq \kappa_{n,M}, \;\frac{|\sum_{l=1}^{L} (\widehat{\gamma}^{[m^*]}_{\delta}]_{l}-[{\gamma}^{*}_{\delta}]_{l})\cdot\widehat{\xnew^{\intercal}b^{(l)}}|}{\sqrt{{\sum_{l=1}^{L}[\widehat{\gamma}^{[m^*]}_{\delta}]_{l}^{2}\widehat{\rm V}^{(l)}_{\xnew}}}}\leq \eta_M\cdot z_{\alpha/2}\right\}
\end{equation*}
}
We have 
\begin{equation}
\PP\left(\mathcal{E}_9\right)\geq \PP\left(\mathcal{G}_3\cap  \mathcal{E}_3\cap \mathcal{E}_5\cap \mathcal{E}_6\right).
\label{eq: event lower}
\end{equation}
We apply the union bound and control the probability in \eqref{eq: type I bound} as
{\footnotesize
\begin{equation}
\begin{aligned}
&\PP\left(\frac{|\widehat{\xnew^{\intercal}\beta}^{[m^*]}_{\delta}-\xnew^{\intercal}\bp_{\delta}|}{\sqrt{\sum_{l=1}^{L}[{\gamma}^{*}_{\delta}]_{l}^{2}{{\rm V}^{(l)}_{\xnew}}}}\geq z_{\alpha/2}\sqrt{\frac{{\sum_{l=1}^{L}[\widehat{\gamma}^{[m^*]}_{\delta}]_{l}^{2}\widehat{\rm V}^{(l)}_{\xnew}}}{{\sum_{l=1}^{L}[{\gamma}^{*}_{\delta}]_{l}^{2}{{\rm V}^{(l)}_{\xnew}}}}}\right)\\
&=\PP\left(\mathcal{E}_9^c\cap\left\{\frac{|\widehat{\xnew^{\intercal}\beta}^{[m^*]}_{\delta}-\xnew^{\intercal}\bp_{\delta}|}{\sqrt{\sum_{l=1}^{L}[{\gamma}^{*}_{\delta}]_{l}^{2}{{\rm V}^{(l)}_{\xnew}}}}\geq z_{\alpha/2}\sqrt{\frac{{\sum_{l=1}^{L}[\widehat{\gamma}^{[m^*]}_{\delta}]_{l}^{2}\widehat{\rm V}^{(l)}_{\xnew}}}{{\sum_{l=1}^{L}[{\gamma}^{*}_{\delta}]_{l}^{2}{{\rm V}^{(l)}_{\xnew}}}}}\right\}\right)\\
&+\PP\left(\mathcal{E}_9\cap\left\{\frac{|\widehat{\xnew^{\intercal}\beta}^{[m^*]}_{\delta}-\xnew^{\intercal}\bp_{\delta}|}{\sqrt{\sum_{l=1}^{L}[{\gamma}^{*}_{\delta}]_{l}^{2}{{\rm V}^{(l)}_{\xnew}}}}\geq z_{\alpha/2}\sqrt{\frac{{\sum_{l=1}^{L}[\widehat{\gamma}^{[m^*]}_{\delta}]_{l}^{2}\widehat{\rm V}^{(l)}_{\xnew}}}{{\sum_{l=1}^{L}[{\gamma}^{*}_{\delta}]_{l}^{2}{{\rm V}^{(l)}_{\xnew}}}}}\right\}\right)\\
&\leq \PP\left(\mathcal{E}_9^c\right)+\PP\left(\frac{|\sum_{l=1}^{L} [{\gamma}^{*}_{\delta}]_{l} \cdot (\widehat{\xnew^{\intercal}b^{(l)}}-{\xnew^{\intercal}b^{(l)}})|}{\sqrt{\sum_{l=1}^{L}[{\gamma}^{*}_{\delta}]_{l}^{2}{{\rm V}^{(l)}_{\xnew}}}}\geq (1-\eta_M)(1-\kappa_{n,M})z_{\alpha/2}\right),
\end{aligned}
\label{eq: decomp upper prob}
\end{equation}
}
where the last inequality follows from the definition of $\mathcal{E}_9.$ We combine the above bound with \eqref{eq: type I bound} and establish 
$$\PP\left(\xnew^{\intercal}\bp_{\delta}\in {\rm CI}_{\alpha}\left(\xnew^{\intercal}\bp_{\delta}\right)\right)\geq \PP\left(\mathcal{E}_9\right)-\PP\left(\frac{|\sum_{l=1}^{L} [{\gamma}^{*}_{\delta}]_{l} \cdot (\widehat{\xnew^{\intercal}b^{(l)}}-{\xnew^{\intercal}b^{(l)}})|}{\sqrt{\sum_{l=1}^{L}[{\gamma}^{*}_{\delta}]_{l}^{2}{{\rm V}^{(l)}_{\xnew}}}}\geq (1-\eta_M)(1-\kappa_{n,M})z_{\alpha/2}\right)$$
It follows from Lemma \ref{lem: linear functional} that 
$
\frac{\sum_{l=1}^{L} [{\gamma}^{*}_{\delta}]_{l} [\widehat{\xnew^{\intercal}b^{(l)}}-{\xnew^{\intercal}b^{(l)}}]}{\sqrt{\sum_{l=1}^{L}[{\gamma}^{*}_{\delta}]_{l}^{2}{{\rm V}^{(l)}_{\xnew}}}}\cid N(0,1).
$
We apply the bounded convergence theorem and establish  
\begin{equation*}
\lim_{n,p\rightarrow \infty}\lim_{M\rightarrow \infty}\PP\left(\frac{|\sum_{l=1}^{L} [{\gamma}^{*}_{\delta}]_{l} \cdot (\widehat{\xnew^{\intercal}b^{(l)}}-{\xnew^{\intercal}b^{(l)}})|}{\sqrt{\sum_{l=1}^{L}[{\gamma}^{*}_{\delta}]_{l}^{2}{{\rm V}^{(l)}_{\xnew}}}}\geq (1-\eta_M)(1-\kappa_{n,M})z_{\alpha/2}\right)=\alpha.
\end{equation*}
By \eqref{eq: control of event 1}, \eqref{eq: event lower} and Theorem \ref{thm: sampling accuracy ridge}, we establish \eqref{eq: coverage ridge}. 
\subsection{Precision Property: Proof of \eqref{eq: length bound ridge}}  
Regarding the length of the confidence interval, we notice that 
\begin{equation}
\mathbf{Leng}\left({\rm CI}\left(\xnew^{\intercal}\bp_{\delta}\right)\right)\leq 2\max_{m\in \MM}\left(\left|\widehat{\xnew^{\intercal}\beta}^{\m}_{\delta}-\widehat{\xnew^{\intercal}\bp_{\delta}}\right|+z_{\alpha/2} \widehat{{\rm se}}^{\m}(\xnew)\right),
\label{eq: length decomp}
\end{equation}
where $\widehat{\xnew^{\intercal}\bp_{\delta}}=\sum_{l=1}^{L} [\widehat{\gamma}_{\delta}]_{l}\cdot \widehat{\xnew^{\intercal}b^{(l)}}$ is defined in \eqref{eq: point estimator ridge} and $\widehat{\xnew^{\intercal}\beta}^{\m}=\sum_{l=1}^{L} [\widehat{\gamma}^{\m}_{\delta}]_{l}\cdot \widehat{\xnew^{\intercal}b^{(l)}}.$
Note that 
\begin{equation}
\begin{aligned}
\max_{m\in \MM}\left|\widehat{\xnew^{\intercal}\beta}^{\m}_{\delta}-\widehat{\xnew^{\intercal}\bp_{\delta}}\right|&=\max_{m\in \MM}\left|\sum_{l=1}^{L}\left([\widehat{\gamma}_{\delta}]_{l}-[\widehat{\gamma}^{\m}_{\delta}]_{l}\right)\cdot \widehat{\xnew^{\intercal}b^{(l)}}\right|\\
&\leq \max_{m\in \MM}\|\widehat{\gamma}_{\delta}-\widehat{\gamma}^{\m}_{\delta}\|_2 \cdot \sqrt{\sum_{l=1}^{L} (\widehat{\xnew^{\intercal}b^{(l)}})^2}.
\end{aligned}
\label{eq: length key bound}
\end{equation}
For  $N_{\T}\gtrsim\max\{n,p\}$ and $\lambda_{\min}(\Gamma^{\T})+\delta\gg \sqrt{\log p/\min\{n,N_{\T}\}}$, we have $$\lambda_{\min}(\Gamma^{\T})+\delta \gg \sqrt{\frac{\log p}{\min\{n,N_{\T}\}}}+\frac{p\sqrt{\log p}}{\sqrt{n}N_{\T}}.$$
On the event $\mathcal{E}_7,$ we establish \begin{equation}
\lambda_{\min}(\widehat{\Gamma})+\delta\geq \frac{1}{2}\left(\lambda_{\min}({\Gamma}^{\T})+\delta\right).
\label{eq: lambda minimum bound}
\end{equation}

We now apply Lemma \ref{lemma: gamma accuracy} with $\widehat{\Gamma}=(\widehat{\Gamma}^{[m]}+\delta\cdot {\rm I})_{+}$ and $\Gamma=\widehat{\Gamma}+\delta\cdot {\rm I}$ and establish that 
\begin{equation*}
\|\widehat{\gamma}_{\delta}-\widehat{\gamma}^{\m}_{\delta}\|_2\leq \frac{\|(\widehat{\Gamma}^{[m]}+\delta\cdot {\rm I})_{+}-(\widehat{\Gamma}+\delta\cdot {\rm I})\|_F}{\lambda_{\min}(\widehat{\Gamma})+\delta}\leq \frac{\|\widehat{\Gamma}^{[m]}-\widehat{\Gamma}\|_F}{\lambda_{\min}(\widehat{\Gamma})+\delta},
\end{equation*}
where the last inequality follows from Lemma \ref{lem: positive inequality} together with $\lambda_{\min}(\widehat{\Gamma})+\delta>0$ on the event $\mathcal{E}_7.$ Combining the above inequality and \eqref{eq: length key bound}, we establish 
\begin{equation}
\max_{m\in \MM}\left|\widehat{\xnew^{\intercal}\beta}^{\m}_{\delta}-\widehat{\xnew^{\intercal}\bp_{\delta}}\right|\leq \max_{m\in \MM}\frac{\|\widehat{\Gamma}^{[m]}-\widehat{\Gamma}\|_F}{\lambda_{\min}(\widehat{\Gamma})+\delta} \cdot \sqrt{\sum_{l=1}^{L} (\widehat{\xnew^{\intercal}b^{(l)}})^2}.
\label{eq: length key bound 2}
\end{equation}
By the definition of $\MM$ in \eqref{eq: generating condition} and the definition of $d_0$ in \eqref{eq: resampling}, we have
$$\max_{m\in \MM}\|\widehat{\Gamma}^{[m]}-\widehat{\Gamma}\|_F\lesssim  L\cdot\sqrt{{d_0}/n}\cdot  1.1 \cdot z_{\alpha_0/[L(L+1)]}.$$ Together with \eqref{eq: length key bound 2}, we establish 
\begin{equation}
\max_{m\in \MM}\left|\widehat{\xnew^{\intercal}\beta}^{\m}-\widehat{\xnew^{\intercal}\bp_{\delta}}\right|\lesssim \frac{1.1 L\cdot \sqrt{{d_0}}}{\sqrt{n}[\lambda_{\min}(\widehat{\Gamma})+\delta]} \cdot \sqrt{\sum_{l=1}^{L} (\widehat{\xnew^{\intercal}b^{(l)}})^2}\cdot  z_{\alpha_0/[L(L+1)]}.
\label{eq: length decomp a}
\end{equation}

On the event $\mathcal{E}_3$, we apply \eqref{eq: first upper bound} and establish 
\begin{equation}
\max_{m\in \MM}\widehat{\rm se}^{\m}(\xnew)\lesssim \frac{\|\xnew\|_2}{\sqrt{n}}.
\label{eq: length decomp b}
\end{equation}
We combine \eqref{eq: length decomp}, \eqref{eq: length decomp a}, \eqref{eq: lambda minimum bound} and \eqref{eq: length decomp b} and establish that, on the event $\mathcal{E}_3\cap \mathcal{E}_7,$ we have 
\begin{equation}
\mathbf{Leng}\left({\rm CI}_{\alpha}\left(\xnew^{\intercal}\bp_{\delta}\right)\right)\lesssim \frac{L\cdot \sqrt{{d_0}}}{\sqrt{n}[\lambda_{\min}({\Gamma}^{\T})+\delta]} \cdot \sqrt{\sum_{l=1}^{L} (\widehat{\xnew^{\intercal}b^{(l)}})^2}\cdot  z_{\alpha_0/[L(L+1)]}+\frac{\|\xnew\|_2}{\sqrt{n}}.
\label{eq: inter length}
\end{equation}

On the event $\mathcal{E}_4,$ we apply the Condition (A2) and establish that
\begin{equation}
\frac{1}{\|\xnew\|_2}\sqrt{\sum_{l=1}^{L} (\widehat{\xnew^{\intercal}b^{(l)}})^2}\lesssim \sqrt{L}\left(\sqrt{\frac{\log n}{n}}+\frac{\left|\xnew^{\intercal}b^{(l)}\right|}{\|\xnew\|_2}\right)\leq C,
\label{eq: constant upper bound}
\end{equation} 
for some positive constant $C>0.$ 
For a finite $L$, ${\rm Vol}(L(L+1)/2)$ and $z_{\alpha_0/[L(L+1)]}$ are bounded from above. If $N_{\T}\gtrsim\max\{n,p\}$, $s\log p/n\rightarrow 0$ and Condition (A2) holds, we apply \eqref{eq: variance 1} and show that 
\begin{equation}
n \cdot \lambda_i({\bf V}) \lesssim n \cdot \|{\bf V}\|_{\infty}\lesssim 1, \quad \text{and}\quad d_0\lesssim 1.
\label{eq: constant upper bound 2}
\end{equation}
 Hence, if $N_{\T}\gtrsim\max\{n,p\}$ and $\lambda_{\min}(\Gamma^{\T})+\delta\gg \sqrt{\log p/\min\{n,N_{\T}\}}$, we establish \eqref{eq: length bound ridge} by combining \eqref{eq: inter length} with \eqref{eq: lambda minimum bound}, \eqref{eq: constant upper bound} and \eqref{eq: constant upper bound 2}.
%

\section{Proofs of Extra Lemmas}
\label{sec: additional proof}

\subsection{Proof of Lemma \ref{lem: bias general}}
\label{sec: bias general}
On the event $\mathcal{G}_1\cap \mathcal{G}_6(\widehat{b}^{(l)}_{init}-b^{(l)},\widehat{b}^{(l)}_{init}-b^{(l)},\sqrt{\log p})$, we have 
$\frac{1}{|B|}\sum_{i\in B}[(X^{\T}_{i})^{\intercal}(\widehat{b}^{(l)}_{init}-b^{(l)})]^2\lesssim \frac{\|{b}^{(l)}\|_0 \log p}{n_l}\sigma_l^2.$
Then we have 
\begin{align*}
\left|(\widehat{b}_{init}^{(l)}-b^{(l)})^{\intercal}\widehat{\Sigma}^{\T}(\widehat{b}_{init}^{(k)}-b^{(k)})\right| \leq \frac{1}{|B|}\|X^{\T}_{B}(\widehat{b}_{init}^{(l)}-b^{(l)})\|_2\|X^{\T}_{B}(\widehat{b}_{init}^{(k)}-b^{(k)})\|_2
 \lesssim \sqrt{\frac{\|b^{(l)}\|_0 \|b^{(k)}\|_0(\log p)^2}{{\nl n_k}}}
\end{align*}
and establish \eqref{eq: error bound 1}. We decompose 
\begin{equation}
\begin{aligned}
&(\widehat{\Sigma}^{\T}\widehat{b}_{init}^{(k)}-\widehat{\Sigma}^{(l)} \widehat{u}^{(l,k)})^{\intercal}(\widehat{b}_{init}^{(l)}-b^{(l)})\\
&=(\widetilde{\Sigma}^{\T}\widehat{b}_{init}^{(k)}-\widehat{\Sigma}^{(l)} \widehat{u}^{(l,k)})^{\intercal}(\widehat{b}_{init}^{(l)}-b^{(l)})+[\widehat{b}_{init}^{(k)}]^{\intercal}(\widehat{\Sigma}^{\T}-\widetilde{\Sigma}^{\T})^{\intercal}(\widehat{b}_{init}^{(l)}-b^{(l)}).
\end{aligned}
\label{eq: bias decomp}
\end{equation}
Regarding the first term of \eqref{eq: bias decomp}, we apply H{\"o}lder's inequality and establish
\begin{equation*}
\left|(\widetilde{\Sigma}^{\T}\widehat{b}_{init}^{(k)}-\widehat{\Sigma}^{(l)} \widehat{u}^{(l,k)})^{\intercal}(\widehat{b}_{init}^{(l)}-b^{(l)})\right|\leq \|\widetilde{\Sigma}^{\T}\widehat{b}_{init}^{(k)}-\widehat{\Sigma}^{(l)} \widehat{u}^{(l,k)}\|_{\infty}\|\widehat{b}_{init}^{(l)}-b^{(l)}\|_1.
\end{equation*}
By the optimization constraint \eqref{eq: constraint 1}, on the event $\mathcal{G}_2,$ we have
\begin{equation}
\left|(\widetilde{\Sigma}^{\T}\widehat{b}_{init}^{(k)}-\widehat{\Sigma}^{(l)} \widehat{u}^{(l,k)})^{\intercal}(\widehat{b}_{init}^{(l)}-b^{(l)})\right|\lesssim\|\omega^{(k)}\|_2 \sqrt{\frac{\log p}{\nl}} \cdot \|b^{(l)}\|_0 \sqrt{\frac{ \log p}{\nl}}.
\label{eq: term 1 bound}
\end{equation}
Regarding the second term of \eqref{eq: bias decomp}, conditioning on $\widehat{b}_{init}^{(k)}$ and $\widehat{b}_{init}^{(l)},$ on the event $\mathcal{G}_6(\widehat{b}_{init}^{(k)},\widehat{b}_{init}^{(l)}-b^{(l)},\sqrt{\log p}),$ we have 
$
\left|[\widehat{b}_{init}^{(k)}]^{\intercal}(\widehat{\Sigma}^{\T}-\widetilde{\Sigma}^{\T})^{\intercal}(\widehat{b}_{init}^{(l)}-b^{(l)})
\right|\lesssim\frac{\sqrt{\log p}}{\sqrt{N_{\T}}}\|\widehat{b}_{init}^{(k)}\|_2\|\widehat{b}_{init}^{(l)}-b^{(l)}\|_2.
$
On the event $\mathcal{G}_1,$ we further have 
$
\left|[\widehat{b}_{init}^{(k)}]^{\intercal}(\widehat{\Sigma}^{\T}-\widetilde{\Sigma}^{\T})^{\intercal}(\widehat{b}_{init}^{(l)}-b^{(l)})
\right|\lesssim\|\widehat{b}^{(k)}_{init}\|_2\sqrt{\frac{\|b^{(l)}\|_0 (\log p)^2}{\nl N_{\T}}}.
$
Combined with \eqref{eq: term 1 bound}, we establish \eqref{eq: error bound 2}.
We establish  \eqref{eq: error bound 3} through applying the similar argument for \eqref{eq: error bound 2} by exchanging the role of $l$ and $k.$ Together with \eqref{eq: control of event 1}, \eqref{eq: control of event 2} and \eqref{eq: control of event 3} with $t=\sqrt{\log p},$ we establish the lemma.

\subsection{Proof of Lemma \ref{lem: event lemma 1}}
\label{sec: event lemma 1}

For ${\bf V}_{\pi(l_1,k_1),\pi(l_2,k_2)}$ defined in \eqref{eq: cov def}, we express it as 
\begin{equation}
{\bf V}_{\pi(l_1,k_1),\pi(l_2,k_2)}={\bf V}^{(a)}_{\pi(l_1,k_1),\pi(l_2,k_2)}+{\bf V}^{(b)}_{\pi(l_1,k_1),\pi(l_2,k_2)}
\label{eq: cov decomp}
\end{equation}
where ${\bf V}^{(a)}_{\pi(l_1,k_1),\pi(l_2,k_2)}$ and ${\bf V}^{(b)}_{\pi(l_1,k_1),\pi(l_2,k_2)}$ defined in \eqref{eq: cov def a} and \eqref{eq: cov def b}, respectively.


The control of the event $\mathcal{E}_1$ follows from the following high probability inequalities:
with probability larger than $1-\exp(-cn)-\min\{N_{\T},p\}^{-c}$ for some positive constant $c>0,$ 
\begin{equation}
n\cdot \left|\widehat{\V}^{(a)}_{\pi(l_1,k_1),\pi(l_2,k_2)}-\V^{(a)}_{\pi(l_1,k_1),\pi(l_2,k_2)}\right|\leq C d_0 \left(\frac{s\log p}{n}+\sqrt{\frac{\log p}{n}}\right)\leq \frac{d_0}{4}.
\label{eq: target 1}
\end{equation}
\begin{equation}
N_{\T}\cdot \left|\widehat{\V}^{(b)}_{\pi(l_1,k_1),\pi(l_2,k_2)}-\V^{(b)}_{\pi(l_1,k_1),\pi(l_2,k_2)}\right|\lesssim \log \max\{N_{\T},p\} \sqrt{\frac{s \log p \log N_{\T}}{n}}
+{\frac{\left(\log N_{\T}\right)^{5/2}}{\sqrt{N_{\T}}}}.
\label{eq: target 2}
\end{equation}
The proofs of \eqref{eq: target 1} and \eqref{eq: target 2} are presented in Sections \ref{sec: target 1 proof} and \ref{sec: target 2 proof}, respectively.
 
We combine \eqref{eq: target 1} and \eqref{eq: target 2} and establish 
\begin{equation*}
\begin{aligned}
\|\widehat{\Cov}-\Cov\|_2&\lesssim \max_{(l_1,k_1),(l_2,k_2)\in \mathcal{I}_{L}}\left|\widehat{\Cov}_{\pi(l_1,k_1),\pi(l_2,k_2)}-\Cov_{\pi(l_1,k_1),\pi(l_2,k_2)}\right|\\
&\leq n\cdot \max_{(l_1,k_1),(l_2,k_2)\in \mathcal{I}_{L}}\left|\widehat{\V}^{(a)}_{\pi(l_1,k_1),\pi(l_2,k_2)}-\V^{(a)}_{\pi(l_1,k_1),\pi(l_2,k_2)}\right|\\
&+n\cdot \max_{(l_1,k_1),(l_2,k_2)\in \mathcal{I}_{L}}\left|\widehat{\V}^{(b)}_{\pi(l_1,k_1),\pi(l_2,k_2)}-\V^{(b)}_{\pi(l_1,k_1),\pi(l_2,k_2)}\right|\\
&\leq \frac{d_0}{4}+\frac{\sqrt{n\cdot s}[\log \max\{N_{\T},p\}]^2}{N_{\T}}+\frac{n\cdot (\log N_{\T})^{5/2}}{N_{\T}^{3/2}}\leq d_0/2,
\end{aligned}
\end{equation*}
where the first inequality holds for a finite $L$ and the last inequality follows from Condition (A2).
\subsubsection{Proof of \eqref{eq: target 1}}
\label{sec: target 1 proof}
\begin{equation}
\begin{aligned}
&n\cdot \left|\widehat{\V}^{(a)}_{\pi(l_1,k_1),\pi(l_2,k_2)}-\V^{(a)}_{\pi(l_1,k_1),\pi(l_2,k_2)}\right|\\
&\lesssim \left|\widehat{\sigma}_{l_1}^2-{\sigma}_{l_1}^2\right|(\widehat{u}^{(l_1,k_1)})^{\intercal}\widehat{\Sigma}^{(l_1)} \left[\widehat{u}^{(l_2,k_2)} {\bf 1}(l_2=l_1)+\widehat{u}^{(k_2,l_2)} {\bf 1}(k_2=l_1)\right]\\
&+\left|\widehat{\sigma}_{k_1}^2-{\sigma}_{k_1}^2\right|(\widehat{u}^{(k_1,l_1)})^{\intercal}\widehat{\Sigma}^{(k_1)} \left[\widehat{u}^{(l_2,k_2)} {\bf 1}(l_2=k_1)+\widehat{u}^{(k_2,l_2)} {\bf 1}(k_2=k_1)\right]
\end{aligned}
\end{equation}
Since 
\begin{equation}
\begin{aligned}
&\left|(\widehat{u}^{(l_1,k_1)})^{\intercal}\widehat{\Sigma}^{(l_1)} \left[\widehat{u}^{(l_2,k_2)} {\bf 1}(l_2=l_1)+\widehat{u}^{(k_2,l_2)} {\bf 1}(k_2=l_1)\right]\right|\\
&\leq \sqrt{(\widehat{u}^{(l_1,k_1)})^{\intercal}\widehat{\Sigma}^{(l_1)} \widehat{u}^{(l_1,k_1)}\cdot (\widehat{u}^{(l_1,k_2)})^{\intercal}\widehat{\Sigma}^{(l_1)} \widehat{u}^{(l_1,k_2)}} \\
&+\sqrt{(\widehat{u}^{(l_1,k_1)})^{\intercal}\widehat{\Sigma}^{(l_1)} \widehat{u}^{(l_1,k_1)}\cdot (\widehat{u}^{(l_1,l_2)})^{\intercal}\widehat{\Sigma}^{(l_1)} \widehat{u}^{(l_1,l_2)}}
\end{aligned}
\end{equation}
we have 
$\left|(\widehat{u}^{(l_1,k_1)})^{\intercal}\widehat{\Sigma}^{(l_1)} \left[\widehat{u}^{(l_2,k_2)} {\bf 1}(l_2=l_1)+\widehat{u}^{(k_2,l_2)} {\bf 1}(k_2=l_1)\right]\right|\lesssim n \max_{(l,k)\in \mathcal{I}_{L}}\V^{(a)}_{\pi(l,k),\pi(l,k)}\lesssim d_0.$
Similarly, we have 
$\left|(\widehat{u}^{(k_1,l_1)})^{\intercal}\widehat{\Sigma}^{(k_1)} \left[\widehat{u}^{(l_2,k_2)} {\bf 1}(l_2=k_1)+\widehat{u}^{(k_2,l_2)} {\bf 1}(k_2=k_1)\right]\right|\lesssim d_0.$
Hence, on the event $\mathcal{G}_3,$ we establish \eqref{eq: target 1}.
\subsubsection{Proof of \eqref{eq: target 2}}
\label{sec: target 2 proof}

Define $W_{i,1}=[b^{(l_1)}]^{\intercal}X_{i}^{\T},\; W_{i,2}=[b^{(k_1)}]^{\intercal}X_{i}^{\T},\; W_{i,3}=[b^{(l_2)}]^{\intercal}X_{i}^{\T}, \; W_{i,4}=[b^{(k_2)}]^{\intercal}X_{i}^{\T},$
and 
$\widehat{W}_{i,1}=(\widehat{b}_{init}^{(l_1)})^{\intercal}X_{i}^{\T}, \;  \widehat{W}_{i,2}=(\widehat{b}_{init}^{(k_1)})^{\intercal}X_{i}^{\T},\;  \widehat{W}_{i,3}=(\widehat{b}_{init}^{(l_2)})^{\intercal} X_{i}^{\T}, \;  \widehat{W}_{i,4}=(\widehat{b}_{init}^{(k_2)})^{\intercal}X_{i}^{\T}.$
With the above definitions, we have 
\begin{equation}
\begin{aligned}
&\E [b^{(l_1)}]^{\intercal}X_{i}^{\T}[b^{(k_1)}]^{\intercal}X_{i}^{\T}[b^{(l_2)}]^{\intercal}X_{i}^{\T}[b^{(k_2)}]^{\intercal}X_{i}^{\T}-(b^{(l_1)})^{\intercal}\Sigma^{\T} b^{(k_1)}(b^{(l_2)})^{\intercal}\Sigma^{\T} b^{(k_2)}\\
&=\E \prod_{t=1}^{4}{W}_{i,t}-\E {W}_{i,1}{W}_{i,2}\cdot \E {W}_{i,3}{W}_{i,4}
\end{aligned}
\label{eq: shorthand 1}
\end{equation} 
\begin{equation}
\begin{aligned}
&{\frac{1}{N_{\T}}\sum_{i=1}^{N_{\T}} \left((\widehat{b}_{init}^{(l_1)})^{\intercal} X_{i}^{\T} (\widehat{b}_{init}^{(k_1)})^{\intercal}X_{i}^{\T}(\widehat{b}_{init}^{(l_2)})^{\intercal} X_{i}^{\T} (\widehat{b}_{init}^{(k_2)})^{\intercal}X_{i}^{\T} -(\widehat{b}_{init}^{(l_1)})^{\intercal}\bar{\Sigma}^{\T}\widehat{b}_{init}^{(k_1)}(\widehat{b}_{init}^{(l_2)})^{\intercal}\bar{\Sigma}^{\T}\widehat{b}_{init}^{(k_2)}\right)}\\
&=\frac{1}{N_{\T}}\sum_{i=1}^{N_{\T}}\prod_{t=1}^{4}\widehat{W}_{i,t}-\frac{1}{N_{\T}}\sum_{i=1}^{N_{\T}}\widehat{W}_{i,1}\widehat{W}_{i,2}\cdot \frac{1}{N_{\T}}\sum_{i=1}^{N_{\T}}\widehat{W}_{i,3}\widehat{W}_{i,4}
\end{aligned}
\label{eq: shorthand 2}
\end{equation}
Hence, it is sufficient to control the following terms.
{\small
\begin{equation*}
\frac{1}{N_{\T}}\sum_{i=1}^{N_{\T}}\prod_{t=1}^{4}\widehat{W}_{i,t}-\E \prod_{t=1}^{4}{W}_{i,t}=\frac{1}{N_{\T}}\sum_{i=1}^{N_{\T}}\prod_{t=1}^{4}\widehat{W}_{i,t}-\frac{1}{N_{\T}}\sum_{i=1}^{N_{\T}}\prod_{t=1}^{4}{W}_{i,t}+\frac{1}{N_{\T}}\sum_{i=1}^{N_{\T}}\prod_{t=1}^{4}{W}_{i,t}-\E \prod_{t=1}^{4}{W}_{i,t}
\label{eq: forth moment}
\end{equation*}
\begin{equation*}
\frac{1}{N_{\T}}\sum_{i=1}^{N_{\T}}\widehat{W}_{i,1}\widehat{W}_{i,2}-\E {W}_{i,1}{W}_{i,2}=\frac{1}{N_{\T}}\sum_{i=1}^{N_{\T}}\widehat{W}_{i,1}\widehat{W}_{i,2}-\frac{1}{N_{\T}}\sum_{i=1}^{N_{\T}}{W}_{i,1}{W}_{i,2}+\frac{1}{N_{\T}}\sum_{i=1}^{N_{\T}}{W}_{i,1}{W}_{i,2}-\E {W}_{i,1}{W}_{i,2}
\label{eq: second moment}
\end{equation*}
\begin{equation*}
\frac{1}{N_{\T}}\sum_{i=1}^{N_{\T}}\widehat{W}_{i,3}\widehat{W}_{i,4}-\E {W}_{i,3}{W}_{i,4}=\frac{1}{N_{\T}}\sum_{i=1}^{N_{\T}}\widehat{W}_{i,3}\widehat{W}_{i,4}-\frac{1}{N_{\T}}\sum_{i=1}^{N_{\T}}{W}_{i,3}{W}_{i,4}+\frac{1}{N_{\T}}\sum_{i=1}^{N_{\T}}{W}_{i,3}{W}_{i,4}-\E {W}_{i,3}{W}_{i,4}
\label{eq: second moment copy}
\end{equation*}
}
Specifically, we will show that, with probability larger than $1-\min\{N_{\T},p\}^{-c},$ 
\begin{equation}
\left|\frac{1}{N_{\T}}\sum_{i=1}^{N_{\T}}{W}_{i,1}{W}_{i,2}-\E {W}_{i,1}{W}_{i,2}\right|\lesssim \|b^{(l_1)}\|_2 \|b^{(k_1)}\|_2\sqrt{\frac{\log N_{\T}}{N_{\T}}},
\label{eq: concentration error 1}
\end{equation}
\begin{equation}
\left|\frac{1}{N_{\T}}\sum_{i=1}^{N_{\T}}{W}_{i,3}{W}_{i,4}-\E {W}_{i,3}{W}_{i,4}\right|\lesssim \|b^{(l_2)}\|_2 \|b^{(k_2)}\|_2\sqrt{\frac{\log N_{\T}}{N_{\T}}},
\label{eq: concentration error 2}
\end{equation}
\begin{equation}
\frac{1}{N_{\T}}\sum_{i=1}^{N_{\T}} \left(\prod_{t=1}^{4}W_{i,t}-\E \prod_{t=1}^{4}W_{i,t}\right)\lesssim \|b^{(l_1)}\|_2\|b^{(k_1)}\|_2\|b^{(l_2)}\|_2\|b^{(k_2)}\|_2 {\frac{\left(\log N_{\T}\right)^{5/2}}{\sqrt{N_{\T}}}},
\label{eq: concentration error 3}
\end{equation}
{\small
\begin{equation}
\left|\frac{1}{N_{\T}}\sum_{i=1}^{N_{\T}}\widehat{W}_{i,1}\widehat{W}_{i,2}-\frac{1}{N_{\T}}\sum_{i=1}^{N_{\T}}{W}_{i,1}{W}_{i,2}\right|\lesssim \sqrt{\frac{s \log p}{n}}\left(\sqrt{\log N_{\T}}(\|b^{(l_1)}\|_2+\|b^{(k_1)}\|_2)+\sqrt{\frac{s \log p}{n}}\right),
\label{eq: approximation error 1}
\end{equation}
}
{\small
\begin{equation}
\left|\frac{1}{N_{\T}}\sum_{i=1}^{N_{\T}}\widehat{W}_{i,3}\widehat{W}_{i,4}-\frac{1}{N_{\T}}\sum_{i=1}^{N_{\T}}{W}_{i,3}{W}_{i,4}\right|\lesssim\sqrt{\frac{s \log p}{n}}\left(\sqrt{\log N_{\T}}(\|b^{(l_2)}\|_2+\|b^{(k_2)}\|_2)+\sqrt{\frac{s \log p}{n}}\right).
\label{eq: approximation error 2}
\end{equation}}
If we further assume that $\|b^{(l)}\|_2\leq C$ for $1\leq l\leq L$ and $s^2(\log p)^2/n\leq c$ for some positive constants $C>0$ and $c>0,$ then with probability larger than $1-\min\{N_{\T},p\}^{-c},$ 
\begin{equation}
\left|\frac{1}{N_{\T}}\sum_{i=1}^{N_{\T}}\prod_{t=1}^{4}\widehat{W}_{i,t}-\frac{1}{N_{\T}}\sum_{i=1}^{N_{\T}}\prod_{t=1}^{4}{W}_{i,t}\right|\lesssim \log \max\{N_{\T},p\} \sqrt{\frac{s \log p \log N_{\T}}{n}}.
\label{eq: approximation error 3}
\end{equation}
By the expression \eqref{eq: shorthand 1} and \eqref{eq: shorthand 2}, we establish \eqref{eq: target 2} by applying \eqref{eq: concentration error 1}, \eqref{eq: concentration error 2}, \eqref{eq: concentration error 3}, \eqref{eq: approximation error 1}, \eqref{eq: approximation error 2}, \eqref{eq: approximation error 3}. In the following, we prove 
\eqref{eq: concentration error 1}, \eqref{eq: concentration error 2} and \eqref{eq: concentration error 3}. Then we will present the proofs of \eqref{eq: approximation error 1}, \eqref{eq: approximation error 2}, \eqref{eq: approximation error 3}.

\paragraph{Proofs of \eqref{eq: concentration error 1}, \eqref{eq: concentration error 2} and \eqref{eq: concentration error 3}.}
We shall apply the following lemma to control the above terms, which re-states the Lemma 1 in \cite{cai2011adaptive}. 
\begin{Lemma} 
	\label{lem: fund concentration lemma}
	Let $\xi_1,\cdots,\xi_n$ be independent random variables with mean 0. Suppose that there exists some $c>0$ and $U_{n}$ such that $\sum_{i=1}^{n}\E\xi_{i}^2 \exp\left(c|\xi_i|\right)\leq U_n^2.$ Then for $0<t\leq U_n$, 
$
	\PP\left(\sum_{i=1}^{n} \xi_i\geq C U_n t\right)\leq \exp(-t^2),
$
	where $C=c+c^{-1}.$
\end{Lemma}

Define $W^{0}_{i,1}=\frac{[b^{(l_1)}]^{\intercal}X_{i}^{\T}}{\sqrt{[b^{(l_1)}]^{\intercal}\Sigma^{\T}b^{(l_1)}}}$, $W^{0}_{i,2}=\frac{[b^{(k_1)}]^{\intercal}X_{i}^{\T}}{\sqrt{[b^{(k_1)}]^{\intercal}\Sigma^{\T}b^{(k_1)}}},$
$W^{0}_{i,3}=\frac{[b^{(l_2)}]^{\intercal}X_{i}^{\T}}{{\sqrt{[b^{(l_2)}]^{\intercal}\Sigma^{\T}b^{(l_2)}}}}$ and $W^{0}_{i,4}=\frac{[b^{(k_2)}]^{\intercal}X_{i}^{\T}}{{\sqrt{[b^{(k_2)}]^{\intercal}\Sigma^{\T}b^{(k_2)}}}}.$ Since $X_{i}^{\T}$ is sub-gaussian, $W^0_{i,t}$ is sub-gaussian and both $W^0_{i,1}W^0_{i,2}$ and $W^0_{i,3}W^0_{i,4}$ are sub-exponetial random variables, which follows from Remark 5.18   in \cite{vershynin2010introduction}. By Corollary 5.17 in \cite{vershynin2010introduction}, we have 
\begin{equation*}
\PP\left(\left|\frac{1}{N_{\T}}\sum_{i=1}^{N_{\T}} \left(W^0_{i,1}W^0_{i,2}-\E W^0_{i,1}W^0_{i,2} \right)\right|\geq C\sqrt{\frac{\log N_{\T}}{N_{\T}}}\right)\leq 2 N_{\T}^{-c}
\end{equation*} 
and 
\begin{equation*}
\PP\left(\left|\frac{1}{N_{\T}}\sum_{i=1}^{N_{\T}} \left(W^0_{i,3}W^0_{i,4}-\E W^0_{i,3}W^0_{i,4} \right)\right|\geq C\sqrt{\frac{\log N_{\T}}{N_{\T}}}\right)\leq 2 N_{\T}^{-c}
\end{equation*} 
where $c$ and $C$ are positive constants. The above inequalities imply \eqref{eq: concentration error 1} and \eqref{eq: concentration error 2} after rescaling.

For $1\leq t\leq 4$, since $W^0_{i,t}$ is a sub-gaussian random variable, there exist positive constants $C_1>0$ and $c>2$ such that the following concentration inequality holds,
\begin{equation}
\sum_{i=1}^{N_{\T}}\PP\left(\max_{1\leq t\leq 4}|W^0_{i,t}|\geq C_1\sqrt{\log N_{\T}}\right)\leq N_{\T} \max_{1\leq i\leq N_{\T}}\PP\left(\max_{1\leq t\leq 4}|W^0_{i,t}|\geq C_1\sqrt{\log N_{\T}}\right)\lesssim N_{\T}^{-c}
\label{eq: concentration of subgaussian}
\end{equation}
Define $H_{i,a}=\prod_{t=1}^{4}W^0_{i,t} \cdot \mathbf{1}\left(\max_{1\leq t\leq 4}|W^0_{i,t}|\leq C_1\sqrt{\log N_{\T}}\right)$ for $1\leq t\leq 4,$ and $H_{i,b}=\prod_{t=1}^{4}W^0_{i,t} \cdot \mathbf{1}\left(\max_{1\leq t\leq 4}|W^0_{i,t}|\geq C_1\sqrt{\log N_{\T}}\right)$ for $1\leq t\leq 4.$ Then we have 
\begin{equation}
\frac{1}{N_{\T}}\sum_{i=1}^{N_{\T}} \prod_{t=1}^{4}W^0_{i,t}-\E \prod_{t=1}^{4} W^0_{i,t}=\frac{1}{N_{\T}}\sum_{i=1}^{N_{\T}} \left(H_{i,a}-\E H_{i,a}\right)+\frac{1}{N_{\T}}\sum_{i=1}^{N_{\T}} \left(H_{i,b}-\E H_{i,b}\right)
\label{eq: error decomposition}
\end{equation}
By applying the Cauchy-Schwarz inequality, we bound $\E H_{i,b}$ as
\begin{equation}
\begin{aligned}
\left|\E H_{i,b}\right|&\leq  \sqrt{\E\left( \prod_{t=1}^{4}W^0_{i,t}\right)^2 \PP\left(\max_{1\leq t\leq 4}|W^0_{i,t}|\geq C_1\sqrt{\log N_{\T}}\right)}\\
&\lesssim \PP\left(|W^0_{i,t}|\geq C_1\sqrt{\log N_{\T}}\right)^{1/2}\lesssim N_{\T}^{-1/2},
\end{aligned}
\label{eq: bound of expectation}
\end{equation}
where the second and the last inequalities follow from the fact that $W^0_{i,t}$ is a sub-gaussian random variable.
Now we apply Lemma \ref{lem: fund concentration lemma} to bound $\frac{1}{N_{\T}}\sum_{i=1}^{N_{\T}} \left(H_{i,a}-\E H_{i,a}\right)$. 
By taking $c={c_1}/{\left(C_1^2 \log N_{\T}\right)^2}$ for some small positive constant $c_1>0$,  we have 
$$\sum_{i=1}^{N_{\T}}\E \left(H_{i,a}-\E H_{i,a}\right)^2\exp\left(c \left|H_{i,a}-\E H_{i,a}\right|\right)\leq C \sum_{i=1}^{N_{\T}}\E \left(H_{i,a}-\E H_{i,a}\right)^2 \leq C_2 N_{\T}.$$
By applying Lemma \ref{lem: fund concentration lemma} with $U_n=\sqrt{C_2 N_{\T}}$, $c={c_1}/{\left(C_1^2 \log N_{\T}\right)^2}$ and $t=\sqrt{\log N_{\T}}$, we have 

\begin{equation}
\PP \left(\frac{1}{N_{\T}}\sum_{i=1}^{N_{\T}} \left(H_{i,a}-\E H_{i,a}\right)\geq C {\frac{(\log N_{\T})^{5/2}}{\sqrt{N_{\T}}}}\right)\lesssim N_{\T}^{-c}.
\label{eq: control of the truncated sum}
\end{equation}
Note that 
\begin{equation}
\begin{aligned}
&\PP\left(\left|\frac{1}{N_{\T}}\sum_{i=1}^{N_{\T}} H_{i,b}\right| \geq C {\frac{\left(\log N_{\T}\right)^{5/2}}{\sqrt{N_{\T}}}}\right)\leq \PP\left(\frac{1}{N_{\T}}\sum_{i=1}^{N_{\T}} \left|H_{i,b}\right| \geq C {\frac{\left(\log N_{\T}\right)^{5/2}}{\sqrt{N_{\T}}}}\right)\\
&\leq \sum_{i=1}^{N_{\T}} \PP\left(\left|H_{i,b}\right| \geq C {\frac{\left(\log N_{\T}\right)^{5/2}}{\sqrt{N_{\T}}}}\right)\leq \sum_{i=1}^{N_{\T}}\PP\left(\max_{1\leq t\leq 4}|W^0_{i,t}|\geq C_1\sqrt{\log N_{\T}}\right)\lesssim N_{\T}^{-c}
\end{aligned}
\label{eq: concentration of subgaussian 2}
\end{equation}
where the last inequality follows from \eqref{eq: concentration of subgaussian}.

By the decomposition \eqref{eq: error decomposition}, we have
\begin{equation*}
\begin{aligned}
&\PP\left(\left|\frac{1}{N_{\T}}\sum_{i=1}^{N_{\T}} \left(\prod_{t=1}^{4}W^0_{i,t}-\E \prod_{t=1}^{4}W^0_{i,t}\right)\right|\geq 3C {\frac{\left(\log N_{\T}\right)^{5/2}}{\sqrt{N_{\T}}}}\right)\\
& \leq  \PP \left(\left|\frac{1}{N_{\T}}\sum_{i=1}^{N_{\T}} \left(H_{i,a}-\E H_{i,a}\right)\right|\geq C {\frac{\left(\log N_{\T}\right)^{5/2}}{\sqrt{N_{\T}}}}\right)\\
&+\PP\left(\left|\E H_{i,b}\right| \geq C {\frac{\left(\log N_{\T}\right)^{5/2}}{\sqrt{N_{\T}}}}\right)+\PP\left(\left|\frac{1}{N_{\T}}\sum_{i=1}^{N_{\T}} H_{i,b}\right| \geq C {\frac{\left(\log N_{\T}\right)^{5/2}}{\sqrt{N_{\T}}}}\right)
\lesssim N_{\T}^{-c}.\\
\end{aligned}
\end{equation*}
where the final upper bound follows from \eqref{eq: bound of expectation}, \eqref{eq: control of the truncated sum} and \eqref{eq: concentration of subgaussian 2}.
Hence, we establish that \eqref{eq: concentration error 3} holds with probability larger than $1-N_{\T}^{-c}.$

\paragraph{Proofs \eqref{eq: approximation error 1}, \eqref{eq: approximation error 2} and \eqref{eq: approximation error 3}.}
It follows from the definitions of $\widehat{W}_{i,t}$ and $W_{i,t}$ that 
\begin{equation}
\begin{aligned}
&\frac{1}{N_{\T}}\sum_{i=1}^{N_{\T}}\widehat{W}_{i,1}\widehat{W}_{i,2}-\frac{1}{N_{\T}}\sum_{i=1}^{N_{\T}}{W}_{i,1}{W}_{i,2}=[\widehat{b}_{init}^{(l_1)}-b^{(l_1)}]^{\intercal}\frac{1}{N_{\T}}\sum_{i=1}^{N_{\T}}X_{i}^{\T}[X_{i}^{\T}]^{\intercal}[\widehat{b}_{init}^{(k_1)}-b^{(k_1)}]\\
&+[b^{(l_1)}]^{\intercal}\frac{1}{N_{\T}}\sum_{i=1}^{N_{\T}}X_{i}^{\T}[X_{i}^{\T}]^{\intercal}[\widehat{b}_{init}^{(k_1)}-b^{(k_1)}]+[\widehat{b}_{init}^{(l_1)}-b^{(l_1)}]^{\intercal}\frac{1}{N_{\T}}\sum_{i=1}^{N_{\T}}X_{i}^{\T}[X_{i}^{\T}]^{\intercal}b^{(k_1)}
\end{aligned}
\end{equation}
On the event $\mathcal{G}_2\cap\mathcal{G}_5$ with $\mathcal{G}_2$ defined in \eqref{eq: high prob event 1} and $\mathcal{G}_5$ defined in \eqref{eq: high prob event 2}, we establish \eqref{eq: approximation error 1}. By a similar argument, we establish \eqref{eq: approximation error 2}.
Furthermore, we define the event
\begin{equation}
\begin{aligned}
\mathcal{G}_7=\left\{\max_{1\leq l\leq L}\max_{1\leq i\leq N_{\T}} \left|X_{i}^{\T}{b}^{(l)}\right|\lesssim (\sqrt{C_0}+\sqrt{\log N_{\T}})\|b^{(l)}\|_2\right\}\\
\mathcal{G}_8=\left\{\max_{1\leq i\leq N_{\T}} \|X_{i}^{\T}\|_{\infty}\lesssim (\sqrt{C_0}+\sqrt{\log N_{\T}+\log p})\right\}
\end{aligned}
\end{equation}
It follows from the assumption (A1) that $\PP\left(\mathcal{G}_7\right)\geq 1- N_{\T}^{-c}$ and $\PP\left(\mathcal{G}_8\right)\geq 1-\min\{N_{\T},p\}^{-c}$ for some positive constant $c>0.$ Note that
\begin{equation}
\begin{aligned}
&\frac{1}{N_{\T}}\sum_{i=1}^{N_{\T}}\left|\widehat{W}_{i,1}\widehat{W}_{i,2}-{W}_{i,1}{W}_{i,2}\right|\leq \frac{1}{N_{\T}} \sum_{i=1}^{N_{\T}}\left|[\widehat{b}_{init}^{(l_1)}-b^{(l_1)}]^{\intercal}X_{i}^{\T}[X_{i}^{\T}]^{\intercal}[\widehat{b}_{init}^{(k_1)}-b^{(k_1)}]\right|\\
&+\frac{1}{N_{\T}}\sum_{i=1}^{N_{\T}}\left|[b^{(l_1)}]^{\intercal}X_{i}^{\T}[X_{i}^{\T}]^{\intercal}[\widehat{b}_{init}^{(k_1)}-b^{(k_1)}]\right|+\frac{1}{N_{\T}}\sum_{i=1}^{N_{\T}}\left|[\widehat{b}_{init}^{(l_1)}-b^{(l_1)}]^{\intercal}X_{i}^{\T}[X_{i}^{\T}]^{\intercal}b^{(k_1)}\right|
\end{aligned}
\label{eq: upper decomposition}
\end{equation}
By the Cauchy-Schwarz inequality, we have
\begin{align*}
&\frac{1}{N_{\T}} \sum_{i=1}^{N_{\T}}\left|[\widehat{b}_{init}^{(l_1)}-b^{(l_1)}]^{\intercal}X_{i}^{\T}[X_{i}^{\T}]^{\intercal}[\widehat{b}_{init}^{(k_1)}-b^{(k_1)}]\right|\\
&\leq \frac{1}{N_{\T}} \sqrt{\sum_{i=1}^{N_{\T}}\left([\widehat{b}_{init}^{(l_1)}-b^{(l_1)}]^{\intercal}X_{i}^{\T}\right)^2 \sum_{i=1}^{N_{\T}}\left([X_{i}^{\T}]^{\intercal}[\widehat{b}_{init}^{(k_1)}-b^{(k_1)}]\right)^2}\\
\end{align*}
Hence, on the event $\mathcal{G}_1\cap \mathcal{G}_6(\widehat{b}_{init}^{(k_1)}-b^{(k_1)},\widehat{b}_{init}^{(k_1)}-b^{(k_1)},\sqrt{\log p})\cap  \mathcal{G}_6(\widehat{b}_{init}^{(l_1)}-b^{(l_1)},\widehat{b}_{init}^{(l_1)}-b^{(l_1)},\sqrt{\log p})$,
\begin{equation}
\frac{1}{N_{\T}} \sum_{i=1}^{N_{\T}}\left|[\widehat{b}_{init}^{(l_1)}-b^{(l_1)}]^{\intercal}X_{i}^{\T}[X_{i}^{\T}]^{\intercal}[\widehat{b}_{init}^{(k_1)}-b^{(k_1)}]\right|\lesssim \frac{s\log p}{n}
\label{eq: part 1}
\end{equation}
On the event $\mathcal{G}_7,$ we have 
\begin{align*}
&\frac{1}{N_{\T}}\sum_{i=1}^{N_{\T}}\left|[b^{(l_1)}]^{\intercal}X_{i}^{\T}[X_{i}^{\T}]^{\intercal}[\widehat{b}_{init}^{(k_1)}-b^{(k_1)}]\right|\\
&\lesssim (\sqrt{C_0}+\sqrt{\log N_{\T}})\|b^{(l_1)}\|_2\frac{1}{N_{\T}}\sum_{i=1}^{N_{\T}}\left|[X_{i}^{\T}]^{\intercal}[\widehat{b}_{init}^{(k_1)}-b^{(k_1)}]\right|\\
&\leq (\sqrt{C_0}+\sqrt{\log N_{\T}})\|b^{(l_1)}\|_2\frac{1}{\sqrt{N_{\T}}}\sqrt{\sum_{i=1}^{N_{\T}}\left([X_{i}^{\T}]^{\intercal}[\widehat{b}_{init}^{(k_1)}-b^{(k_1)}]\right)^2}\\
\end{align*}
where the last inequality follows from the Cauchy-Schwarz inequality. Hence, on the event $\mathcal{G}_1\cap  \mathcal{G}_7\cap \mathcal{G}_6(\widehat{b}_{init}^{(k_1)}-b^{(k_1)},\widehat{b}_{init}^{(k_1)}-b^{(k_1)},\sqrt{\log p}),$ we establish 
\begin{equation}
\frac{1}{N_{\T}}\sum_{i=1}^{N_{\T}}\left|[b^{(l_1)}]^{\intercal}X_{i}^{\T}[X_{i}^{\T}]^{\intercal}[\widehat{b}_{init}^{(k_1)}-b^{(k_1)}]\right|\lesssim (\sqrt{C_0}+\sqrt{\log N_{\T}})\|b^{(l_1)}\|_2\sqrt{\frac{s\log p}{n}}.
\label{eq: part 2}
\end{equation}
Similarly, we establish 
$$\frac{1}{N_{\T}}\sum_{i=1}^{N_{\T}}\left|[\widehat{b}_{init}^{(l_1)}-b^{(l_1)}]^{\intercal}X_{i}^{\T}[X_{i}^{\T}]^{\intercal}b^{(k_1)}\right|\lesssim (\sqrt{C_0}+\sqrt{\log N_{\T}})\|b^{(k_1)}\|_2\sqrt{\frac{s\log p}{n}}.
$$
Combined with \eqref{eq: upper decomposition}, \eqref{eq: part 1} and \eqref{eq: part 2}, we establish 
\begin{equation}
\frac{1}{N_{\T}}\sum_{i=1}^{N_{\T}}\left|\widehat{W}_{i,1}\widehat{W}_{i,2}-{W}_{i,1}{W}_{i,2}\right|\lesssim \left(\sqrt{\log N_{\T}}(\|b^{(l_1)}\|_2+\|b^{(k_1)}\|_2)+\sqrt{\frac{s\log p}{n}}\right)\sqrt{\frac{s\log p}{n}}
\label{eq: absolute bound 1}
\end{equation}
Similarly, we establish  
\begin{equation}
\frac{1}{N_{\T}}\sum_{i=1}^{N_{\T}}\left|\widehat{W}_{i,3}\widehat{W}_{i,4}-{W}_{i,3}{W}_{i,4}\right|\lesssim \left(\sqrt{\log N_{\T}}(\|b^{(l_2)}\|_2+\|b^{(k_2)}\|_2)+\sqrt{\frac{s\log p}{n}}\right)\sqrt{\frac{s\log p}{n}}
\label{eq: absolute bound 2}
\end{equation}
Define 
$
{H}_{i,1}={W}_{i,1}{W}_{i,2}, {H}_{i,2}={W}_{i,3}{W}_{i,4}, \widehat{H}_{i,1}=\widehat{W}_{i,1}\widehat{W}_{i,2}$  and $\widehat{H}_{i,2}=\widehat{W}_{i,3}\widehat{W}_{i,4}.
$
Then we have 
\begin{equation}
\begin{aligned}
&\frac{1}{N_{\T}}\sum_{i=1}^{N_{\T}}\prod_{t=1}^{4}\widehat{W}_{i,t}-\frac{1}{N_{\T}}\sum_{i=1}^{N_{\T}}\prod_{t=1}^{4}{W}_{i,t}=\frac{1}{N_{\T}}\sum_{i=1}^{N_{\T}}\widehat{H}_{i,1}\widehat{H}_{i,2}-\frac{1}{N_{\T}}\sum_{i=1}^{N_{\T}}{H}_{i,1}{H}_{i,2}\\
&=\frac{1}{N_{\T}}\sum_{i=1}^{N_{\T}}\left(\widehat{H}_{i,1}-{H}_{i,1}\right){H}_{i,2}+\frac{1}{N_{\T}}\sum_{i=1}^{N_{\T}}\left(\widehat{H}_{i,2}-{H}_{i,2}\right){H}_{i,1}+\frac{1}{N_{\T}}\sum_{i=1}^{N_{\T}}\left(\widehat{H}_{i,1}-{H}_{i,1}\right)\left(\widehat{H}_{i,2}-{H}_{i,2}\right)\\
\end{aligned}
\label{eq: error bound decomposition}
\end{equation}
On the event $\mathcal{G}_7,$ we have 
\begin{equation}
\left|{H}_{i,1}\right|\lesssim({C_0}+{\log N_{\T}})\|b^{(l_1)}\|_2\|b^{(k_1)}\|_2
\quad \text{and}\quad 
\left|{H}_{i,2}\right|\lesssim({C_0}+{\log N_{\T}})\|b^{(l_2)}\|_2\|b^{(k_2)}\|_2
\label{eq: bounded verification}
\end{equation}
On the event $\mathcal{G}_7\cap\mathcal{G}_8,$ we have 
\begin{equation}
\begin{aligned}
&\left|\widehat{H}_{i,2}-{H}_{i,2}\right|\leq \left|[\widehat{b}_{init}^{(l_2)}-b^{(l_2)}]^{\intercal}X_{i}^{\T}[X_{i}^{\T}]^{\intercal}[\widehat{b}_{init}^{(k_2)}-b^{(k_2)}]\right|\\
&+\left|[b^{(l_2)}]^{\intercal}X_{i}^{\T}[X_{i}^{\T}]^{\intercal}[\widehat{b}_{init}^{(k_2)}-b^{(k_2)}]\right|+\left|[\widehat{b}_{init}^{(l_2)}-b^{(l_2)}]^{\intercal}X_{i}^{\T}[X_{i}^{\T}]^{\intercal}b^{(k_2)}\right|\\
&\lesssim (C_0+\log N_{\T}+\log p) \left(s^2\frac{\log p}{n}+s\sqrt{\frac{\log p}{n}}\|b^{(k_2)}\|_2+s\sqrt{\frac{\log p}{n}}\|b^{(l_2)}\|_2\right)
\end{aligned}
\label{eq: single upper decomposition}
\end{equation}
By the decomposition \eqref{eq: error bound decomposition}, we combine \eqref{eq: bounded verification}, \eqref{eq: single upper decomposition}, \eqref{eq: absolute bound 1} and \eqref{eq: absolute bound 2} and establish 
\begin{equation}
\begin{aligned}
&\left|\frac{1}{N_{\T}}\sum_{i=1}^{N_{\T}}\prod_{t=1}^{4}\widehat{W}_{i,t}-\frac{1}{N_{\T}}\sum_{i=1}^{N_{\T}}\prod_{t=1}^{4}{W}_{i,t}\right|\\
&\leq({C_0}+{\log N_{\T}})\|b^{(l_2)}\|_2\|b^{(k_2)}\|_2\left(\sqrt{\log N_{\T}}(\|b^{(l_1)}\|_2+\|b^{(k_1)}\|_2)+\sqrt{\frac{s\log p}{n}}\right)\sqrt{\frac{s\log p}{n}}\\
&+({C_0}+{\log N_{\T}})\|b^{(l_1)}\|_2\|b^{(k_1)}\|_2\left(\sqrt{\log N_{\T}}(\|b^{(l_2)}\|_2+\|b^{(k_2)}\|_2)+\sqrt{\frac{s\log p}{n}}\right)\sqrt{\frac{s\log p}{n}}\\
&+\left(\sqrt{\log N_{\T}}(\|b^{(l_1)}\|_2+\|b^{(k_1)}\|_2)+\sqrt{\frac{s\log p}{n}}\right)\sqrt{\frac{s\log p}{n}}\\
&\cdot(C_0+\log N_{\T}+\log p) \left(s^2\frac{\log p}{n}+s\sqrt{\frac{\log p}{n}}\|b^{(k_2)}\|_2+s\sqrt{\frac{\log p}{n}}\|b^{(l_2)}\|_2+\|b^{(l_2)}\|_2\|b^{(k_2)}\|_2\right)
\end{aligned}
\end{equation}
If we further assume that $\|b^{(l)}\|_2\leq C$ for $1\leq l\leq L$ and $s^2(\log p)^2/n\leq c$ for some positive constants $C>0$ and $c>0,$ then we establish \eqref{eq: approximation error 3}.

\section{Additional Numerical Results}
\label{sec: additional num}

We consider additional settings to evaluate the finite-sample performance of our proposed method. The results for these extra settings are similar to those presented in Section \ref{sec: sim} in the main paper. Our proposed CIs achieve the desired coverage level and the intervals become shorter with a larger $n$ or $\delta$.
\subsection{Additional Simulation Results}
\label{sec: additional simulation}
\paragraph{Setting 2 with covariate shift and a higher dimension.} Set $L=2$. $b^{(1)}_{1:500}$ and $b^{(2)}_{1:500}$ are the same as setting 1, except for $b^{(1)}_{498}=0.5$, $b^{(1)}_j = -0.5$ for $j=499, 500,$ and $b^{(2)}_{500}=1$. Set $b^{(1)}_{j}=b^{(2)}_{j}=0$ for $501\leq j\leq p.$
$[{\xnew}]_{j}=1$ for $498\leq j\leq 500$, and $[{\xnew}]_{j}=0$ otherwise.
$\Sigma^{\T}_{i,i}=1.5$ for $1\leq i\leq p$, $\Sigma^{\T}_{i,j}=0.9$ for $1\leq i\neq j\leq 5,$ $\Sigma^{\T}_{i,j}=0.9$ for $499\leq i\neq j\leq 500$ and $\Sigma^{\T}_{i,j}=\Sigma_{i,j}$ otherwise.

\begin{figure}[H]
\centering
\includegraphics[width=0.9\linewidth]{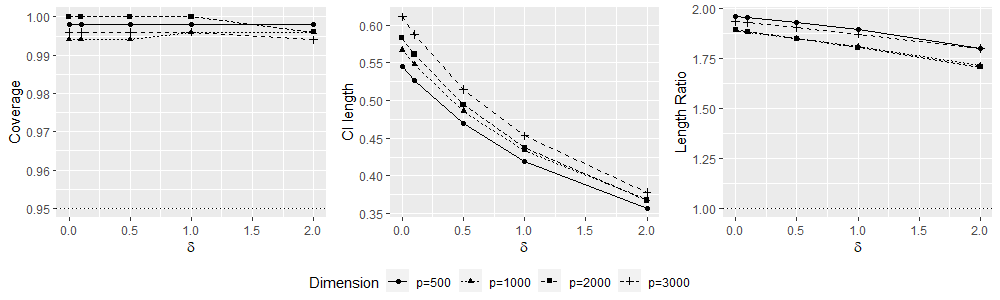}
\caption{\small Dependence on $\delta$ and $p$: setting 2 (covariate shift) with $n=500$. ``Coverage" and ``CI Length" stand for the empirical coverage and the average length of our proposed CI, respectively; ``Length Ratio" represents the ratio of the average length of our proposed CI to the normality CI in \eqref{eq: normality CI}.}
\label{fig: setting 2 higher dim}
\end{figure}

 In Figure \ref{fig: setting 2 higher dim}, we have explored our proposed method for a lager $p$ value and our proposed CIs are still valid for $p=1000,2000$ and $3000.$

\vspace{-10mm}

\begin{figure}[H]
\centering
\begin{subfigure}[b]{0.9\textwidth}
\centering
    \includegraphics[scale=0.6]{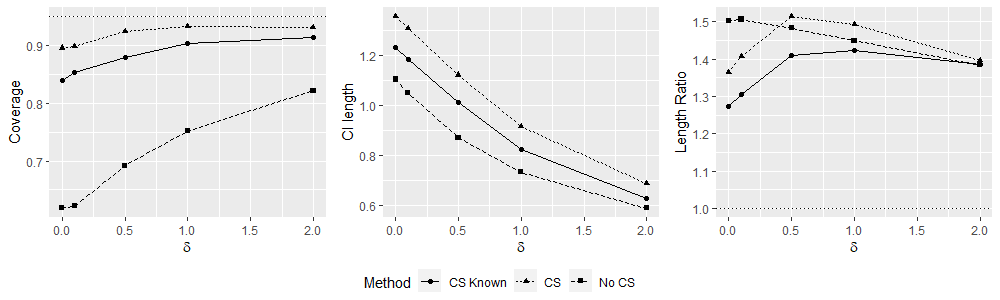}
   \caption{setting 3(a) with covariate shift}
\end{subfigure}
\begin{subfigure}[b]{0.9\textwidth}
\centering
    \includegraphics[scale=0.6]{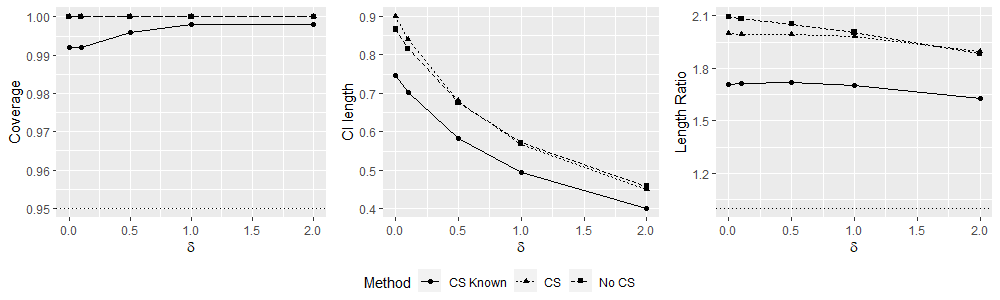}
  \caption{setting 3(b) with no covariate shift}
\end{subfigure}
\caption{\small Comparison of covariate shift and no covariate shift algorithms with $n=200$. {``CS Known", ``CS" and ``No CS" represent Algorithm \ref{algo: SAR} with known $\Sigma^{\T}$,  Algorithm \ref{algo: SAR}  with covariate shift but unknown $\Sigma^{\T}$, and Algorithm \ref{algo: SAR}  with no covariate shift, respectively.}
 ``Coverage" and ``CI Length" stand for the empirical coverage and the average length of our proposed CI, respectively; ``Length Ratio" represents the ratio of the average length of our proposed CI to the normality CI in \eqref{eq: normality CI}.}
\label{fig: method comparison n=200}
\end{figure}

\paragraph{Setting 3 with $n=200$.} We report the comparison of the covariate-shift algorithm and no covariate-shift algorithm for setting 3 with $n=200$ in Figure \ref{fig: method comparison n=200}. 

\vspace{-5mm}

\paragraph{Setting 4 with varying $L$.} Vary $L$ across $\{2,5,10\}$, denoted as (4a), (4b) and (4c), respectively. $b^{(1)}_j = j/40$ for $1\leq j \leq 10$, $b^{(1)}_{498} = 0.5$, $b^{(1)}_j=-0.5$ for $j=499,500$, and $b^{(1)}_j=0$ otherwise.
For $2\leq l \leq L$, $b^{(l)}_{10\cdot l+j}=b^{(1)}_j$ for $1\leq j \leq 10$ and $b^{(l)}_j = b^{(1)}_j/2^{l-1}$ for $j=498\leq j \leq 500$, and $b^{(l)}_j=0$ otherwise. 
$[{\xnew}]_{j}=1$ for $498\leq j\leq 500$, and $[{\xnew}]_{j}=0$ otherwise; 
$\Sigma^{\T}_{i,i}=1.5$ for $1\leq i\leq 500$, $\Sigma^{\T}_{i,j}=0.9$ for $1\leq i\neq j\leq 5$ and $499\leq i\neq j\leq 500,$ and $\Sigma^{\T}_{i,j}=\Sigma_{i,j}$ otherwise.

\begin{figure}[H]
    \centering
     \begin{subfigure}[b]{0.9\textwidth}
     \centering
        \includegraphics[scale=0.6]{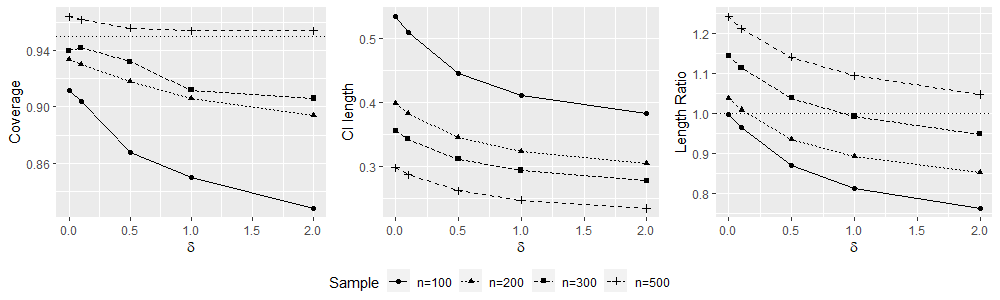}
        \caption{setting 4(a) with $L=2$}
    \end{subfigure}
     \begin{subfigure}[b]{0.9\textwidth}
     \centering
        \includegraphics[scale=0.6]{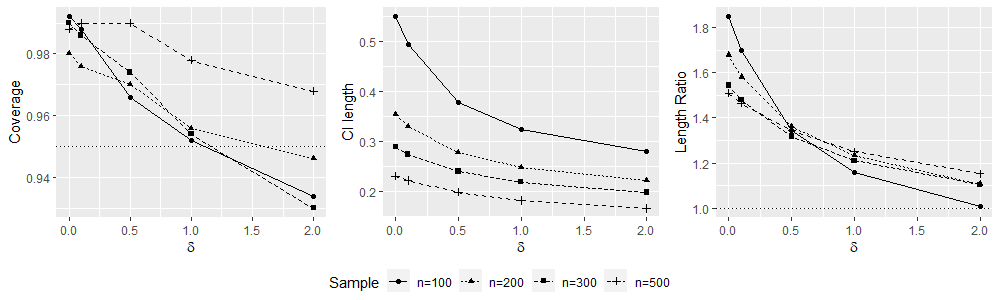}
        \caption{setting 4(b) with $L=5$}
    \end{subfigure}    \begin{subfigure}[b]{0.9\textwidth}
       \centering
        \includegraphics[scale=0.6]{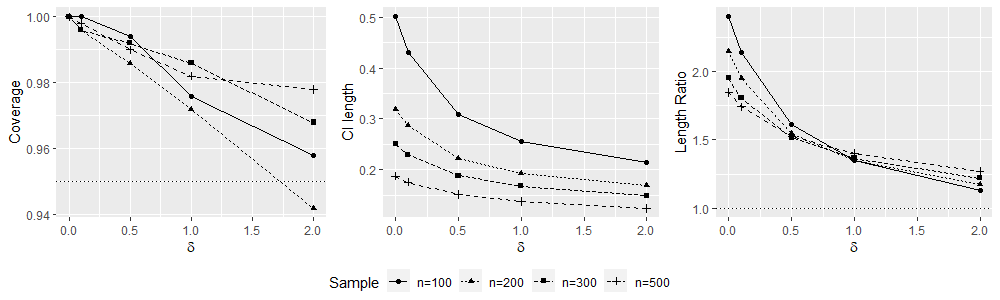}
        \caption{setting 4(c) with $L=10$}
    \end{subfigure}
    \caption{\small Dependence on $\delta$ and $n$. ``Coverage" and ``CI Length" stand for the empirical coverage and the average length of our proposed CI, respectively; ``Length Ratio" represents the ratio of the average length of our proposed CI to the normality CI in \eqref{eq: normality CI}.}
   \label{fig: setting 4}
\end{figure}


\paragraph{Setting 5 with coefficient perturbation.} 
We consider the no covariate shift setting with $L=2$. Set
$b^{(1)}_j=j/40$ for $1\leq j\leq 10,$ $b^{(1)}_j=(10-j)/40$ for $11\leq j\leq 20$, $b^{(1)}_j=0.2$ for $j=21$, $b^{(1)}_j=1$ for $j=22,23$;
$b^{(2)}_j=b^{(1)}_j+{\rm perb}/\sqrt{300}$ for $1\leq j\leq 10,$ $b^{(2)}_j=0$ for $11\leq j\leq 20$, $b^{(2)}_{j}=0.5$ for $j=21$, $b^{(2)}_j=0.2$ for $j=22,23$ .
We vary the values of ${\rm perb}$ across $\{1,1.125,1.25,1.5,3.75,4,5,7,10,12\}.$
$\xnew_{j}=j/5$ for $1\leq j\leq 5$.

\vspace{-3mm}
\begin{figure}[H]
\centering
\includegraphics[width=0.9\linewidth]{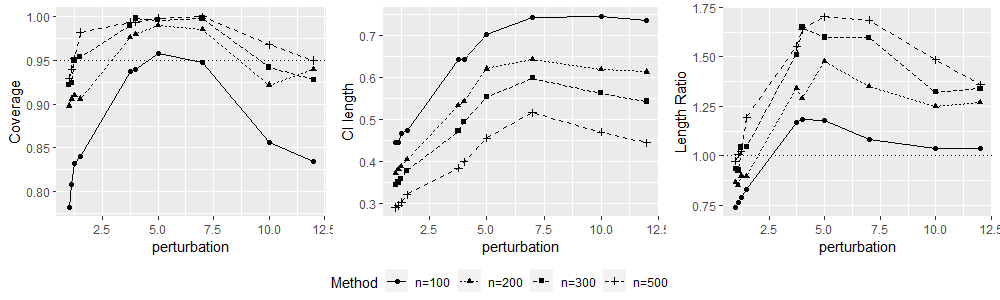}
\caption{\small Dependence on the value ${\rm perb}$ for setting 5. ``Coverage" and ``CI Length" stand for the empirical coverage and the average length of our proposed CI, respectively; ``Length Ratio" represents the ratio of the average length of our proposed CI to the normality CI in \eqref{eq: normality CI}. }
\label{fig: setting 5}
\end{figure}

\vspace{-10mm}

\begin{figure}[htp!]
\centering
\begin{subfigure}[b]{\textwidth}
    \includegraphics[width=0.9\linewidth]{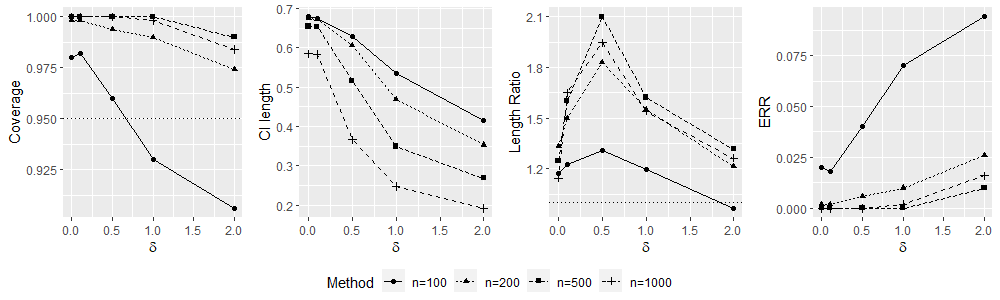}
   \caption{setting6(a)}
\end{subfigure}
\begin{subfigure}[b]{\textwidth}
    \includegraphics[width=0.9\linewidth]{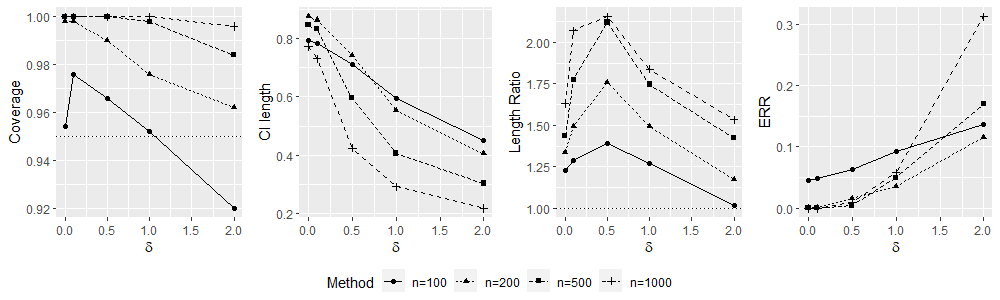}
  \caption{setting6(b)}
\end{subfigure}
\caption{\small Setting 6 with opposite effects. ``Coverage" and ``CI Length" stand for the empirical coverage and the average length of our proposed CI, respectively; ``Length Ratio" represents the ratio of the average length of our proposed CI to the normality CI in \eqref{eq: normality CI}; ``ERR" represents the empirical rejection rate out of 500 simulations.}
\label{fig: opposite effect}
\end{figure}

\paragraph{Setting 6 with opposite effects.} We investigate the cancellation of opposite effects in Figure \ref{fig: opposite effect}. We consider two no covariate shift settings with $L=2$: (6a) $b^{(l)}$ for $1\leq l \leq 2$ are the same as setting 1, except for $b^{(1)}_j=0$ for $j=499$, $b^{(1)}_j=0.2$ for $j=500$, $b^{(2)}_j=-0.2$ for $j=500$. 
$[{\xnew}]_{j}=1$ for $j=500$; (6b) Same as 6(a) except for $b^{(2)}_{j}=-0.4$ for $j=500$.
Since $b^{(1)}_{500}$ and $b^{(2)}_{500}$ have opposite signs, the maximin effect is zero for $\delta=0$.  We use the Empirical Rejection Rate (ERR) to denote the proportion of rejecting the null hypothesis out of 500 simulations. In Figure \ref{fig: opposite effect}, we observe that ERR is below 5\% for $\delta=0$, which indicate that the corresponding maximin effect is not significant.

\vspace{-2mm}

\vspace{-2.5mm}

\subsection{Instability Measure: dependence on $\delta$}
\label{sec: delta choice sim}

\vspace{-2.5mm}

Recall the stability measure $\mathbb{I}(\delta)$ is introduced in Section \ref{sec: ridge-type} in the main paper, \begin{equation*}
\mathbb{I}(\delta)=\frac{\sum_{m=1}^{M}\|\widehat{\gamma}^{\m}_{\delta}-\widehat{\gamma}_{\delta}\|^2_2}{\sum_{m=1}^{M} \|\widehat{\Gamma}^{\m}-\widehat{\Gamma}^{\T}\|^2_2},\label{eq: instability measure}
\end{equation*}
where a larger $\mathbb{I}(\delta)$ indicates an unstable integration. In Table \ref{tab: complete instability}, we report the instability measure $\mathbb{I}(\delta)$ for all simulation settings except for setting 5, which is not included since it consists of more than 10 subsettings. We observe settings 1 and 6 are settings with instability. For both settings, the penalty $\delta$ is instrumental in decreasing the instability measure.  For settings 2, 3, and 4, the small instability measure $\mathbb{I}(\delta)$ indicates that the standard maximin effect (without adding the ridge penalty) is already a stable integration. 
\begin{table}[H]
\centering
\begin{tabular}[t]{|c|c|c|c|c|c|c||c|}
\hline
setting & $L$&$\mathbb{I}(0)$ & $\mathbb{I}(0.1)$ & $\mathbb{I}(0.5)$ & $\mathbb{I}(1)$ & $\mathbb{I}(2)$ & $\Gamma^{\T}_{11}+\Gamma^{\T}_{22}-2\Gamma^{\T}_{12}$\\
\hline
1 & 2&5.464 & 1.966 & 0.264 & 0.072 & 0.019 & 0.026\\
\hline
2 & 2& 0.023 & 0.021 & 0.014 & 0.010 & 0.006 & 4.635\\
\hline
3(a) &2& 0.094 & 0.082 & 0.044 & 0.023 & 0.010 & 1.935\\
\hline
3(b) &2& 0.058 & 0.050 & 0.029 & 0.017 & 0.008 & 2.810\\
\hline
4(a) & 2&0.108 & 0.087 & 0.044 & 0.024 & 0.011 & 2.007\\
\hline
4(b) &5& 0.076 & 0.059 & 0.027 & 0.014 & 0.006 & -\\
\hline
4(c) &10& 0.052 & 0.039 & 0.016 & 0.008 & 0.003 & -\\
\hline
6(a) &2& 3.305 & 1.449 & 0.221 & 0.065 & 0.018 & 0.160\\
\hline
6(b) & 2&1.451 & 0.816 & 0.168 & 0.056 & 0.017 & 0.360\\
\hline
\end{tabular}
\caption{The instability measure $\mathbb{I}(\delta)$ for $\delta\in \{0,0.1,0.5,1,2\}$. The reported values are averaged over 100 repeated simulations. For the column indexed with $\Gamma^{\T}_{11}+\Gamma^{\T}_{22}-2\Gamma^{\T}_{12}$, we only report their values for $L=2$ since $\Gamma^{\T}_{11}+\Gamma^{\T}_{22}-2\Gamma^{\T}_{12}$ is only a measure of instability for $L=2$. }
\label{tab: complete instability}
\end{table}

\subsection{Sample Splitting Comparison}
\label{sec: sample splitting}
We compare the algorithm with and without sample splitting and report the results in Table \ref{tab: splitting comparison}. For the sample splitting algorithm, we split the samples into two equal size sub-samples. For $n=100,$ no sample splitting algorithm is slightly under-coverage (the empirical coverage level is still above 90\%). When $n\geq 200,$ both the algorithm with and without sample splitting achieve the desired coverage levels. As expected, the CIs with sample splitting are longer than those without sample splitting. In Table \ref{tab: splitting comparison}, under the column indexed with ``Length ratio", we report the ratio of the average length of CI with sample splitting to that without sample splitting.

\begin{table}[H]
\centering
\begin{tabular}[t]{|c|c|c c|c c|c|}
\hline
\multicolumn{1}{|c|}{ } & \multicolumn{1}{c|}{ } & \multicolumn{2}{c|}{Coverage} & \multicolumn{2}{c|}{Length} & \multicolumn{1}{c|}{} \\
\hline
$\delta$ & $n$ & Splitting & No Splitting & Splitting & No Splitting & Length ratio\\
\hline
 & 100 & 0.978 & 0.920 & 1.921 & 1.062 & 1.809\\
\cline{2-7}
 & 200 & 0.994 & 0.972 & 1.618 & 0.898 & 1.802\\
\cline{2-7}
 & 300 & 0.994 & 0.990 & 1.336 & 0.789 & 1.693\\
\cline{2-7}
\multirow{-4}{*}{\centering\arraybackslash 0.0} & 500 & 1.000 & 0.992 & 0.999 & 0.651 & 1.534\\
\cline{1-7}
 & 100 & 0.980 & 0.922 & 1.904 & 1.023 & 1.861\\
\cline{2-7}
 & 200 & 0.994 & 0.972 & 1.567 & 0.863 & 1.816\\
\cline{2-7}
 & 300 & 0.994 & 0.990 & 1.286 & 0.759 & 1.695\\
\cline{2-7}
\multirow{-4}{*}{\centering\arraybackslash 0.1} & 500 & 1.000 & 0.992 & 0.952 & 0.629 & 1.514\\
\cline{1-7}
 & 100 & 0.982 & 0.918 & 1.814 & 0.890 & 2.038\\
\cline{2-7}
 & 200 & 0.996 & 0.972 & 1.373 & 0.748 & 1.836\\
\cline{2-7}
 & 300 & 0.996 & 0.992 & 1.099 & 0.666 & 1.651\\
\cline{2-7}
\multirow{-4}{*}{\centering\arraybackslash 0.5} & 500 & 1.000 & 0.994 & 0.800 & 0.559 & 1.433\\
\cline{1-7}
 & 100 & 0.986 & 0.914 & 1.639 & 0.769 & 2.131\\
\cline{2-7}
 & 200 & 0.998 & 0.968 & 1.155 & 0.655 & 1.764\\
\cline{2-7}
 & 300 & 0.996 & 0.992 & 0.911 & 0.589 & 1.546\\
\cline{2-7}
\multirow{-4}{*}{\centering\arraybackslash 1.0} & 500 & 1.000 & 0.994 & 0.681 & 0.499 & 1.364\\
\cline{1-7}
 & 100 & 0.986 & 0.902 & 1.291 & 0.630 & 2.049\\
\cline{2-7}
 & 200 & 0.996 & 0.968 & 0.871 & 0.549 & 1.587\\
\cline{2-7}
 & 300 & 0.998 & 0.988 & 0.706 & 0.499 & 1.415\\
\cline{2-7}
\multirow{-4}{*}{\centering\arraybackslash 2.0} & 500 & 0.998 & 0.992 & 0.549 & 0.426 & 1.290\\
\hline
\end{tabular}
\caption{Comparison of algorithm with/without sample-splitting in setting 2 with $p=500.$}
\label{tab: splitting comparison}
\end{table}

\subsection{Additional results for real data analysis.}
\begin{table}[H]
\centering
\resizebox{0.7\linewidth}{!}{
\begin{tabular}[t]{|c|c|c|c|c|c|c|}
\hline
SNP index & 420 & 423 & 424 & 437 & 442 & 443\\
\hline
Gene & KRE33 & TCB2 & IMP4 & HPF1 & SPO21 & ATP19\\
\hline
\end{tabular}}
\caption{Gene names for several SNPs with maximin significant effects.}
\label{table: gene names}
\end{table}

\putbib
\end{bibunit}

\end{document}